\documentclass[fleqn,usenatbib]{mnras}

\usepackage{array, makecell}

\usepackage[T1]{fontenc}
\usepackage{ae,aecompl}
\usepackage{grffile}

\usepackage{longtable}
\usepackage{color}
\usepackage[table]{xcolor}

\newcommand{\Eq}[1]{Eq.~(\ref{#1})}

\newcommand{\Fig}[1]{Fig.~\ref{#1}}
\newcommand{\Figs}[1]{Figs.~\ref{#1}}
\newcommand{\Sec}[1]{\S\ref{#1}}
\newcommand{\Tab}[1]{Table~\ref{#1}}

\newcommand{\jzd}[1]{}

\newcommand{\rew}{\hbox{$EW_0$}}
\newcommand{\rewmgii}{\hbox{$EW_0^{\lambda 2796}$}}%
\newcommand{\incl}{i}

\newcommand{\NeV}{\textsc{[{\rm Ne}\kern 0.1em{\sc v}}]}
\newcommand{\NII}{\textsc{[{\rm N}\kern 0.1em{\sc ii}}]}
\newcommand{\OIII}{\textsc{[{\rm O}\kern 0.1em{\sc iii}}]}
\newcommand{\OII}{\textsc{[{\rm O}\kern 0.1em{\sc ii}}]}
\newcommand{\OI}{\textsc{{\rm O}\kern 0.1em{\sc i}}}
\newcommand{\MgI}{\textsc{{\rm Mg}\kern 0.1em{\sc i}}}
\newcommand{\MgII}{\textsc{{\rm Mg}\kern 0.1em{\sc ii}}}
\newcommand{\FeII}{\textsc{{\rm Fe}\kern 0.1em{\sc ii}}}

\newcommand{\HI}{\textsc{{\rm H}\kern 0.1em{\sc i}}}
\newcommand{\HII}{\textsc{{\rm H}\kern 0.1em{\sc ii}}}
\newcommand{\lya}{\textsc{{\rm Ly}\kern 0.1em$\alpha$}}
\newcommand{\Ly}{\textsc{{\rm Ly}\kern 0.1em$\alpha$}}
\newcommand{\Ha}{\textsc{{\rm H}\kern 0.1em$\alpha$}}
\newcommand{\Hb}{\textsc{{\rm H}\kern 0.1em$\beta$}}
\newcommand{\Hg}{\textsc{{\rm H}\kern 0.1em$\gamma$}}
\newcommand{\SII}{\hbox{[{\rm S}\kern 0.1em{\sc ii}}]}
\newcommand{\Ne}{\hbox{[{\rm Ne}\kern 0.1em{\sc v}}]}
\newcommand{\kms}{\hbox{km~s$^{-1}$}}
\newcommand{\kpc}{\hbox{kpc}}
\newcommand{\uerglf}{\mathrm{erg}\,\mathrm{s}^{-1}\,\mathrm{cm}^{-2}}
\newcommand{\uerglum}{\mathrm{erg}\,\mathrm{s}^{-1}}
\newcommand{\flux}{\uerglf}
\newcommand{\mpy}{\hbox{M$_{\odot}$~yr$^{-1}$}}

\definecolor{GalColJ0103}{HTML}{FF0000} 
\definecolor{GalColJ0145}{HTML}{FFBC00} 
\definecolor{GalColJ0800}{HTML}{85FF00} 
\definecolor{GalColJ1039}{HTML}{00FF37} 
\definecolor{GalColJ1107}{HTML}{00FFF3} 
\definecolor{GalColJ1236}{HTML}{004FFF} 
\definecolor{GalColJ1358}{HTML}{339CFF} 
\definecolor{GalColJ1509}{HTML}{6e00FF} 
\definecolor{GalColJ2152}{HTML}{FF00D4} 

\newcommand{\Colgale}{cyan}

\newcommand{\Colgali}{violet}

\newcommand{\vr}{v_\mathrm{r}}
\newcommand{\vrot}{v_\mathrm{\phi}}
\newcommand{\vcirc}{v_\mathrm{circ}}
\newcommand{\vvir}{v_\mathrm{vir}}

\newcommand{\rvir}{r_\mathrm{vir}}

\newcommand{\Rgal}{R}
\newcommand{\vlos}{v_\mathrm{los}}

\newcommand{\vmax}{v_\mathrm{max}}
\newcommand{\ebv}{E(B-V)}

\newcommand{\Mvir}{M_\mathrm{vir}}
\newcommand{\Mvirabund}{M_\mathrm{vir;abund.}}
\newcommand{\rhalf}{r_{\rm half}}
\newcommand{\rturn}{r_{\rm turn}}

\newcommand{\msun}{M$_{\odot}$}

\newcommand{\gpk}{\emph{$GalPak^{3D}$}}{}
\newcommand{\mfl}{MEGAFLOW}{}
\newcommand{\sext}{\emph{SExtractor}}{}
\newcommand{\galfit}{\emph{GALFIT}}{}
\newcommand{\camel}{\emph{CAMEL}}{}
\newcommand{\mpdaf}{{\it MPDAF}}{}
\newcommand{\pampelmuse}{{\it Pampelmuse}}{}

\newcommand{\fluxLimit}{4\times10^{-18}$~$\flux}

\newcommand{\EWThresTot}{0.3}

\newcommand{\NabsTot}{79} 
\newcommand{\NabsNoneMore}{59}
\newcommand{\NabsNoneMorePercent}{75}
\newcommand{\NabsNone}{41}
\newcommand{\NabsNtwo}{10}
\newcommand{\NabsNtwoPrim}{4}
\newcommand{\NabsNoneTwo}{51}

\newcommand{\nprimary}{45}{} 
{} 
{}
\newcommand{\stnprimarynprobustalpha}{Two}{} 
\newcommand{\stnprimaryDiffFlagThreeFive}{Four}{} 
{} 
\newcommand{\nsample}{9}{}
\newcommand{\nsampletext}{nine}{}
\newcommand{\cnsampletextNTwo}{Two}{}

\newcommand{\colcorot}{salmon}{}
\newcommand{\colantirot}{light-blue}{}

\newcommand{\figskin}{\Fig{fig:img_and_kin:J0145p1056_0554} and \Figs{fig:img_and_kin:J0103p1332_0788} to \ref{fig:img_and_kin:J2152p0625_1053} of the Supplementary Appendix}{}

\newcommand{\figsspec}{\Fig{fig:spec:J0145p1056_0554} and \Figs{fig:img_and_kin:J0103p1332_0788} to \ref{fig:img_and_kin:J2152p0625_1053} of the Supplementary Appendix}{}

\newcommand{\figrotcurvesanduves}{\Fig{fig:rotcurve_vs_abs} and \Fig{fig:all_rotcurve_vs_abs} of the Supplementary Appendix}{}

\usepackage{graphicx}	
\usepackage{amsmath}	
\usepackage{amssymb}	

\title[Gas accretion in MEGAFLOW]{MusE GAs FLOw and Wind (MEGAFLOW) II. A study of gas accretion around $z\approx1$ star-forming galaxies with background quasars~\thanks{Based on observations made with ESO Telescopes at the La Silla Paranal Observatory under programme IDs 
094.A-0211,
095.A-0365,
096.A-0609, 
096.A-0164,
097.A-0138,
097.A-0144, 
098.A-0216,
098.A-0310, 
099.A-0059,
293.A-5038 
}}

\author[J. Zabl et al.]{
Johannes Zabl,$^{1,2}$\thanks{E-mail: johannes.zabl@univ-lyon1.fr}
Nicolas F. Bouch\'e,$^{1,2}$ 
Ilane Schroetter,$^{3,1}$ 
Martin Wendt,$^{4,5}$
\newauthor
Hayley Finley,$^{6,1}$
Joop Schaye,$^{7}$
Simon Conseil,$^{2}$
Thierry Contini,$^{1}$
Raffaella A. Marino,$^{8}$
\newauthor
Peter Mitchell,$^{2}$ 
Sowgat Muzahid,$^{7}$
Gabriele Pezzulli,$^{8}$
Lutz Wisotzki$^{5}$
 \\
$^{1}$ Institut de Recherche en Astrophysique et Plan\'etologie (IRAP), Universit\'e de Toulouse, CNRS, UPS, F-31400 Toulouse, France \\
$^{2}$ Univ Lyon, Univ Lyon1, Ens de Lyon, CNRS, Centre de Recherche Astrophysique de Lyon UMR5574, F-69230 Saint-Genis-Laval, France\\
$^{3}$ GEPI, Observatoire de Paris, CNRS-UMR8111, PSL Research University, Univ. Paris Diderot, 5 place Jules Janssen, 92195 Meudon, France \\
$^{4}$ Institut f\"ur Physik und Astronomie, Universit\"at Potsdam, Karl-Liebknecht-Str. 24/25, 14476 Golm, Germany \\
$^{5}$ Leibniz-Institut f\"ur Astrophysik Potsdam (AIP), An der Sternwarte 16, 14482 Potsdam, Germany \\
$^{6}$  Stockholm University, Department of Astronomy and Oskar Klein Centre for Cosmoparticle Physics, AlbaNova, University Centre\\
\hspace{1Ex} SE-10691, Stockholm, Sweden \\
$^{7}$  Leiden Observatory, Leiden University, PO Box 9513, NL-2300 RA Leiden, the Netherlands\\
$^{8}$ Department of Physics, ETH Z\"urich,Wolfgang-Pauli-Strasse 27, 8093 Z\"urich, Switzerland \\
}

\date{Accepted XXX. Received YYY; in original form ZZZ}

\pubyear{2018}

\begin{document}
\label{firstpage}
\pagerange{\pageref{firstpage}--\pageref{lastpage}}
\maketitle

\begin{abstract}
    We use the MusE GAs FLOw and Wind (MEGAFLOW) survey to study the kinematics
    of extended disk-like structures of cold gas around $z\approx1$ star-forming
    galaxies. The combination of VLT/MUSE and VLT/UVES observations allows us
    to connect the kinematics of the gas measured through \MgII{} quasar
    absorption spectroscopy to the kinematics and orientation of the associated
    galaxies constrained through integral field spectroscopy. Confirming
    previous results, we find that the galaxy-absorber pairs of the MEGAFLOW
    survey follow a strong bimodal distribution, consistent with a picture of
    \MgII{} absorption being predominantly present in outflow cones and
    extended disk-like structures. This allows us to select a {\it bona-fide} sample
    of galaxy-absorber pairs probing these disks for impact parameters of $10\mbox{--}70\,\kpc$. We test the hypothesis that
    the disk-like gas is co-rotating with the galaxy disks, and find that for 7
    out of 9 pairs the absorption velocity shares the sign of the disk
    velocity, disfavouring random orbits. We further show that the data are
    roughly consistent with inflow velocities and angular momenta predicted by
simulations, and that the corresponding mass accretion rates are sufficient to
balance the star formation rates. \end{abstract}




\section{Introduction}

A number of arguments (theoretical and observational) indicate that galaxies
cannot be closed boxes with an {\it ab-initio} fixed reservoir of gas. Indeed,
numerical simulations show that galaxies grow from the accretion of cool
intergalactic gas (via the cosmic web), a process most efficient in galaxies
with luminosities lower than $L^*$
\citep{White:1991a,Birnboim:2003a,Keres:2005a,Dekel:2006a,Dekel:2009a,VandeVoort:2011a,LHuillier:2012a}
owing to the short cooling times in these halos \citep{Rees:1977a,Silk:1977a}.
Observationally, a number of indirect arguments support the notion that
galaxies need to continuously replenish their gas, implying that they are
continuously fed by the  accretion of gas from the intergalactic medium (IGM),
as reviewed in \citet{FoxBook2017a}. 

Originally, the most common indirect argument comes from the G-dwarf problem
\citep{van-den-Bergh:1962a,Schmidt:1963a}, which says that the metallicity
distribution of G stars in the solar neighbourhood  is not consistent with the
closed-box expectations, and the data can be reconciled with chemical models
provided that there is a significant amount of metal poor infall
\citep{Larson:1972b,Larson:1972a,Lynden-Bell:1975a,Pagel:1975a,Casuso:2004a}.
Another common indirect argument relies on the observed short  {gas depletion
time-scales} ($\equiv M_{\rm gas}/$SFR), seen in local and distant galaxies to
be typically 0.5--2~Gyr
\citep[e.g.][]{Daddi:2010a,FreundlichJ:2013a,Genzel:2015a,Tacconi:2010a,Tacconi:2013a,Tacconi:2018a,ScovilleN:2016a,ScovilleN:2017a,SchinnererE:2016a,SaintongeA:2013a,SaintongeA:2016a,SaintongeA:2017a}.
These short depletion times imply that the observed amount of gas available is
too low to sustain their star formation rate (SFR) for more than a few Gyr,
i.e. it is not enough to support the galaxies' future star-formation.  A third
indirect argument comes from  the slow decline of the cosmic HI density as a
function of redshift \citep[e.g.][]{Peroux:2003a,Neeleman:2016a} tied to the
gas content of galaxies \citep{WongT:2002a}.

As mentioned, in low mass galaxies hosted by halos below the virial shock mass
threshold ($M_{h} \lesssim 10^{11-12} M_\odot$)  gas accretion is expected to
be very efficient
\citep[e.g.][]{White:1991a,Birnboim:2003a,Keres:2005a,FaucherGiguere:2011a,Nelson:2015a,Correa:2018a}.
Once inside the galaxy dark matter halo, the accreted gas is expected to orbit
the galaxy, bringing along not just fuel for star formation but also angular
momentum
\citep[e.g.][]{Stewart:2011a,Stewart:2013a,Stewart:2017a,Danovich:2015a}.  In
this context, the accreting material coming from the large-scale filamentary
structure should form a warped, extended  gaseous structure
\citep[e.g.][]{Pichon:2011a,Kimm:2011a,ShenS:2013a,Danovich:2015a,Stewart:2017a},
which co-rotates with the central disk and is sometimes referred to as a
`cold-flow disk' \citep{Stewart:2011a,Stewart:2013a,Stewart:2017a}.  

This `cold-flow disk' scenario leads to large gaseous  ($T\sim10^{4}$~K)
structures, which could in part become the large disks  often seen around
galaxies in \HI\ 21cm surveys  and extending 2--3 times beyond the stellar
radius \citep[e.g.][]{Bosma:1981a,PutmanM:2009a, BigielF:2010a, KreckelK:2011a,
Wang:2016a, Ianjamasimanana:2018a}. 

At higher redshifts, the `cold-flow disk' scenario is expected to lead to
distinct signatures in absorption systems with  $N_{\HI}$ of $10^{17}$ to
$10^{21}$ cm$^{-2}$ seen in background quasar sight-lines
\citep{Dekel:2009a,Kimm:2011a,Fumagalli:2011a,Stewart:2011b,Stewart:2013a,Goerdt:2012a,vandeVoort:2012a}.
In particular, some of the infalling gas kinematics is expected to be offset
from the galaxy's systemic velocity when observed in absorption along the
sight-lines of background quasars \citep{Stewart:2011b}, because the gas is
partly rotationally supported.

These expected signatures are testable against observations with suitably
located background sources such as quasars \citep{Barcons:1995a, Steidel:2002a,
    Chen:2005a, Kacprzak:2010a, Kacprzak:2011a, Bouche:2013a, Turner:2014a,
    Bouche:2016a,Ho:2017a, Rahmani:2018a}, bright galaxies \citep{
DiamondStanic:2016a}, or directly in redshifted
absorption lines in galaxy spectra ({\it down-the-barrel}; \citealt{Coil:2011a,
Rubin:2012a, Martin:2012a}; for review see \citealt{Rubin:2017b}).  

Among background sources, background galaxies are more numerous, but their
usefulness is usually limited by the typically low S/N unless one reverts to a
stacking approach as in \citet{Bordoloi:2011a}.  By contrast,  background
quasars are rarer, but allow one to obtain more informations, such as the gas
location from the host, gas ionization properties
\citep[e.g.][]{Muzahid:2015a,Lehner:2016a,Prochaska:2017a} and most importantly
the gas kinematics \citep[e.g.][]{Barcons:1995a,Steidel:2002a, Kacprzak:2010a,
Bouche:2013a, Bouche:2016a, Ho:2017a}.  Among those, \citet{Ho:2017a}
demonstrated the existence of co-rotating structures at $z\approx0.2$ in a
sample of half-dozen galaxies, a step forward from the individual analyses of
\citet{Bouche:2013a} and \citet{Bouche:2016a}

Progress in sample size has been slow in spite of decades of research with
galaxy-quasar pairs, as studies investigating the connections between the host
galaxy kinematics and the low-ionization absorption line kinematics were
limited to $\sim 50$ pairs \citep[see][for a recent review]{Kacprzak:2017a}.
Less than half of these have orientations favourable to study extended gas
disks (accretion cases) \citep{Barcons:1995a, Steidel:2002a, Chen:2005a,
    Kacprzak:2010a, Kacprzak:2011a, Bouche:2013a,Bouche:2016a,Ho:2017a,
Rahmani:2018a}.

Thanks to the MUSE (Multi Unit Spectroscopic Explorer;
\citealt{Bacon:2006a,Bacon:2010a}) instrument on the VLT (Very Large Telescope)
with its unprecedented field-of-view (1'$\times$1') and sensitivity, the
situation is about to change significantly by taking advantage of the
combination of MUSE kinematics and high-resolution UVES (Ultraviolet and Visual
Echelle Spectrograph; \citealt{Dekker:2000a}) data. Indeed, we recently started
the MuseE GAs FLOw and Wind (\mfl{}) survey (Bouch{\'e} et al. in prep),
which consists of MUSE$+$UVES observations of 22 quasar fields, each with
multiple (three or more) strong ($>0.5$--0.8\AA) \MgII\ absorbers at redshifts
$0.3<z_{\rm abs}<1.5$ selected from the JHU-SDSS catalog \citep{Zhu:2013a}.
The \mfl{} survey leads to one of the largest surveys of \MgII{}
absorber-galaxy pairs with spectroscopic and kinematic information, with about
80$+$ galaxy-quasar pairs suitable to study either outflows \citep[as
in][hereafter Paper I]{Schroetter:2016a} or accretion \citep[as in][this
work]{Bouche:2016a} depending on the apparent location of the quasar with
respect to the galaxy major-axis.

In this paper, we present results on \nsampletext{} galaxy-quasar pairs
suitable for characterising the kinematics of accreting gas, while the wind
cases will be presented in a companion paper (Schroetter et al. in prep). After
briefly introducing the \mfl{} survey (\Sec{sect:megaflow}), we discuss in
\Sec{sect:data} the observation and reduction strategy both for the MUSE and
UVES data.  The selection of the \nsampletext{} galaxy-quasar pairs of this
study from the $\approx 80$ absorbers pairs in the \mfl{} survey is discussed
in \Sec{sec:sample}.  Then, we infer kinematical and physical properties of the
selected galaxies and their host halos in \Sec{sec:gal_parameters}. As the main
result of our work, we compare galaxy to absorber kinematics in
\Sec{sec:result}, with a focus on testing for co-rotation and potential radial
infall. 

Throughout this paper, we use the $\Lambda$CDM standard cosmological
parameters: $H_0=70$~\kms, $\Omega_{\Lambda} = 0.7$, and $\Omega_{\rm m}=0.3$.
All distances are proper. Further, we assume a \citet{Chabrier:2003a} stellar
Initial Mass Function (IMF) and all stated magnitudes are on the AB system
\citep{Oke:1974a}.  When we refer in the following to \OII{} without additional
wavelength qualifier, we refer to the \OII{}$\lambda\lambda 3727, 3729$
doublet. The prefix `pseudo' in pseudo-filter and pseudo-spectrum refers to the
fact that these were created from the MUSE data cube. All stated uncertainties
are 68\% confidence intervals.  The \nsampletext{} galaxy-absorber pairs can be
identified throughout by uniquely assigned colours.

\section {The Megaflow survey}
\label{sect:megaflow}

\subsection{Motivation}

Since the initial work of \citet{BergeronJ:1988a,Bergeron:1991a}  and
\citet{SteidelC:1995b,Steidel:2002a}, there is a well established association
between the cool ($\mathrm{T}\sim10^4$ K) component of the CGM traced by the
low-ionization \MgII\ doublet seen in absorption in background quasar spectra
and star-forming galaxies.  Large samples of galaxy-quasar pairs are rare and
difficult to construct owing to the difficulty in  finding the host galaxy
responsible for the \MgII\ absorption, which is often a painstaking process
requiring deep imaging (preferably from the {\it Hubble Space Telescope} (HST))
and multi-object spectroscopy, with the added problems of the quasar point
spread function (PSF) blocking the view directly along the line-of-sight.  One
of the largest samples of \MgII\ selected galaxy-quasar pairs with
morphological data is the MAGIICAT sample
\citep{Churchill:2013a,Nielsen:2013a,Nielsen:2013b,Nielsen:2015a,Nielsen:2016a},
which consists of 123 isolated foreground galaxies with associated \MgII{}
detections at $0.07 \leq z \leq 1.1$.

Surveys of galaxy-quasar pairs such as the MAGIICAT sample suffer from two
major limitations, namely that they must rely on photometric pre-selection
(i.e. suffer from redshift incompleteness) and that they lack kinematical
information on the host galaxies. Both of these limitations must be overcome
using expensive follow-up spectroscopic campaigns. This can be partially
by-passed with integral field unit (IFU) surveys as described in
\citet{Bouche:2017a} because IFU surveys can simultaneously (i) locate the host
galaxy; (ii) determine the host photometric and kinematics properties; (iii)
determine the host morphological properties in most cases; and (iv) allow for
proper PSF subtraction in case of small impact parameters.

\subsection{The survey}

With the field-of-view ($1\arcmin\times1\arcmin$), sensitivity, and wavelength coverage of
the VLT/MUSE instrument ($\sim4800\,\text{\AA}\mbox{--}9300\,\text{\AA}$),
building large samples of absorber-galaxy pairs is now feasible with only a
handful of observing nights.  In particular, we started the \mfl{} survey of
22 quasar fields, which aims at building a sample of $\sim100$
galaxy-quasar pairs.  In order to reach this goal, we selected quasars from the
JHU-SDSS \MgII\ absorber catalogue \citep{Zhu:2013a} which had at least three
(or more) \MgII\ absorbers within the redshift range from 0.4 to 1.5, suitable
for \OII\ based identification of star-forming galaxies in the MUSE wavelength
range.\footnote{\OII{} can be observed with MUSE starting from $z\approx0.3$, but the JHU-SDSS \MgII{} catalog does not include absorbers below z=0.4.} In addition, we imposed that the rest-frame equivalent width of
$\MgII\,\lambda2796$, \rewmgii, of the three required absorbers be greater than
$0.5 \textnormal{\AA}$, with a preference given to sight-lines with multiple
$\rewmgii> 0.8 \textnormal{\AA}$ absorbers.  The restriction on \MgII{} rest
equivalent width $\rewmgii> 0.5 \textnormal{\AA}$ was chosen because the host
galaxy is then expected to be within $\approx$100~kpc of the quasar
line-of-sight, i.e. matching the field-of-view of MUSE, given the well known
anti-correlation between the impact parameter and \rewmgii\
\citep{LanzettaK:1990a,Bergeron:1991a,SteidelC:1995b,Bordoloi:2011a,Chen:2010a,Nielsen:2013a,Werk:2013a}.
A slightly less stringent equivalent width threshold of $\rewmgii> 0.3
\textnormal{\AA}$ is often used in the literature to separate strong \MgII{}
absorbers from weak \MgII{} absorbers \citep[e.g.][]{Churchill:1999a}. Our 22
quasar sight-lines serendipitously include several (10) absorbers with $0.3
\textnormal{\AA} < \rewmgii < 0.5\textnormal{\AA}$ in the right redshift range.
We included these absorbers in the analysis. Even so their equivalent widths
are slightly below our survey threshold, the galaxy-absorber association for
absorbers of this strength is still expected to be sufficiently robust.  In
total, the 22 quasar sight-lines contain 79 strong $\MgII$ absorbers with
$\rewmgii>0.3 \textnormal{\AA}$ with $0.51<z<1.45$.

\section{Data}

\label{sect:data}

Each quasar field was observed with MUSE and each quasar was followed up with
the high-resolution spectrograph UVES at the VLT.  

\subsection{MUSE observations}

We observed all 22 quasar fields with the VLT/MUSE instrument over the period
September 2014 to May 2017\footnote{Observations are still ongoing. This was
the period of data used for our first full analysis of the data, which is the
basis for this paper and a companion paper discussing outflows (Schroetter et
al. in prep).} as part of guaranteed time observations (GTO).  A full
description of the data for all 22 fields will be given in a future paper
describing the full survey (Bouch\'e et al., in prep.).  Briefly, all except
two fields were observed for at least 2hr, i.e. the resulting exposure times
are $>6$k sec.  Observation details are listed in \Tab{tab:obs}.

\subsubsection{Data reduction}
\label{sec:muse:data_reduction}

We reduced the data using the ESO MUSE pipeline version v1.6
\citep{WeilbacherP:2012a,WeilbacherP:2014a,WeilbacherP:2016a}. 
First, each individual exposure was processed by the {\it scibasic} recipe to
produce a table (hereafter called pixtable) containing relative locations,
wavelength, counts, and an estimate of the variance. This recipe removes the
instrumental signatures by applying daily calibrations (lamp flat-fields, bias,
twilight-flat illumination corrections) and calibrates the wavelength scale
(based on daily arc-lamps). Further, {\it scibasic} also applies the geometric
solution (determined once per GTO run) for each of the 24 IFUs. Bad pixels
corresponding to known CCD defects (columns or pixels) are also masked at this
time. For each exposure we also used an `illumination' exposure, which are short
flats, to correct for flux variations on the slices due to small temperature
changes between the daily calibration exposures and the science exposures.

Second, the individual pixtables were flux-calibrated (using the response from
daily standards), telluric corrected (using a telluric absorption estimate from
the flux-standard), sky-subtracted, astrometrically calibrated, and resampled
onto a cube (using the drizzle algorithm) with the pipeline's {\it scipost}
recipe.  However, clear variations of the residual background level between
individual slices were visible in white-light images created from the cube,
caused by imperfections from the flat-fielding/illumination correction. To
mitigate these imperfections we used a self-calibration strategy, as in
\citet{Bacon:2015a, BaconR:2017a}, which is conceptually similar to the {\it
CubeFix} method developed by Cantalupo (in prep., see also
\citealt{Borisova:2016a} and \citealt{Marino:2018a}). Essentially, it consists
of normalizing the background in all slices to the overall background level.

In practice \footnote{The current version of the MUSE DRS (v2.4), which was not
    available at the time when we reduced the data for this work, has the
    self-calibration directly implemented. The steps described in this
    paragraph would no longer be necessary when using DRS v2.4.}, we were using
    the `selfcalibrate' method in the python package \mpdaf\  (MUSE Python
    Data Analysis Framework) v2.3dev~\footnote{Available at
\url{https://git-cral.univ-lyon1.fr/MUSE/mpdaf}.} \citep{Piqueras:2017a}.
 This method computes the
multiplicative corrections necessary to bring each slice to the reference
background level, which is determined by the mean sky background across the
field.  Consequently, it requires as input a `positioned'\footnote{A
    `\emph{pixtable\_positioned}' is a pixtable where the spatial position
information for each pixel is given in absolute R.A. and Dec.} pixtable with the
sky subtraction turned off, and an object mask.  We used \sext{} on the white
light images (as described above) to produce the object masks and we reran the
scipost from the 24 scibasic pixtables to produce a `positioned' scipost
pixtable per exposure.  In this rerun of scipost, sky subtraction and
correction for barycentric velocity were switched off.  Because the
self-calibration does successfully remove the slice-to-slice variations but
fails to remove the sharp flat-field imperfections visible at the edges of the
IFUs, we simply masked the affected regions in the {\it scibasic} pixtables used
as input.

After performing the self-calibration, we resampled the corrected positioned
pixtables to datacubes using again the {\it scipost} recipe. Here, we performed
the sky subtraction, barycentric correction, and use the same 3D output world
coordinate system (WCS) grid for each of the cubes. We then used the
software-package {\it Zurich Atmosphere Purge} ({\it ZAP})
\citep{SotoK:2016a,SotoK:2016b} to remove skyline residuals from each datacube,
which makes use of a principal component analysis PCA
analysis~\footnote{Available at \url{https://github.com/musevlt/zap}.} using
{\it cftype=`fit'} using an improved object mask created from
the white-light pseudo-images. 
After manual inspection of the individual cubes and masking of visible
satellites tracks, we combined the cubes. For those fields where the seeing
between individual exposures was strongly differing, we weighted each exposure
with the inverse of the full width at half maximum (FWHM) of the PSF.

\subsubsection{Data characterization}
\label{sec:data:quality}

In order to assess the image quality, we measured the PSF on the quasar itself
in  the combined cubes by fitting an elliptical 2D Moffat profile
\citep{Moffat:1969a}.  Due to the large wavelength range covered by the MUSE
data (from 4800 to 9300\,\AA), we measured the PSF as a function of wavelength
using   $100\,\textnormal{\AA}$ wide pseudo-filter images   at five different
wavelengths separated by $1000\,\textnormal{\AA}$, and interpolated these
measurements for other wavelengths.  We first performed the PSF measurement on
each of these images using a  Moffat profile with $\beta$ set to 2.5.  The
Moffat FWHM values at $7050\,\textnormal{\AA}$ range from 0\arcsec{}.53 to
0\arcsec{}.98 across the 22 fields, with a median value of 0\arcsec{}.76.
Second, we also determined the wavelength dependence of the PSF  with the
\pampelmuse{} code \citep{Kamann:2013a} using a Moffat profile with the $\beta$
parameter free. Overall, the difference between the fixed-$\beta$  values and
the free-$\beta$ \pampelmuse{} values are different by a median of 5\% and at
most 14\%.

In order to obtain a realistic estimate for our sensitivity to \OII{} emitters,
we estimated the $5\sigma$ point source detection limit in a pseudo-NB filter
with an appropriate width of $400\,\kms$. A filter width of $400\,\kms$ gives
the optimal S/N for the  $\OII{}\,\lambda\lambda 3727,3729$ doublet when
assuming a line-width of $\textnormal{FWHM}\approx50\,\kms$. In the spatial
direction, we assumed a circular detection aperture with radius of
$1.5\times$FWHM$_\text{Moffat}$.  This aperture size gives the optimal S/N for
a point source convolved with a Moffat PSF with $\beta=2.5$ in the background
limited case.  By using an estimate for the per-pixel noise and scaling it to
the number of pixels spanned by the assumed spatial size and filter width, we
derived an estimate for the total noise within the aperture.  Subsequently, we
multiplied this noise estimate by $1/0.52$ in order to correct for aperture
losses both in the spatial and the wavelength directions. 

The wavelength dependent per-pixel noise was estimated from the pipeline's
variance map of a cube in source free regions. Using this estimate we derive a
typical $5 \sigma$ \OII{} detection limit
$\sim4\times10^{-18}\times(FWHM_\text{Moffat}/0\arcsec.6)\times(\text{T}_{\rm
exp}/6ks)^{-0.5}\,\uerglf$ in MUSE's most sensitive wavelength region around
$7000\,\text{\AA}$, which corresponds to a \OII{} redshift of $z\approx0.9$.
This derived \OII{} flux limit corresponds to an unobscured SFR limit of
$0.07\,\mpy$ (cf.  \Eq{eq:sfr_oii}).  The line flux sensitivity both
short-wards and long-wards of this wavelength decreases somewhat, with the ends
of the relevant wavelength range having about a factor 1.5 lower sensitivity.
The SFR sensitivity further changes according to the change of the luminosity
distance with redshift. The above estimates assume sky-line free regions. While
this means that the sensitivity can in practice be lower, \OII{} is a doublet
with a separation larger than the spectral resolution of MUSE and hence usually
a substantial part of the doublet is in sky-line free regions. Finally, the
presence of the background quasar can impact the \OII{} detection limit, if a
galaxy happens to be right in front of the quasar. Our quasars have r-band
magnitudes between 19.5 and 17.5, with a median of 18.5. The detection limit
would increase to $\sim11\times10^{-18}\,\uerglf$ for a galaxy exactly in front
of a $18.5\,\text{mag}$ quasar, assuming the same seeing and exposure time as
above. In addition, there might remain systematic residuals after the quasar
light was subtracted, which are difficult to generalise.  However, as the
wavelength range covered by the \OII{} doublet is small, neither the PSF nor
the quasar continuum change much over the relevant wavelength range. Therefore,
a continuum subtraction with two well chosen off-band filters typically leaves
very small quasar residuals.

\begin{table*}
    \begin{tabular}{ccccccccc}
    \multicolumn{9}{c}{MUSE Observations} \\
    \hline
        Quasar & R.A. & Dec. &  T$_{\rm exp}$ & Seeing (G.) & Seeing (M.) & Date-Obs & Prog. IDs  & Refs\\
        (1) & (2) & (3) & (4) &  (5) & (6) & (7) & (8) & (9) \\
        \hline
    J0145p1056 & 01:45:13.1 & +10:56:27 & 6.0 & 1.03 & 0.85 & \makecell[t]{2015-11-12,\\2016-08-29} & \makecell[t]{096.A-0164(A),\\097.A-0138(A)} & This work \\
     \hline
    \end{tabular}
    \caption{Details of MUSE observations for the 22 \mfl{} quasar fields as used in this study. \textbf{The full table with all 22 fields is in Table \ref{sup:tab:obs_full} of the Supplementary Appendix.} 
    (1) Quasar/Field identifier;
    (2) Right ascension of the QSO [hh:mm:ss; J2000];
    (3) Declination of the QSO [dd:mm:ss; J2000];
    (4) Total MUSE exposure time [s];
    (5) Seeing FWHM measured at $7050\text{\AA}$ by fitting a Gaussian [arcsec];
    (6) Seeing FWHM measured at $7050\text{\AA}$ by fitting a Moffat profile with $\beta=2.5$ [arcsec];
    (7) Date of Observations;
    (8) ESO Program IDs;
    (9) Reference.
    \label{tab:obs}}
\end{table*}

\subsection{UVES observations}

\subsubsection{Observations}

Each quasar was also observed with the VLT high-resolution spectrograph UVES
with settings chosen in order to cover $\MgII\,\lambda\lambda2796,2803$,
$\MgI\,\lambda2852$, $\FeII\,\lambda2600$, and when possible other elements such as
Ti, Zn. We used a slit width of 1\arcsec{}.0, resulting in a spectral
resolution of R$\approx$38000 ($FWHM\approx8\kms$).  Further, we chose a 2x2
readout binning for all observations.  The UVES observations are presented in
Table~\ref{tab:uvesobs}.

\subsubsection{Data reduction}

The Common Pipeline Language (CPL version 6.3) of the UVES pipeline was used to
bias correct and flat-field the exposures and then to extract the wavelength
and flux calibrated spectra.  After the standard reduction, the custom software
UVES POst PipeLine Echelle Reduction (POPLER) \citep{MurphyM_16a} version 0.66
was used.  The processing of the spectra, including the air-to-vacuum
correction, was carried out with this software.  The spectra of echelle orders
were  re-dispersed and combined onto a common vacuum heliocentric wavelength
scale and a pixel width of 1.3\,\kms. Left-over cosmic rays were removed by
$\sigma$-clipping.  The automatic procedure of cosmic ray clipping was verified
by visual inspection and the continuum was fitted with fourth order Chebyshev
polynomials and adjusted manually whenever deemed necessary.

\begin{table*}
    \begin{tabular}{cccccccccc}
    \multicolumn{9}{c}{UVES Observations} \\
    \hline
        Quasar & R.A. & Dec. & $z_{\rm em}$ & T$_{\rm exp}$  & Seeing & Date-Obs & Setting & Prog. IDs  & Refs\\
        (1) & (2) & (3) & (4) &  (5) & (6) & (7) & (8) & (9) & (10) \\
        \hline
        J0145p1056 & 01:45:13.1  & $+$10:56:27 & 0.94  &12020 & 0.6 & \makecell[t]{2015-11-11,\\2016-09-03,\\2016-10-28} &  HER\_5\&SHP700 & \makecell[t]{096.A-0609,\\097.A-0144,\\098.A-0310}  &  This work\\
             \hline
    \end{tabular}
    \caption{Details of UVES observations for the 22 \mfl{} Quasars. \textbf{The full table with all 22 quasars is in \Tab{sup:tab:uvesobs_full} of the Supplementary Appendix.} 
        (1) Quasar identifier;
        (2) Right ascension of QSO [hh:mm:ss; J2000];
        (3) Declination of QSO [dd:mm:ss; J2000];
        (4) Emission redshift of the QSO;
        (5) Total UVES exposure time split into settings [s];
        (6) Seeing FWHM measured by DIMM split into settings [arcsec] ;  
        (7) Date of Observations;
        (8) UVES settings
        (9) ESO Program IDs;
    \label{tab:uvesobs}}
\end{table*}

\section{Sample}
\label{sec:sample}


As motivated in \Sec{sect:megaflow},  \mfl{} is a \MgII{} absorber-selected
survey and as such the first step is to identify the galaxies whose CGM is
associated with the selected strong \MgII{} absorption.  In this section, we
describe how we carefully identify all galaxies within the MUSE field-of-view
(FoV) down to the deepest limits (in \Sec{sec:sample:abs_gal_assoc}), a
critical step since \MgII{} absorbers could be associated with multiple
galaxies.  In \Sec{sec:sample:abs_gal_assoc:primary}, we describe how we assign
a primary galaxy to the \MgII{} absorbers.  In
\Sec{sec:sample:accretionsample}, we describe the sub-sample of galaxy-absorber
pairs suitable for this paper, whose focus is on the extended gaseous disks
around star-forming galaxies.

\subsection{Galaxy detections}
\label{sec:sample:abs_gal_assoc}

Our main identification strategy is based on narrowband (NB) images
constructed at the redshift of each absorber. Aside from a  visual inspection
of \OII{} NB images using QFitsView~\footnote{Available  at
\url{http://www.mpe.mpg.de/~ott/QFitsView/}.}, we performed an automatic source
detection designed to detect the lowest SNR galaxies (from both emission lines
and absorption lines).

The automatic detection algorithm is based on `optimized' multi-NB images. The
`optimized' means that we weighted at each spaxel the pixels in the wavelength
direction with the squared S/N of the respective pixels. This
efficiently filters out sky-lines and gives most weight to wavelengths where
the source signal is strong.  The `multi-NB' means that the pseudo-NB filter
has transmittance not only around a single emission line but at multiple lines
simultaneously with the individual passbands matched in velocity width. Each of
the passbands was continuum subtracted by using the median flux density in two
off-band NB filters to the blue and red, respectively.

The multi NB images are created by combining NB-images for multiple
emission lines (each over the same velocity range). This included \OII{} and
depending on the redshift \Hb{} and/or \OIII{}$\lambda5007$. Each NB image is
created with a width of ~\footnote{The width of the \OII{} NB filters was
extended by the width of the separation of the \OII{} doublet.} $400\,\kms$. 
For comparison, a virial velocity of $400\,\kms$
corresponds to a virial mass of $\sim10^{13} M_\odot$, which is the typical
halo mass for a galaxy with stellar mass, $M_*$ of $10^{11} M_\odot$.  For each
absorber redshift, we created three NB pseudo-images  at three different
velocity offsets from the absorber redshifts, namely at -250, 0, 250~\kms{}.
We then performed source detection with \sext{} \citep{Bertin:1996a} on each of
these three images, centred at -250, 0, 250~\kms{}.  We optimize \sext{} to
detect low SNR objects in order to reduce the possibility of missing real
candidates, but this leads to a number of false positives, which have to be
removed manually.

We also searched for quiescent galaxies
specifically, by creating an `optimized' multi NB filter including both lines
of the {\textsc{{\rm Ca}} H\&K doublet.  Quiescent galaxies at the right
redshift have negative fluxes in the continuum-subtracted NB filter.
Therefore, we ran \sext{} in this case on inverted images. Again, we checked
for all candidates that the signal is indeed coming from {\textsc{{\rm Ca}}
    H\&K, hence confirming them to be at the right redshift.

In summary, our algorithm is able to detect both emission line galaxies and
galaxies with mere H\&K absorption.

\subsection{\MgII{} host association}
\label{sec:sample:abs_gal_assoc:primary}

With its $60\arcsec$ wide FoV, MUSE covers at redshift $z=1$ about $480\kpc$,
so $\sim240\kpc$ in each direction from the quasar.  To put this into
perspective, the virial radius of a $z=1$ galaxy with $M^*$ and its
corresponding halo mass of $\log(M_{\rm{h}}/\text{M}_\odot) \approx 12.4$ is
$\approx200 \kpc$.  Consequently, the MUSE observations allow us to identify
the galaxies associated with the absorption, even if the associated absorption
would be all the way out at the virial radius.  However, due to the
anti-correlation between impact parameter and $\rewmgii$
\citep{LanzettaK:1990a,Chen:2010a,Nielsen:2013a}, we expect most of the strong
$\MgII$ absorbers to originate from gas at impact parameters, b,  smaller than
the virial radius. This justifies to focus in the \MgII{} host association on
galaxy-absorber pairs which have $b\lesssim100\,\kpc$.

From our \mfl{} survey of \NabsTot\ \MgII\ absorbers with
$\rewmgii\gtrsim\EWThresTot\,\text{\AA}$, we detect one or more galaxies in
\NabsNoneMorePercent\%\ (\NabsNoneMore/\NabsTot) of the cases within
$100\,\kpc$. When there is at least one galaxy, we find that \NabsNone\
(\NabsNtwo) absorbers have one (two) galaxies within 100~\kpc, respectively,
accounting together for the majority (\NabsNoneTwo/\NabsNoneMore) of the
sample. We choose the absorbers with a maximum of two galaxies within
$100\,\kpc$, in order to study isolated galaxies, and avoid groups where a
unique host association becomes not practicable. However, when there are two
galaxies within 100~\kpc, a decision needs to be made whether one of the two
galaxies should be identified with the absorption. We decide that this is the
case if the galaxy with the smaller impact parameter has also the higher \OII{}
flux (\NabsNtwoPrim{} out of the \NabsNtwo{}). This decision is motivated by
the anti-correlations of \rewmgii with $b$ and the correlation with SFR
($\propto \OII{}$ luminosity, see \Sec{sec:oiiflux}) \citep{Lan:2018a}. This
results in a final sample of \nprimary{} galaxy-absorber associations, which we
refer to in the following as `primary' associations.

While one potential caveat with this quasar-galaxy pair identification is that
it depends on the depth of the data (down to $f_{\OII}\gtrsim\fluxLimit$), the
final sub-sample used for this study (in \Sec{sec:sample:accretionsample}) will
happen to have $f_{\OII}>4\times10^{-17}~\flux$, implying that the satellite
missed by our selection ought to be $\approx10$ times fainter than these
primary galaxies.

\subsection{Geometrical classification and sub-sample selection}
\label{sec:sample:accretionsample}

Since the main goal of our present work is to study kinematics of the
approximately co-planar, possibly co-rotating and accreting gas, we selected
galaxy-absorber pairs with orientations where the quasar sight-line is most
favourable for intersecting the presumed extended gaseous disk
\citep[e.g.][]{Stewart:2017a} and is the least favourable to galactic winds.
This can be ensured using the azimuthal angle $\alpha$ \citep[as
in][]{Bordoloi:2011a, BoucheN:2012a, Kacprzak:2012a,
Schroetter:2015a,Ho:2017a}, since outflows are expelling baryons from the
galaxy in the direction of least resistance/density, i.e. more or less
perpendicularly to the star-forming disk. The azimuthal angle $\alpha$ is the
angle between the apparent quasar location and the galaxy major axis, as
indicated in \Fig{fig:geometry}.

Determining $\alpha$ does require a robust measurement of the galaxies'
position angles, and to a lesser extent inclinations, $i$, in order to remove
face-on galaxies where $\alpha$ is undefined. The position angles were
determined by fitting the morphological and kinematic parameters jointly from
the \OII{} doublet using the \gpk{} \citep{Bouche:2015a} algorithm (see
\Sec{sec:physprop:kinematics}). We also checked the morphological parameters
obtained directly from the continuum 2D flux maps with \galfit{} (See section
\ref{sup:sec:alpha-incl-uncerts} of the Supplementary Appendix).

\Fig{fig:sel:alphavsincl} shows the distribution of the primary galaxies in the
$\alpha-i$ plane, where the top panel shows the $\alpha$ histogram,
demonstrating a strong bimodal distribution of strong \MgII{} absorption around
galaxies. Therefore, strong \MgII{} absorption is preferably found either along
the minor-axis or the major-axis of \nprimary\ primary galaxies, which confirms
the earlier results of \citet{BoucheN:2012a} and \citet{Kacprzak:2012a}. In
addition, one should note that this non-random distribution arises without
making any pre-selection on the orientation of the galaxies and also supports
our primary galaxy identification (\Sec{sec:sample:abs_gal_assoc:primary})
because the $\alpha$'s would be randomly distributed if our primary galaxies
were unrelated to the absorption.

From this result, the galaxy-quasar pairs used in this paper are selected with
the following criteria:

\begin{enumerate} 
    
\item A primary galaxy identification was possible
    (see sec. \ref{sec:sample:abs_gal_assoc:primary}), i.e.  we excluded cases
    where the identification with a single galaxy is ambiguous or not possible;

\item The primary galaxy has an \OII{} flux $>3\times10^{-17}\,\uerglf$, i.e.
    we did not include galaxies that are too faint to obtain robust kinematics
    (and position angles (PA) \&\ inclinations) at the depth of the data;

\item The orientation is favourable for extended gaseous disks, i.e. the
    azimuthal angle is $\vert \alpha \vert < 40^\circ$ and the inclination is
    $i > 40^\circ $ (see Fig.~\ref{fig:sel:alphavsincl});

\item The primary galaxy is not a clear merger and does not have strong AGN
    signatures.   

\end{enumerate}

After applying (i) we are left with \nprimary{} galaxy-absorber pairs. Removing
faint galaxies with $f_{\OII}\lesssim3\times10^{-17}~\flux$ with criteria (ii),
leaves 33 galaxies. Applying the main geometric selection (iii) leaves a sample
of 10 galaxies. With one galaxy~\footnote{This galaxy shows clear AGN
    signatures, e.g. a strong $\NeV{} \lambda \lambda 3346,3426$ detection
    \citep{Mignoli:2013a}. While this doublet is not detected in any of the
    \nsampletext{} remaining galaxies, we cannot rule out AGN contribution with
certainty for the sample based on the available data.} excluded by criterion
(iv) results in a final sample of \nsampletext{} galaxy-absorber pairs, which
are listed in \Tab{tab:sample:galidenti}. \cnsampletextNTwo{} of the selected
primary galaxy-absorber pairs have a second galaxy within $100\,\kpc$.
Incidentally, all of the selected primary galaxies happen to have
$f_{\OII}>4\times10^{-17}\,\uerglf$.

The \nsample{} galaxies selected for this accretion study are indicated in the
$\alpha-i$ plot (\Fig{fig:sel:alphavsincl}) as thick green circles. In this
figure, the points are colour-coded according to the \OII{} flux. Similarly,
\Fig{fig:sel:bvsalpha} shows the distribution of the accretion sample galaxies
compared to all \mfl{} primary galaxies in the $\alpha-b$ plane, showing that
the we probe a range of impact parameters (b) from a few to $100\,kpc$.

As for none of the quasar sight-lines more than one absorber ended up in the
final sample of the present study, we choose to refer in the following for
brevity to the absorber simply by a shortened field ID, e.g. \emph{J0103}
stands for the absorber at $z=0.788$ in the field J0103p1332.

\begin{figure}
\begin{center}
\includegraphics[width=0.95\columnwidth]{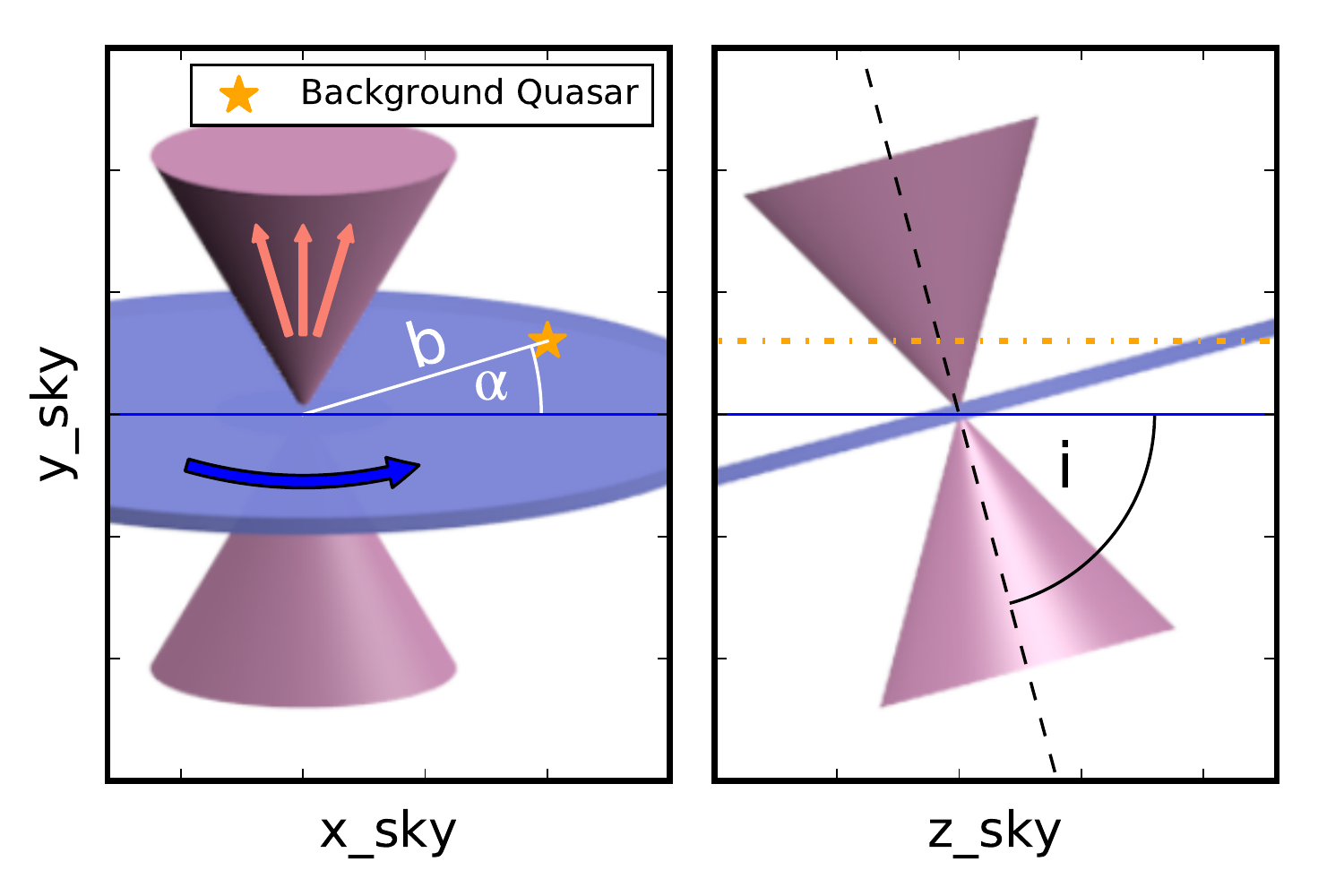}
\end{center}
\caption{\label{fig:geometry} Assumed geometry of the cold CGM.  \textbf{Left:}
    CGM geometry as observed on the sky-plane, where $x_\mathrm{sky}$ is
    without loss of generality aligned with the disk's projected major axis.
    The impact parameter $b$ and the azimuthal angle $\alpha$ are the polar
coordinates of the background quasar (orange)  on the sky-plane.
\textbf{Right}: Same geometry as seen from the side, where $z_\mathrm{sky}$ is
along the line-of-sight. $i$ is the inclination of the disk on the sky.} 
\end{figure}

\begin{figure}
    \centering
    \includegraphics[width=0.95\columnwidth]{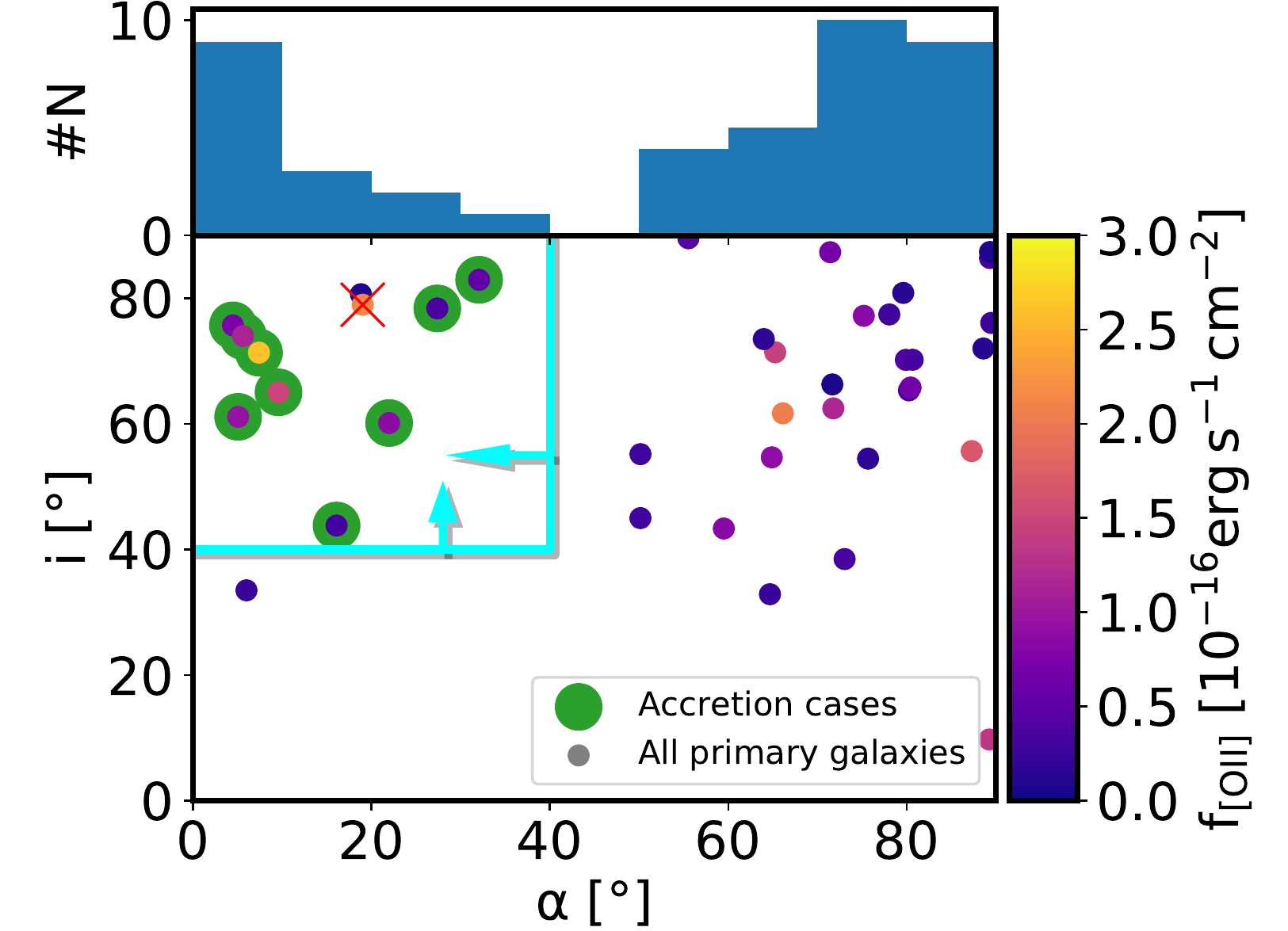}
    \caption{\label{fig:sel:alphavsincl} Distribution of the \mfl{} primary
    \MgII{} host galaxies (see \Sec{sec:sample:abs_gal_assoc:primary}).
    In the upper part the
$\alpha$ histogram is shown for the points in the $\alpha-i$ plane below.
The colour-scale indicates the \OII{} flux of the galaxies. The \nsampletext{} points circled in green
within the cyan boundaries are suitable (as in \Sec{sec:sample:accretionsample})
candidates for this study.
One galaxy in the selection region is excluded as it is an AGN (red cross).
	\stnprimarynprobustalpha{} of the \nprimary{} primary galaxies are omitted in the $\alpha$ histogram, as we could not obtain robust $\alpha$ for those. \stnprimaryDiffFlagThreeFive{} further primary galaxies are not included in the lower panel, as we could not obtain robust inclinations.}
\end{figure}

\begin{figure}
	\includegraphics[width=0.8\columnwidth]{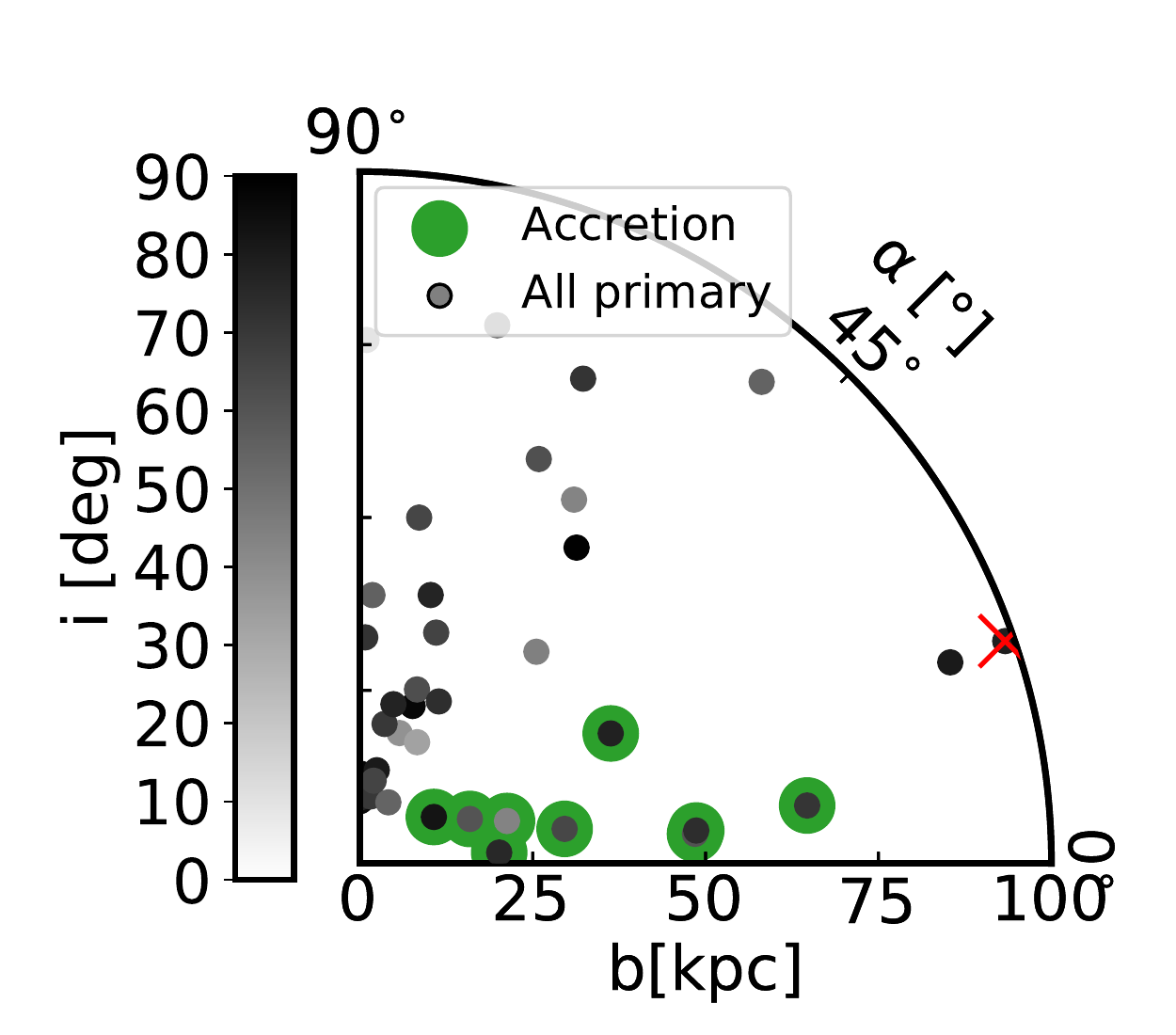}
    \caption{\label{fig:sel:bvsalpha} Same galaxies as in \Fig{fig:sel:alphavsincl}, but here shown in the impact parameter vs $\alpha$ diagram. The grey-scale indicates here the inclination. The two objects at $b\approx 50\kpc$ are overlapping. For an explanation of the red cross see  \Fig{fig:sel:alphavsincl}.}
\end{figure}

\subsection{Discussion of individual cases}

In Figure~\ref{fig:img_and_kin:J0145p1056_0554}, and
Figs.~\ref{fig:img_and_kin:J0103p1332_0788} \mbox{--}
\ref{fig:img_and_kin:J2152p0625_1053}, we show the entire MUSE FoV for the
\OII{} NB image centred on the absorber redshift. The images show all galaxies
including primary and secondary galaxies, that we identified to be associated
with the relevant absorbers and are listed in \Tab{tab:sample:galidenti}. The
NB images are made from red, green and blue channels, where each channel is a
slightly different but overlapping NB image. The green channel is a NB filter
of $\pm150\kms$ around the absorber redshift. The blue (red) channel is made at
$-(+)300\kms$ from the absorber redshift using a transmittance of 100\% and
decreases linearly to 0\% at $(+)-150\kms$, respectively (a method motivated
by \citealt{Hayashi:2014a, Zabl:2016a}). Hence, the colour represents the
velocity offset of the galaxy with respect to the absorber, where blue and red
colours represent the corresponding velocity shifts. For galaxies with strong
velocity gradients, also the velocity field of individual galaxies is directly
visible. 

These colour NB images in Figure~\ref{fig:img_and_kin:J0145p1056_0554}, and
Figs.~\ref{fig:img_and_kin:J0103p1332_0788} \mbox{--}
\ref{fig:img_and_kin:J2152p0625_1053} (see also Table
\ref{tab:sample:galidenti}) show that for five out of the \nsampletext{}
absorbers there is exactly one galaxy associated with the respective absorber
over the entire MUSE FoV. For three absorbers (in the fields of \emph{J1039},
\emph{J1358}, and \emph{J2152}), there are two galaxies in the FoV, and for one
field, \emph{J0800}, we identified five galaxies in the FoV.

Among the absorbers with two host galaxy candidates, for one of them,
\emph{J2152}, the second galaxy is at an impact parameter of $190\,\kpc{}$,
four times further away from the quasar sight-line than the primary galaxy, and
is also fainter. For \emph{J1039}, the second galaxy is at $b=72\,\kpc{}$,
which is a factor 1.5$\times$ further away from the quasar than the primary
galaxy. Moreover, this second galaxy is aligned so that a potential outflow
cone would be covered by the quasar sight-line ($\alpha=68^\circ$) and is part
of the wind analysis of Schroetter et al. (in prep.). This situation needs to
be kept in mind for the discussion of the absorption profiles (see
\Sec{sec:physprop:abskin}). In the third field with two galaxies, \emph{J1358},
the second galaxy is only at slightly larger impact parameter than the primary
galaxy ($b=32\,\kpc{}$ vs $b=40\,\kpc{}$). However, the second galaxy has only
about 10\% of the primary galaxy's \OII{} flux.

For \emph{J0800} we identified five galaxies in the FoV, but only one of them
is within $100\,\kpc{}$ ($b=64\,\kpc$) and the second closest galaxy is a
quiescent galaxy that is a factor two further away and at a large velocity
offset of $\approx400\,\kms$ from the absorber.\footnote{Redshift of the
quiescent galaxy was determined with {\it pPXF} \citep{Capellari:2004a,
Capellari:2017a}.} For this absorber, we will assume that all absorption is
associated with the primary galaxy.

\begin{table*}
\begin{tabular}{llrrrrr}
\hline
Fiekd and absorber &   Galaxy ID & Coordinate & b & $\Delta v$ & $f_{\OII}$ & Note \\
                   &   (1) & (2) & (3) & (4) & (5) &  \\
\hline
Field: J0103p1332 $z_\mathrm{abs}=0.788$ & 
{\bf gal\_0788\_3\_25} & 01:03:32.37 +13:32:36.1 & 20 & 61 & $7.2\pm0.1$ &   \\
\hline
Field: J0145p1056 $z_\mathrm{abs}=0.554$ &
{\bf gal\_0554\_3\_52} &  01:45:13.28 +10:56:28.8 & 22 & -97 & $3.2\pm0.1$ &   \\
\hline
Field: J0800p1849 $z_\mathrm{abs}=0.608$ & 
{\bf gal\_0608\_10\_108} & 08:00:05.20  +18:49:32.6&   65 & -12  & $26.0\pm1.0$         & a  \\
 & gal\_608\_19\_140     & 08:00:05.41  +18:49:20.5&  129 & -419 & \mbox{--}            & b \\
 & gal\_608\_23\_163         & 08:00:05.03  +18:49:13.1&  155 & -278 & $1.06\pm0.04$        & \\
 & gal\_608\_27\_322         & 08:00:03.37  +18:49:56.6& 184 & -60 & $0.5\pm0.1$            & \\
 & gal\_608\_30\_144         & 08:00:05.79  +18:49:10.9& 201& 6 &  $0.7\pm0.1$              & \\
\hline
Field: J1039p0714 $z_\mathrm{abs}=0.949$ &
{\bf gal\_0949\_6\_324} & 10:39:36.42  +07:14:32.4  & 49 & 141 & $9.5\pm0.1$   \\
                        & gal\_0949\_9\_344 & 10:39:36.48 +07:14:36.1  & 72 & 111 & $3.1\pm0.1$ & c \\
\hline
Field: J1107p1021 $z_\mathrm{abs}=1.048$ & 
{\bf gal\_1048\_5\_359} & 11:07:42.71 +10:21:31.4 & 41  & -45 & $3.8\pm0.1$ &  \\
\hline
Field: J1236p0725 $z_\mathrm{abs}=0.912$ & 
{\bf gal\_0912\_2\_246} &  12:36:24.25 +07:25:50.8 & 17 & 34 & $8.7\pm0.6$ &  \\
\hline
Field: J1358p1145 $z_\mathrm{abs}=1.418$ & 
{\bf gal\_1418\_3\_291}  &  13:58:09.26 +11:45:59.2 & 30 & -60 & $14.8\pm0.1$ & d  \\
                         & gal\_1418\_5\_238 & 13:58:09.22 +11:45:55.1 & 40 &  -186 & $1.4\pm0.1$  &   \\
\hline
Field: J1509p1506 $z_\mathrm{abs}=1.046$ & 
{\bf gal\_1046\_2\_351} & 15:09:00.10  +15:06:36.5  & 13 & 68 & $5.4\pm0.2$  \\ 
\hline
Field: J2152p0625 $z_\mathrm{abs}=1.053$ & 
{\bf gal\_1053\_6\_57}  &  21:52:00.36 +06:25:19.7 &  49 & -68 & $11.4\pm0.3$ & \\
                        & gal\_1053\_23\_341  & 21:51:59.54 +06:25:38.4 & 187 & 6 & $5.7\pm0.2$ & \\
\hline
\end{tabular}
\caption{\label{tab:sample:galidenti} Absorber-galaxy identification. Primary galaxies are indicated in bold. 
    (1) ID. The first number in the ID indicates the absorber redshift, the second the impact parameter in arcsec, and the third the position angle between quasar and galaxy in degrees.
    (2) Right ascension and Declination of galaxy (hh:mm:ss dd:mm:ss; J2000);   
    (3) Impact parameter [kpc];
        (4) Velocity offset between absorber redshift, $z_\mathrm{abs}$, and redshift of galaxy, $z_\mathrm{gal}$ [\kms];
    (5) \OII{} flux in units of $10^{-17}\;\uerglf$ as obtained from the 1D line flux (fluxes are measured in large \sext{} \emph{MAG\_AUTO} apertures, but not aperture corrected).
        }
\vspace{-2.5Ex}
\begin{flushleft}
 {\it Note.} a) blend w. foreground galaxy; b) passive HK; c) aligned with
    minor axis to quasar; d) At this redshift Ca H\&K falls outside of the MUSE
    wavelength range and our automatic detection would miss quiescent galaxies
    without any residual \OII{} line emission. As an alternative, we checked here
    for stellar \MgII$\,\lambda 2796, 2803$ absorption, but did not find any
    additional candidates.
\end{flushleft}
\end{table*}

\begin{figure*}
\includegraphics[width=\textwidth]{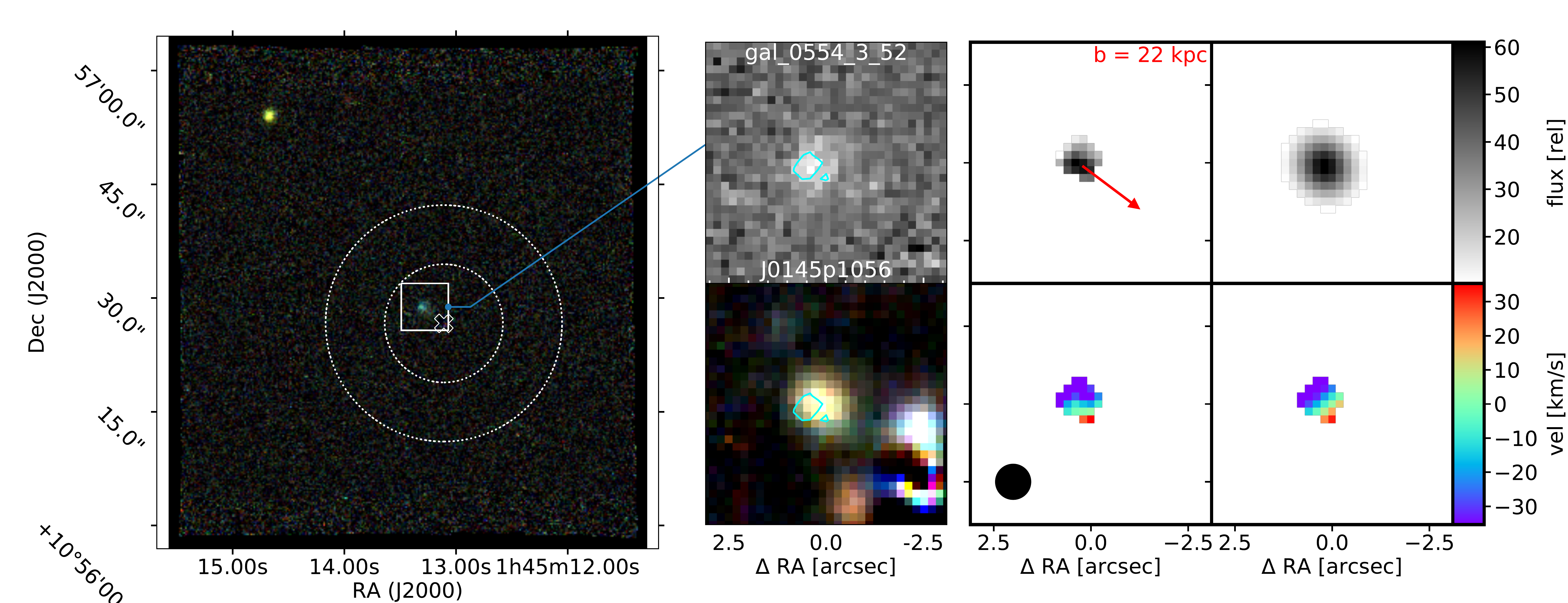}
\caption{\label{fig:img_and_kin:J0145p1056_0554} \textbf{Column 1 (left):}
    Shown is an \OII{} NB image covering $\pm300\kms$ around the redshift of
    the absorber \emph{J0145p1056\_0554}. More precisely, three slightly
    different \OII{} NB images were used for each of the three channels of an
    RGB image. In such an image, emission lines blue-shifted with respect to
    the absorber redshift appear bluer, while those redshifted will appear
    redder.  The position of the quasar is indicated as a white cross and
    circles indicate impact parameters corresponding to $50\,\mathrm{kpc}$ and
    $100\,\mathrm{kpc}$. A white box indicates an emission line galaxies
    associated with the absorber, meaning that the \OII{} emission is in the
    filter. In the shown example there is only the primary galaxy. The
    remaining NB excess sources are either due to other emission lines than
    \OII{} in the NB filter, which means that they are at other redshifts, or
    residuals from bright objects.  \textbf{Column 2:} Top: Simple NB image of
    the primary galaxy optimised for redshift and width of the \OII{} emission.
    Overlaid is a contour of this image. Below a colour image is displayed,
    where pseudo V, R, I broadband images constitute blue, green, and red
    channels, respectively. The same contour as in the NB image is overlaid.
    \textbf{Column 3:} Flux (top) and line-of-sight velocity (bottom) maps
    obtained from direct fitting with \camel{} to the \OII{} cube. More details
    are given in the text. The filled black circle in the lower left corner
    indicates the FWHM of the Moffat PSF at the observed wavelength of \OII{}. 
    \textbf{Column 4:} Similar as in column 3, but here the best-fit model flux
and velocity maps as obtained from fitting with \gpk{} are shown. The zero
velocity in the velocity maps is taken from the \gpk{}
redshift.} 
\end{figure*}

\begin{figure*}
    \begin{tabular}{rr}
        \includegraphics[width=0.25\textwidth]{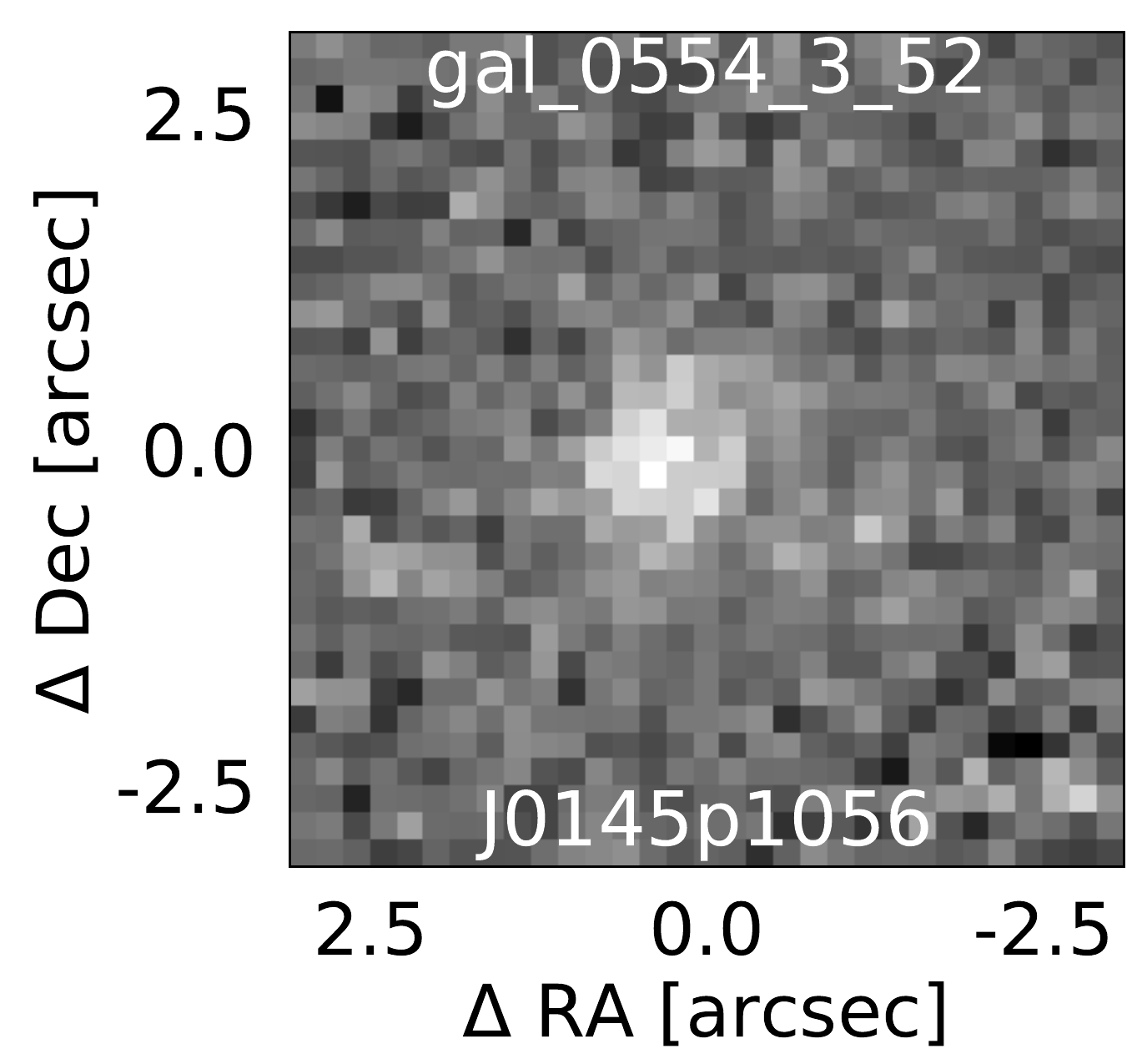}
        & \includegraphics[width=0.53\textwidth]{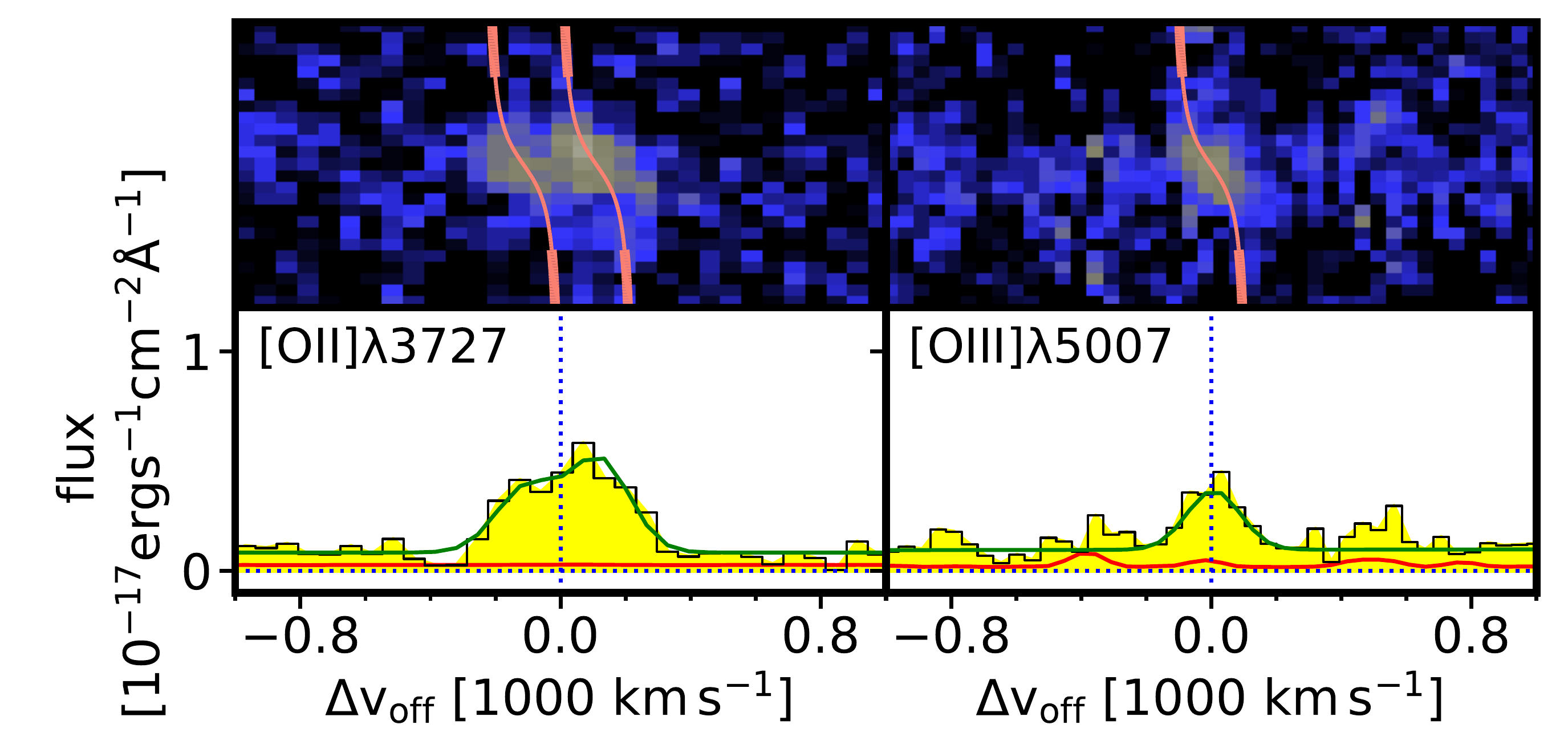} \\
        \multicolumn{2}{c}{\includegraphics[width=0.85\textwidth]{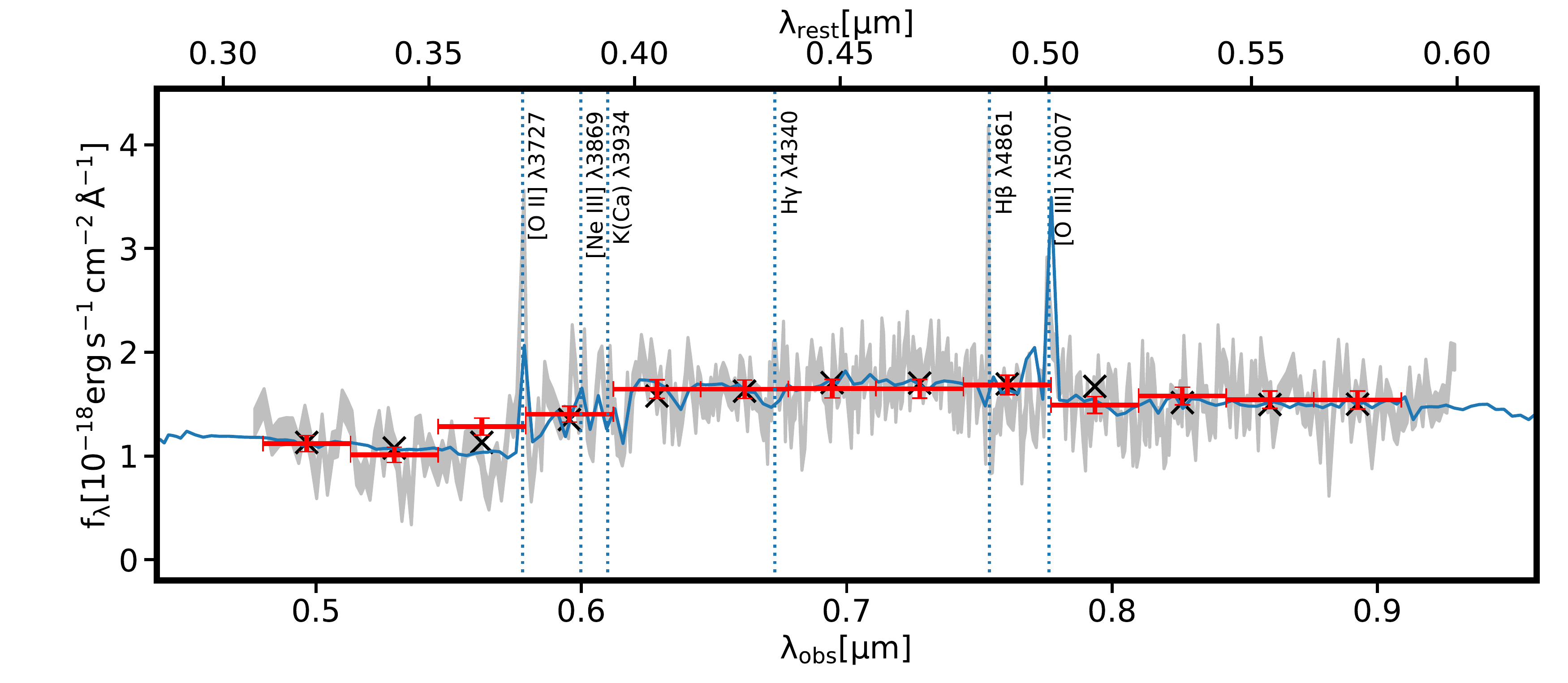}} \\
    \end{tabular}
    \caption{\label{fig:spec:J0145p1056_0554}Spectral information of the galaxy
        gal\_J0145p1056\_0554\_3\_52. \textbf{Upper left}: \OII{} NB image.
        The image is identical to that shown in Fig.
        \ref{fig:img_and_kin:J0145p1056_0554}. \textbf{Upper right}: 1D
        (bottom) and 2D spectra (top) for both the \OII{} doublet and the
        \OIII{} $\lambda 5007$ line. The yellow shaded area in the 1D figures
        is the extracted aperture spectrum, the green line is the best-fit 1D
        spectrum, and the red line is the $1\sigma$ noise spectrum. Zero
        velocity is set to the systemic redshift of the galaxy. Dotted vertical
        and horizontal lines indicate zero velocity and zero flux,
        respectively. The 2D spectra are pseudo 2D spectra with the virtual
        slit aligned along the major axis. Over-plotted is the arctan rotation
        curve as determined from the \gpk{} fit (seeing de-convolved).
        \textbf{Lower}: The red error bars show the flux-densities measured
        with \galfit{} in the 13 boxcar medium-band filters. The horizontal
        width of the bars indicates the width of the filter. The blue curve is
        the best-fit SED obtained from fitting to these filters and the black
        crosses indicate the filter-averaged flux-densities of this SED. The
    1D spectrum extracted from apertures is shown as a grey line, with its
vertical width indicating the $1\sigma$ uncertainty. For this plot, this
spectrum was binned into bins with the same S/N using weighted re-binning (not
flux conserving). In addition, it was corrected to total fluxes using the
ratios between the \galfit{} fluxes. More precisely, we used a straight line
fit through the measured ratios for all 13 filters in order to estimate a
linear wavelength dependence of the aperture loss.}
\end{figure*}

\begin{table*}
\begin{tabular}{ccrrrrrrrrrr}
    \hline
    & Field ID & Gal ID  & z & b & $\alpha$ & i & $\rhalf$  &  $\rturn$ & $\vmax$ & $\sigma_0$  & $f_{\OII}$\\
    &  (1)     & (2) & (3) & (4) & (5) & (6) & (7) & (8) & (9) & (10) & (11)  \\
\hline
\hline
\cellcolor{GalColJ0103} & J0103p1332 & gal\_0788\_3\_25   & 0.7882 & 20 & $4{\scriptstyle \pm7 (7)}$ & $76{\scriptstyle \pm6 (7)}$ & $1.09{\scriptstyle \pm0.27}$ & $0.98{\scriptstyle \pm0.17}$ & $46{\scriptstyle \pm13 (14)}$ & $44{\scriptstyle \pm2}$ & $7.7{\scriptstyle \pm0.1}$ \\
\cellcolor{GalColJ0145} & J0145p1056 & gal\_0554\_3\_52   & 0.5500  & 22 & $16{\scriptstyle \pm7 (7)}$ & $44{\scriptstyle \pm7 (7)}$ & $2.90{\scriptstyle \pm0.27}$ & $0.76{\scriptstyle \pm0.47}$ & $164{\scriptstyle \pm35 (39)}$ & $11{\scriptstyle \pm10}$ & $4.4{\scriptstyle \pm0.2}$ \\
\cellcolor{GalColJ0800} & J0800p1849 & gal\_0608\_10\_108 & 0.6082  & 65 & $7.4{\scriptstyle \pm0.4 (7)}$ & $71.3{\scriptstyle \pm0.5 (7)}$ & $4.51{\scriptstyle \pm0.04}$ & $1.45{\scriptstyle \pm0.12}$ & $108{\scriptstyle \pm3 (11)}$ & $33{\scriptstyle \pm1}$ & $27.0{\scriptstyle \pm0.1}$ \\
\cellcolor{GalColJ1039} & J1039p0714 & gal\_0949\_6\_324  & 0.9494  & 49 & $5{\scriptstyle \pm1 (7)}$ & $61{\scriptstyle \pm1 (7)}$ & $2.80{\scriptstyle \pm0.05}$ & $1.03$ & $158{\scriptstyle \pm3 (16)}$ & $43{\scriptstyle \pm2}$ & $9.2{\scriptstyle \pm0.1}$ \\
\cellcolor{GalColJ1107} & J1107p1021 & gal\_1048\_5\_359  & 1.0481 & 41 & $27{\scriptstyle \pm2 (7)}$ & $78{\scriptstyle \pm3 (7)}$ & $7.00{\scriptstyle \pm0.29}$ & $3.70{\scriptstyle \pm0.44}$ & $189{\scriptstyle \pm8 (21)}$ & $3{\scriptstyle \pm4}$ & $4.4{\scriptstyle \pm0.1}$ \\
\cellcolor{GalColJ1236} & J1236p0725 & gal\_0912\_2\_246  & 0.9128  & 17 & $22{\scriptstyle \pm2 (7)}$ & $60{\scriptstyle \pm3 (7)}$ & $3.41{\scriptstyle \pm0.10}$ & $1.24$ & $232{\scriptstyle \pm7 (24)}$ & $8{\scriptstyle \pm5}$ & $8.8{\scriptstyle \pm0.1}$ \\
\cellcolor{GalColJ1358} & J1358p1145 & gal\_1418\_3\_291  & 1.4171 & 30 & $10{\scriptstyle \pm1 (7)}$ & $65{\scriptstyle \pm1 (7)}$ & $4.02{\scriptstyle \pm0.05}$ & $1.15{\scriptstyle \pm0.91}$ & $8{\scriptstyle \pm2 (2)}$ & $48{\scriptstyle \pm1}$ & $14.1{\scriptstyle \pm0.1}$ \\
\cellcolor{GalColJ1509} & J1509p1506 & gal\_1046\_2\_351  & 1.0469 & 13 & $32{\scriptstyle \pm3 (7)}$ & $83{\scriptstyle \pm5 (7)}$ & $3.29{\scriptstyle \pm0.21}$ & $0.89{\scriptstyle \pm0.30}$ & $134{\scriptstyle \pm10 (17)}$ & $4{\scriptstyle \pm5}$ & $7.0{\scriptstyle \pm0.2}$ \\
\cellcolor{GalColJ2152} & J2152p0625 & gal\_1053\_6\_57   & 1.0530  & 49 & $6{\scriptstyle \pm1 (7)}$ & $74{\scriptstyle \pm1 (7)}$ & $4.88{\scriptstyle \pm0.08}$ & $0.78{\scriptstyle \pm0.08}$ & $177{\scriptstyle \pm3 (18)}$ & $2{\scriptstyle \pm2}$ & $11.0{\scriptstyle \pm0.2}$ \\
\end{tabular}
\caption{\label{tab:kinematics_and_morphology} Kinematical and morphological
    measurements as obtained from fitting to the \OII{}$\,\lambda \lambda 3727,
    3729$ doublet with \gpk{}. (1) Field ID; (2) Galaxy ID; (3) Galaxy redshift; (4) Impact parameter [kpc]; (5)
    Azimuthal angle [$^\circ$]; (6) inclination [$^\circ$]; (7) half-light
    radius [kpc]; (8) Turnover radius [kpc]; for the two galaxies without error
    bar $\rturn$ was fixed to $\approx r_\text{half}/2.7$;  (9) Intrinsic maximum rotation velocity $v_{\rm max}$ [\kms]; 
    (10) Velocity dispersion from turbulence $\sigma_0$ [\kms]; (11)  Integrated \OII\
    flux from the \gpk{}\ model [$10^{-17} \,\flux$].  }
\vspace{-2.5Ex}
\begin{flushleft}
{\it Note.} The errors ($\pm$)
    are the statistical $1\sigma$ Bayesian uncertainties from \gpk{} increased by 20\%. For the three parameters used in this paper
    ($v_{\rm max}$, $i$ and $\alpha$) we list in parenthesis the total uncertainty, which includes systematics (see
    \Sec{sup:sec:alpha-incl-uncerts} of the Supplementary Appendix).
\end{flushleft}
 \end{table*}

 \section{Galaxy physical properties}
 \label{sec:gal_parameters}
 
The MUSE data allows us to determine both photometric and kinematic properties
for each detected galaxy. In the following, we discuss the physical properties
for our sample of \nsampletext{} primary galaxies. In
\Sec{sec:physprop:kinematics}, we describe how we determined the galaxy
kinematics and redshifts. In \Sec{sec:photometry}, we explain our continuum
photometric measurements used for stellar mass estimates. In
\Sec{sec:oiiflux}, we discuss our SFR estimates based on \OII{} fluxes.
In \Sec{sec:physprop:haloprop}, we derive the halo mass properties.
Finally, we describe the absorption properties in \Sec{sec:physprop:abskin}.

\subsection{Galaxy kinematics and redshifts}
\label{sec:physprop:kinematics}

The main ingredient for our study is a robust comparison between galaxy and
absorber kinematics. Recent 3D fitting
codes \citep[e.g.][]{Bouche:2015a,Di-Teodoro:2015a} allow one to take advantage
of the full 3D information provided by IFU data taking into account
the spatial PSF and the spectral line spread
function (LSF). Here, we measured both the
redshift and galaxy kinematics with the 3D algorithm \gpk{}
\citep{Bouche:2015a} and compared the latter to the
traditional 2D  method using the \camel{}\footnote{Available at
\url{https://bitbucket.org/bepinat/camel.git}.} code of \citet{Epinat:2012a}.

\subsubsection{Morpho-kinematical modelling}
\label{sec:kin:galpak}

In order to apply the 3D line fitting tool \gpk{} to the \OII{} data, we
subtracted the continuum by taking the median in each spaxel over a wavelength
window of $\pm1250\kms$ around the centre of the \OII{} doublet and excluding
the central $\pm250\kms$.

In short, \gpk{} creates a mock \OII\ observation~\footnote{We use a fixed line
ratio of 0.8:1.0 per default, except when the observed \OII\ doublet ratio
deviates strongly from 0.8.} from the parametrised 3D model of a disk galaxy,
compares it to the data, and finds the posterior of the parameters through
Markov chain Monte Carlo (MCMC) sampling. In such a parametrised approach, a
choice for rotation curve and light distribution needs to be made. For the
rotation curve, we assume throughout an $\arctan$ function, $v(R) = \vmax
\frac{2}{\pi} \arctan(R/\rturn)$, where the two free parameters are the maximum
velocity $\vmax$ and the turn-over radius, $\rturn$. For the distribution of
the light emitted in \OII, we assumed an exponential disk, $I(R)\propto
\exp(-1.68(R/\rhalf))$.

For compact galaxies, defined as those which have half-light radii smaller than
0.75 times the Moffat's PSF FWHM, we often tested a Gaussian surface-brightness profile (Sercic index
$n = 0.5$) and chose the appropriate Sersic profile based on the lowest
$\chi^2$. For these compact galaxies, we either limited the allowed range of
the turnover radius $\rturn$ (to $<0.8\,\rhalf$) or fixed the turnover radius
to $1/2.7\,\rhalf$ in order to break potential degeneracies. 
This value of 2.7 is motivated by the tight relation between rotation
curve scale length and disk scale length found in local galaxies by
\citet{Amorisco:2010a}\footnote{We converted the \citet{Amorisco:2010a} relation between
the exponential disk scale-length and the radius where the rotation curve
reaches 2/3 of $\vmax$ to a $\rhalf$ and $\rturn$ ratio for an arctan rotation curve.}.

An additional free parameter in our morpho-kinematical model is a radially
constant velocity dispersion, $\sigma_0$, which is meant to describe a
turbulence component added in quadrature to the disk model, i.e. $\sigma_0$ is
not the total velocity dispersion \citep[see][for details]{Bouche:2015a}. All
inferred parameters for all \nsampletext{} galaxies are listed in Table
\ref{tab:kinematics_and_morphology}.

As a consistency check, we created 2D velocity maps from our fitted model,
which can be compared to a map created from a more classical pixel-by-pixel
velocity fit. The latter we performed with the code \camel{}. This code
directly fits the \OII{} doublet in each pixel. To increase the S/N, we
convolved the cube in the spatial direction with a kernel of FWHM=2\,pixels.
Both the \gpk{} and \camel{} based velocity maps are shown in \figskin.
Reassuringly, no strong discrepancies are visible.

\subsubsection{Redshifts}
\label{sec:kin:redshifts}

Our analysis relies heavily on comparing the kinematics of the host galaxy to
that of the absorption in the quasar line-of-sight. Thus, this comparison will
depend critically on the accuracy of the systemic redshift of the galaxy. While
the \gpk{} measurements described in \Sec{sec:kin:galpak} also provided the
redshift of the galaxy (see \Tab{tab:kinematics_and_morphology}), we carefully
tested the robustness of the \gpk{} based redshift through comparison to
redshifts inferred using two other methods.

The first of these two comparison methods makes use of 1D spectra extracted
from the cubes  using the \mpdaf{} routine {\it
extract\_spectra}. The spatial extent used for these extractions was set by the
extent of the sources as determined by \sext{} from the `optimized' NB images
(see \Sec{sec:sample:abs_gal_assoc}). 
From these 1D source spectra, we {\it simultaneously} fit all strong rest-frame
emission lines available in the wavelength range covered by the MUSE spectra
with a custom MCMC based algorithm that takes into account the spectral FWHM as
parametrised by \citet{Guerou:2017a}. The simultaneous fit also allows us to
robustly determine the \OII{} doublet ratio, but we keep the
\OIII{}$\lambda5007$/\OIII{}$\lambda4959$ fixed to 2.98 \citep[as expected
theoretically:][]{Storey:2000a}. The fit results for \OII{} and the second
brightest line in the MUSE wavelength range other than \OII{} are shown in
\figsspec.

The second comparison method is a visual inspection of (pseudo-)2D spectra,
which we refer to as position velocity diagrams (PVDs). These PVDs were
extracted from the MUSE cubes using a pseudo slit, with the slit aligned along
the morpho-kinematic major axis of each galaxy, shown in \figsspec{} for each galaxy. 
 We carried out this visual
redshift determination for \OII$\;\lambda 3729$ and, if available also for
\OIII$\;\lambda 5007$.

For the first (second) method, the velocity difference with respect to the redshifts from \gpk{} 
is -1$\pm12$ $\kms$ (5$\pm15$ \kms), respectively, with a maximum difference of 22 (40) $\kms$. The individual values are listed in
\Tab{sup:tab:redshift_comparison} of the Supplementary Appendix.

\begin{table*}
\begin{tabular}{rrrrrrrrrrrr}
\hline
& ID & $E_\text{{Mass}}(B-V)$ & $E_{\mathrm{SED}}(B-V)$ & $\mathrm{SFR}_{\OII;2}$ & $\mathrm{SFR}_{\OII;3}$ & $\mathrm{SFR}_{SED}$ & $M_*$ & $\delta(MS)$ & $S05$ & B \\
&   (1) & (2) & (3) & (4) & (5) & (6) & (7) & (8) & (9) & (10) \\
\hline
\cellcolor{GalColJ0103} & J0103p1332 & $0.22_{-0.12}^{+0.09}$ & $0.00_{-0.00}^{+0.68}$ & $3.1_{-0.9}^{+0.7}$ & $0.9_{-0.0}^{+1.5}$ & $0.8_{-0.0}^{+26.7}$ & $9.8_{-0.4}^{+0.0}$ & $0.18_{-0.11}^{+0.52}$ & $55{\scriptstyle \pm6}$ & -19.6 \\
\cellcolor{GalColJ0145} & J0145p1056 & $0.23_{-0.09}^{+0.10}$ & $0.26_{-0.09}^{+0.29}$ & $0.8{\scriptstyle \pm0.2}$ & $0.9_{-0.2}^{+0.6}$ & $1.5_{-0.5}^{+4.3}$ & $9.8_{-0.1}^{+0.3}$ & $-0.35_{-0.19}^{+0.17}$ & $117{\scriptstyle \pm27}$ & -19.2 \\
\cellcolor{GalColJ0800} & J0800p1849 & $0.18{\scriptstyle \pm0.09}$ & $0.09_{-0.01}^{+0.20}$ & $4.6{\scriptstyle \pm1.0}$ & $2.7_{-0.1}^{+1.3}$ & $1.4_{-0.0}^{+3.0}$ & $9.5_{-0.0}^{+0.2}$ & $0.60_{-0.16}^{+0.12}$ & $83{\scriptstyle \pm7}$ & -20.0 \\
\cellcolor{GalColJ1039} & J1039p0714 & $0.21{\scriptstyle \pm0.09}$ & $0.77_{-0.10}^{+0.04}$ & $5.5{\scriptstyle \pm1.2}$ & $112.3_{-26.9}^{+9.3}$ & $110.2_{-53.2}^{+20.1}$ & $9.7_{-0.0}^{+0.1}$ & $0.41_{-0.14}^{+0.12}$ & $120{\scriptstyle \pm11}$ & -20.6 \\
\cellcolor{GalColJ1107} & J1107p1021 & $0.28_{-0.10}^{+0.12}$ & $0.26_{-0.03}^{+0.18}$ & $4.9_{-1.1}^{+1.4}$ & $4.3_{-0.3}^{+1.8}$ & $6.5_{-1.0}^{+9.5}$ & $10.0_{-0.1}^{+0.3}$ & $0.06_{-0.21}^{+0.26}$ & $134{\scriptstyle \pm14}$ & -20.6 \\
\cellcolor{GalColJ1236} & J1236p0725 & $0.39_{-0.09}^{+0.18}$ & $0.64_{-0.00}^{+0.70}$ & $12.8_{-2.7}^{+5.4}$ & $48.8_{-0.5}^{+79.8}$ & $7.1_{-0.0}^{+680.4}$ & $10.5_{-0.0}^{+0.5}$ & $0.14_{-0.24}^{+0.15}$ & $164{\scriptstyle \pm17}$ & -20.8 \\
\cellcolor{GalColJ1358} & J1358p1145 & $0.26_{-0.11}^{+0.12}$ & $0.26_{-0.07}^{+0.10}$ & $29.9_{-7.7}^{+8.2}$ & $28.9_{-4.6}^{+6.5}$ & $27.8_{-4.6}^{+44.9}$ & $9.9{\scriptstyle \pm0.3}$ & $0.78_{-0.22}^{+0.39}$ & $48{\scriptstyle \pm1}$ & -22.0 \\
\cellcolor{GalColJ1509} & J1509p1506 & $0.14{\scriptstyle \pm0.09}$ & $0.60_{-0.12}^{+0.05}$ & $3.7{\scriptstyle \pm0.8}$ & $43.2_{-12.1}^{+5.5}$ & $38.6_{-18.9}^{+14.9}$ & $9.3_{-0.1}^{+0.2}$ & $0.51_{-0.17}^{+0.15}$ & $95{\scriptstyle \pm12}$ & -20.2 \\
\cellcolor{GalColJ2152} & J2152p0625 & $0.34_{-0.10}^{+0.12}$ & $0.43_{-0.14}^{+0.15}$ & $16.8_{-3.8}^{+4.8}$ & $27.3_{-8.9}^{+9.6}$ & $22.1_{-11.0}^{+28.6}$ & $10.2_{-0.1}^{+0.3}$ & $0.39_{-0.21}^{+0.23}$ & $125{\scriptstyle \pm13}$ & -21.2 \\
\hline
\end{tabular}
\caption{\label{tab:physproperties} Physical parameters of the galaxies as
    obtained from the [OII] emission line fluxes and SED fitting. 
    (2) Nebular $E(B-V)$ estimated from stellar mass (\Eq{eq:ebv_mass_garnbest10});
    (3) nebular $E(B-V)$ as obtained from SED fitting (see \Sec{sec:photometry});
    (4) \OII{} based SFR [$\mpy$] from \Eq{eq:sfr_oii} and assuming $E_\text{Mass}(B-V)$ as extinction;
    (5) Same as in 4, but using $E_{\rm SED}(B-V)$ as extinction estimate;
    (6) Instantaneous SFR [$\mpy$] directly from SED fit;
    (7) Stellar mass [$\log_{10}(\text{M}_\odot)$] from SED fit;
    (8) Distance from the Main Sequence ($\log(\mathrm{sSFR}(Obs)/\mathrm{sSFR}(MS)$). The observed sSFR was calculated using columns 4) and 7);
    (9) $S_{0.5} = {(0.5 \vmax^2 + \sigma_{0}^2)}^{0.5}$ [\kms{}] (10) rest-frame B absolute magnitude
calculated from best fit-SED model [mag].}
\end{table*}

\subsection{Photometry and stellar masses}
\label{sec:photometry}

In order to determine continuum photometric magnitudes from the MUSE data, and
perform spectral energy distribution (SED) fitting, we determined for each of
the galaxies photometry in 13 pseudo medium bands covering the wavelength range
from $4800\,\text{\AA}$ to $9090\,\text{\AA}$. Here, instead of creating simple
aperture photometry, we determined total magnitudes using \galfit{}
\citep{Peng:2010a}, which provides two advantages. First, \galfit{} can
simultaneously fit neighbouring or blended galaxies (foreground or background
galaxies) and thus remove this contamination, and second it provides a total
flux measurement, i.e. is a natural way to take into account the wavelength
dependence of the PSF.

For the main galaxies, we assumed a fixed Sersic index of $n=1$ (exponential).
Once we had a satisfying model, we ran \galfit{} with this model on the medium
band filters, allowing only the fluxes to vary.
We assumed for each band a Moffat PSF with parameters and
wavelength dependence as determined for the quasar (see
\ref{sec:data:quality}).

The statistical uncertainties on the flux-densities obtained by the \galfit\ fit
are very small. In order to crudely account for systematic
uncertainties in the \galfit{} modelling, we added a somewhat arbitrary
systematic 5\% relative uncertainty to the flux-densities. 

For the SED fitting, we used a custom SED
fitting code \emph{coniecto} \citep{Zabl:2016a}. As input we used BC03 models
\citep{Bruzual:2003a} with exponential SFHs and nebular line and continuum
emission added following the recipe by \citet{Schaerer:2009a} and
\citet{Ono:2010a}. Here, we use a \citet{Chabrier:2003a} IMF and a
\citet{Calzetti:2000a} extinction law. While we used the same extinction law
both for nebular and stellar emission, we assumed higher nebular extinction
$E_\text{N}(B-V)$, than stellar extinction, $E_\text{S}(B-V)$ ($E_\text{S}(B-V)
= 0.7\,E_\text{N}(B-V)$). We omit in the following the suffix `N' and use
$E(B-V)$ for the nebular extinction throughout.

The stellar masses, $M_*$, $E(B-V)$, instantaneous SFRs, and rest-frame B
magnitude as obtained from the SED fitting are listed in
\Tab{tab:physproperties}. The primary galaxies in our sample cover a relatively
small mass range, with all galaxies around $\log(M_*/\text{M}_\odot) \approx
10.0\pm0.5$.

\subsection{\OII{} Fluxes and Star formation rates}
\label{sec:oiiflux}

The only strong emission line we have access to for all of our galaxies is
\OII{} due to the wavelength coverage of MUSE. Therefore, we need to rely on the observed \OII{}
luminosity, $L_{\OII;\rm o}$, as our main SFR indicator. The main problem with
having only $L_{\OII;\rm o}$ as SFR indicator is the lack of knowledge about
the extinction.

In order to get an approximate estimate for the extinction, one could take
advantage of the correlation between
the star-formation indicator $L_{\OII}$ itself and $\ebv$
\citep[e.g.][]{Kewley:2004a} which is equivalently to a SFR$-\ebv$ correlation. However,
given that the \citet{Kewley:2004a} relation was determined at $z=0$ and that the
$M_\star-$SFR main-sequence \citep[e.g.][]{Brinchmann:2004,Noeske:2007,Salim:2007} 
 evolves strongly with redshift \citep[e.g.][]{Elbaz:2007,Whitaker:2014a,
Speagle:2014, Ilbert:2015a,Boogaard:2018a}, 
it might be better to use the $M_\star-\ebv$
relation instead. Indeed, the SFR--$\ebv$ relation does strongly
depend on redshift \citep[e.g.][]{Sobral:2012a}, while the $M_\star-\ebv$ relation seems
to have little or no evolution with redshift \citep[e.g.][]{Sobral:2012a,
Kashino:2013a, Cullen:2017a, McLure:2018a}, indicating $\ebv$ is determined by $M_\star$.

 Hence, we use the $z=0$ $M_*-\ebv$ relation~\footnote{Alternatively, one could correct the $z=0$ \citet{Kewley:2004a}
  $L_{\OII;\rm i}-\ebv$ relation by taking into
  account   the MS redshift evolution ($\mathrm{SFR} \propto
(1+z)^\alpha$;  with   $\alpha\approx2-3$).}
from \citet{Garn:2010a}, corrected  to a \citet{Chabrier:2003a} IMF:
\begin{equation}
    E(B-V) = (0.93 + 0.77\,X + 0.11\,X^2 - 0.09\,X^3)/k_{\Ha} \label{eq:ebv_mass_garnbest10}
\end{equation}
Here  $X=\log(M/M_\odot) - 10$ and $k_{\Ha}=3.326$ for the
\citet{Calzetti:2000a} extinction law, both assumed by \citet{Garn:2010a} and
in this study. \citet{Garn:2010a} state an intrinsic scatter in this relation
of about $0.3\,\text{dex}$ for the extinction at \Ha{} ($A_{\Ha}$). Therefore, we
include a systematic error of $0.3\,\text{dex}/k_\Ha$ in the error budget for
$E(B-V)$.

Another way to get an estimate for the $E(B-V)$ is through SED fitting
(\Sec{sec:photometry}). Both the mass based and the SED based $E(B-V)$
estimates are listed in \Tab{tab:physproperties}. While for most of the
galaxies the two $E(B-V)$ values agree within the uncertainties, there are a
few cases where the SED based estimates are significantly higher
(\emph{J1509}, \emph{J1039}). 

Using the assumed extinction curve and the estimated $E(B-V)$ from
\Eq{eq:ebv_mass_garnbest10} we can then de-redden the observed \OII\
luminosity to estimate   the intrinsic luminosity, $L_{\OII;\rm i}$ assuming a
\citet{Calzetti:2000a} curve. The SFR can then be estimated using the
calibration from \citet{Kewley:2004a}:

\begin{equation}
    \mathrm{SFR}(\OII) = 4.1\times10^{-42} (L_{\OII;\rm i}/\uerglum)\,\mpy \label{eq:sfr_oii}
\end{equation}

The version here is adjusted with respect to the original version in
\citet{Kewley:2004a} to convert from the Salpeter IMF to the
Chabrier IMF assumed here. The obtained SFRs estimates, both using the $\ebv{}$
from \Eq{eq:ebv_mass_garnbest10} and the $\ebv$ from the SED fit are listed in
\Tab{tab:physproperties}.

Based on these SFR and $M_*$ (\Sec{sec:photometry}) estimates we assessed
whether we selected typical star-forming galaxies on the SFR-$M_*$
main-sequence (MS). We list for each of our galaxies in
\Tab{tab:physproperties} the distance from the MS, $\delta(MS)$, which is
defined as the difference of the logarithms of the measured and expected
specific star formation rates (sSFR=SFR/$M_*$) based on the MS parametrisation
by \citet{Boogaard:2018a} (their eq. 11). Further we show the position of the
galaxies in the SFR-$M_*$ plane in \Fig{fig:MS} of the Supplementary Appendix.
While two galaxies have SFRs elevated compared to the
$\approx0.4\,\mathrm{dex}$ scatter of the MS, the seven other galaxies are
within the scatter. In addition, it appears that eight out of the
\nsampletext{} galaxies are slightly above the MS, which might be significant.
However, the assessment of the significance of this trend must take into
account all selection effects and this is beyond the scope of the present paper
and will be part of the \mfl{} survey publication.

\begin{table}
\begin{tabular}{crrrrr}
\hline
ID &  $\vvir$ & $\Mvir$ & $\Mvirabund$ & $\rvir$ & $r_s$ \\
(1) & (2) & (3) & (4) & (5) & (6) \\
\hline
\cellcolor{GalColJ0103}J0103 & $121_{-18}^{+17}$ & $11.1{\scriptstyle \pm0.4}$ & $11.6{\scriptstyle \pm0.2}$ & $128_{-19}^{+18}$ & $20{\scriptstyle \pm4}$ \\
\cellcolor{GalColJ0145}J0145 & $149_{-48}^{+66}$ & $12.0{\scriptstyle \pm0.5}$ & $11.7_{-0.1}^{+0.2}$ & $189_{-60}^{+84}$ & $27_{-11}^{+17}$ \\
\cellcolor{GalColJ0800}J0800 & $98_{-23}^{+38}$ & $11.4{\scriptstyle \pm0.4}$ & $11.5_{-0.1}^{+0.2}$ & $119_{-28}^{+46}$ & $15_{-5}^{+8}$ \\
\cellcolor{GalColJ1039}J1039 & $143_{-34}^{+56}$ & $11.8{\scriptstyle \pm0.4}$ & $11.6_{-0.1}^{+0.2}$ & $136_{-32}^{+53}$ & $23_{-7}^{+12}$ \\
\cellcolor{GalColJ1107}J1107 & $172_{-41}^{+67}$ & $12.0{\scriptstyle \pm0.4}$ & $11.8{\scriptstyle \pm0.2}$ & $153_{-37}^{+60}$ & $29_{-9}^{+15}$ \\
\cellcolor{GalColJ1236}J1236 & $211_{-50}^{+82}$ & $12.3{\scriptstyle \pm0.4}$ & $12.2_{-0.2}^{+0.4}$ & $205_{-49}^{+80}$ & $39_{-12}^{+20}$ \\
\cellcolor{GalColJ1358}J1358 & $152_{-21}^{+29}$ & $10.9_{-0.3}^{+0.4}$ & $11.8{\scriptstyle \pm0.2}$ & $108_{-15}^{+20}$ & $23_{-4}^{+5}$ \\
\cellcolor{GalColJ1509}J1509 & $121_{-30}^{+48}$ & $11.6{\scriptstyle \pm0.4}$ & $11.4{\scriptstyle \pm0.1}$ & $108_{-27}^{+43}$ & $19_{-6}^{+10}$ \\
\cellcolor{GalColJ2152}J2152 & $161_{-38}^{+63}$ & $11.9{\scriptstyle \pm0.4}$ & $12.0_{-0.2}^{+0.3}$ & $142_{-34}^{+55}$ & $27_{-8}^{+14}$ \\
\hline
\end{tabular}
\caption{\label{tab:physprop:haloprop} Properties of the host halos
    (2) Virial velocity [$\kms$]; For all galaxies except \emph{J0103} and \emph{J1358} identical to $\vmax/1.1$; For the latter galaxies derived from 4);
    (3) Viral mass [$\log_{10}(\text{M}_\odot)$] from eq. \ref{eq:vir_mass} using 2); for \emph{J0103} and \emph{J1358} using $\vvir$ estimate based on \citet{Burkert:2010a} correction for pressure support; 
    (4) Halo mass [$\log_{10}(M_\odot)$] estimated using the stellar\mbox{--}halo mass relation \citep{Behroozi:2010a}; The uncertainties include both the uncertainties on the stellar mass and the scatter in the stellar\mbox{--}halo mass relation;
    (5) virial radius [kpc] (cf. \Sec{sec:physprop:haloprop});   
    (6) NFW scale radius [kpc]  (cf. \Sec{sec:physprop:haloprop}.)
}
\end{table}

\subsection{Halo properties}
\label{sec:physprop:haloprop}

The interpretation of the kinematics of the circumgalactic gas requires an
estimate of the properties of the dark matter halos through which the gas
moves. We determine the halo masses of our galaxies using two different
methods. First, we use the stellar\mbox{--}halo mass relation as obtained from
abundance matching by e.g. \citet{Behroozi:2010a}. Second, we derive halo mass
estimates from the galaxy kinematics. From the halo masses, we will then
compute virial radii.

Using the stellar masses derived in \Sec{sec:photometry}, and the $z=1$
stellar\mbox{--}halo mass relation from \citet{Behroozi:2010a}, the halo masses
of our galaxies range from $M_{\rm vir}\approx 3\times 10^{11}-3\times
10^{12}$~\msun, covering a range starting from about 1\,dex smaller than the
halo of a $\text{L}^*$ galaxy. Using an estimate for the halo's virial
velocity $\vvir$ from $\vmax$, $\vvir=\vmax/(1.1\pm0.3)$ as motivated by
\citet{Dutton:2010a} \citep[cf. also][]{Reyes:2012a, Cattaneo:2014a} we
calculate the virial mass of the halos of our galaxies with: 
\begin{equation}
    \Mvir = \vvir^3 \left(\frac{\Delta_\mathrm{vir}}{2}\right)^{-0.5}
    \frac{1}{G H(z)} \label{eq:vir_mass} \end{equation}
where the over-density $\Delta_\mathrm{vir}$ is defined as the ratio between
the average matter density within the halo's virial radius and the critical
density at the considered redshift and can be approximated as
$\Delta_\mathrm{vir} = 18\pi^2 + 82x - 39x^2$ \citep{Bryan:1998a} with
$x=\Omega_m(z) - 1$, for a flat Universe.

Both the abundance matching based halo estimate, $\Mvirabund$, and the
dynamical estimate, $\Mvir$, are listed in \Tab{tab:physprop:haloprop}. Apart
from \emph{J0103} and especially \emph{J1358}, the agreement between the two
estimates is generally good (for a visual comparison see
\Fig{suppl:fig:galprob:mvir_diff_est} in the Supplementary Appendix).

The two outliers can be explained. When using the $\vmax$ measured from the
galaxies, we make the assumption that the rotation velocity $\vrot$ corresponds
to the rotational velocity of the halo $\vcirc$, where $\vcirc$ is defined
through $M_{h}(<r) = \frac{r\,\vcirc^2(r)}{G}$. The assumption $\vcirc=\vrot$
will not be correct if the galaxies have substantial pressure support as
discussed in \citet{Burkert:2010a,Burkert:2016a}. And indeed, the two galaxies
with the largest discrepancy between the two halo estimates, are the two
galaxies in our sample with substantial pressure support, as \emph{J0103} has
$v/\sigma_0 \approx 1$, while \emph{J1358} has a even more extreme
$v/\sigma_0=0.3$. Therefore, the approximation of $\vvir=\vmax/1.1$ might not be
appropriate in these cases.

Using the pressure support correction from \citep{Burkert:2010a} to estimate
$\vcirc$, where $\vcirc^2(r)=\vrot(r)^2 + 3.3567\sigma^2(r)\times(r/\rhalf)$,
evaluated at $\rhalf$ and assuming $\vvir=\vcirc/1.1$, leads indeed to an
estimate of $M_\mathrm{vir}$ which is in much better agreement with the
estimate based on the stellar mass. For the remainder of the analysis, we use
for \emph{J0103} and \emph{J1358} the abundance matching estimates for $\Mvir$
and calculated the corresponding $\rvir$ and $\vvir$. We use the $\vmax$ based
estimates for the other seven galaxies.

Finally, from our virial mass estimates, we determine the virial radius,
$r_\mathrm{vir}$ (and the scale radius $r_s$ for an NFW profile) for the
halos. The virial radius, $r_\mathrm{vir}$, is related to $M_\mathrm{vir}$
through $M_\mathrm{vir} = \frac{4\pi}{3} \Delta_{vir} \rho_\mathrm{crit}
r_\mathrm{vir}^3$. The scale radius, $r_s$, can be obtained from
$r_\mathrm{vir}$, by making use of the tight relation between $M_\mathrm{vir}$
or $r_\mathrm{vir}$ and $r_s$ (e.g. \citealt{Navarro:1996a}; using here the
version of \citealt{Diemer:2015a} and making the conversion with their
\emph{Colossus} code~\footnote{Available at
\url{https://bitbucket.org/bdiemer/colossus}.}). The resulting radii are
listed in Table \ref{tab:physprop:haloprop}.

\subsection{Absorber kinematics}
\label{sec:physprop:abskin}

\begin{figure}
\includegraphics[width=0.5\textwidth]{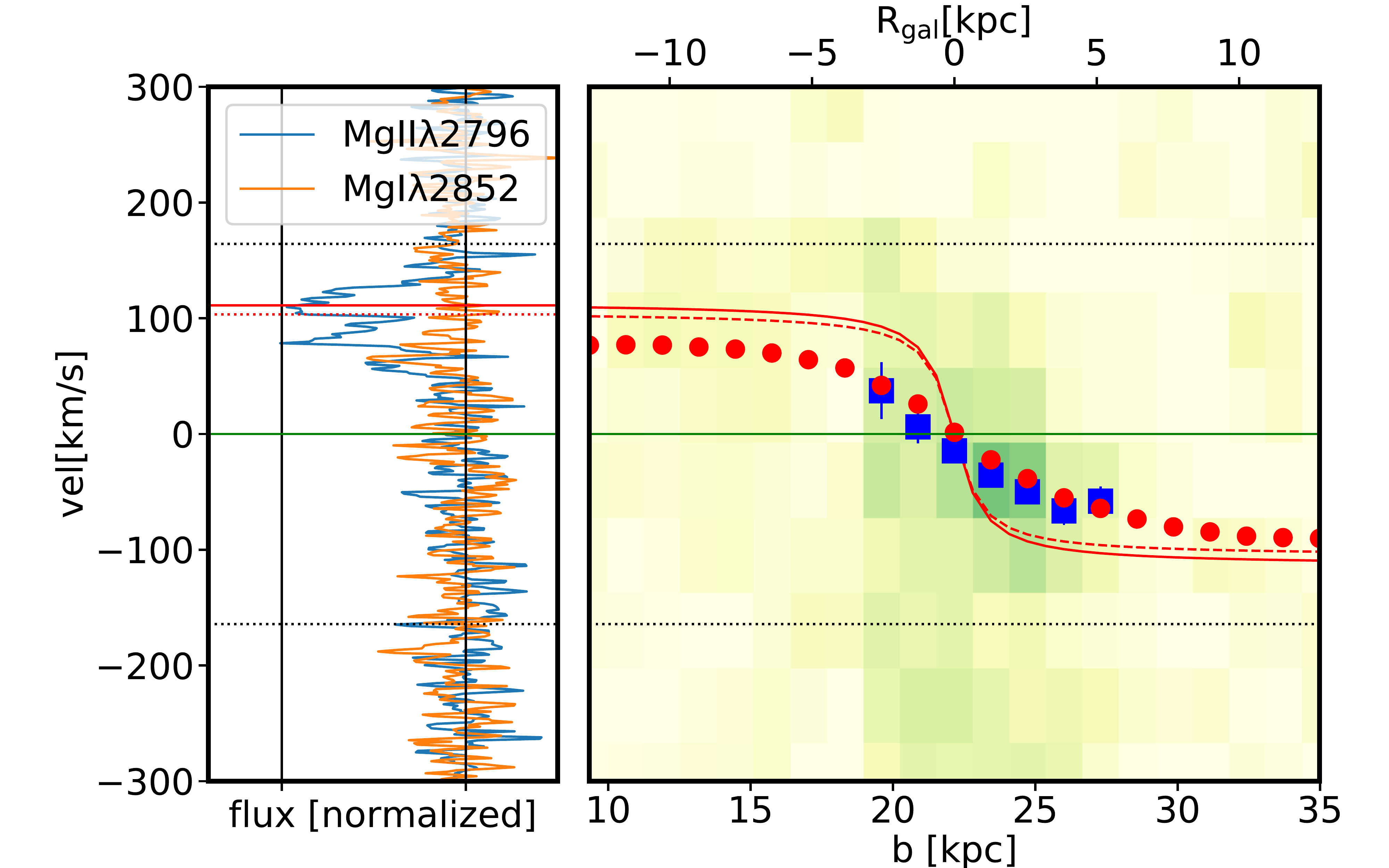}

\caption{\label{fig:rotcurve_vs_abs} Comparison of galaxy and absorber
    kinematics at the example of \emph{J0145}. Identical plots for all
    \nsampletext{} galaxy-absorber pairs are available in
    \Fig{fig:all_rotcurve_vs_abs} of the Supplementary Appendix. The
    \textbf{right panel} shows the 1D galaxy rotation curve (blue points)
    obtained from the 2D PVD diagram (shown as background image) on the \OII{}
    doublet (see \Sec{sec:comparison_gal_abs_kin}). The red points are
    obtained by reproducing this measurement procedure on the seeing convolved
    best-fit \gpk{} model. The solid red line represents the intrinsic \gpk{}
    rotation curve along the galaxy major-axis. The dashed red line represents
    the modeled rotation curve along the line connecting the galaxy and quasar
    positions on the sky. The lower x-axis represents the distance $b$ from
    the quasar along this connecting line. The upper x-axis shows the
    galacto-centric distance along the galaxy's major axis. In the
    \textbf{left panel} the \MgII{}$\,\lambda2796$ and \MgI{}$\,\lambda2852$
    absorption profiles are shown on the same velocity scale as the galaxy
    rotation curve. The solid red line in this panel indicates
    $\vmax$ at the observed inclination, which is a continuation of
    the red curve in the right panel. Similarly, the red dashed line is the
    continuation of the rotation curve along the galaxy-quasar axis. Further,
    the black dotted line shows $\vmax$ at $incl=90^\circ$ and the green line
    is the systemic redshift as obtained from \gpk{} ($v=0\,\kms$).}
\end{figure}

\begin{table}
\centering
\begin{tabular}{cccccc}
\hline
ID & $v_\text{peak}$ & v sign & $\rew_{;\MgII}$ & $\rew_{;\MgI}$ & $\rew_{;\FeII}$ \\ 
(1) & (2) & (3) & (4) & (5) & (6) \\
 \hline
 \hline
\cellcolor{GalColJ0103} J0103 & -52 & -1 & 1.1 & 0.2 & 0.5 \\
\cellcolor{GalColJ0145} J0145 & 112 & 1 & 0.5 & 0.1 & 0.1 \\
\cellcolor{GalColJ0800} J0800 & 23 & 1 & 0.8 & 0.1 & 0.3 \\
\cellcolor{GalColJ1039} J1039 & -144 & -1 & 0.8 & 0.2 & 0.4 \\
\cellcolor{GalColJ1107} J1107 & 60 & -1 & 0.4 & 0.1 & 0.2 \\
\cellcolor{GalColJ1236} J1236 & -41 & -1 & 2.1 & 0.7 & 1.6 \\
\cellcolor{GalColJ1358} J1358 & 62 & 1 & 2.6 & 0.5 & 1.9 \\
\cellcolor{GalColJ1509} J1509 & -116 & -1 & 1.5 & 0.3 & 1.0 \\
\cellcolor{GalColJ2152} J2152 & 63 & -1 & 0.6 & 0.1 & 0.2 \\

\end{tabular}
\caption{\label{tab:abs_ew_vel} Absorber properties. (2) Velocity at peak
absorption with respect to systemic redshift $[\kms]$. For details see
\Sec{sec:physprop:abskin}. (3) Sign of the galaxy rotation field at the
position of the quasar sight-line. (4-6) \rew{} for $\MgII\,\lambda 2796$,
$\MgI\,\lambda 2853$, $\FeII\,\lambda 2600$, respectively $[\text{\AA}]$. }
\end{table}

For the purpose of our work, we need an estimate of a 'characteristic' velocity
of the absorbing gas with respect to the systemic redshift defined by the
primary galaxy. In practice, we use here the velocity where the optical depth
is maximum. The caveat here is that \MgII{} is for most cases saturated and
hence the \MgII{} absorption profiles do not allow us to find the peak
absorption velocity. Therefore, we used the unsaturated $\MgI\,\lambda2852$
line to measure the peak optical depth, except for \emph{J0145}. In this case
the $\MgI$ line is too weak and we could use here the (nearly) unsaturated
$\MgII\lambda\,2796$ line.

The absorption profiles as obtained from the normalised UVES spectra are shown
both for $\MgII\,\lambda2796$ and $\MgI\,\lambda2852$ in the left panels of
\Fig{fig:rotcurve_vs_abs}.
The peak absorption velocities are listed in \Tab{tab:abs_ew_vel}, where we
also list rest-frame equivalent widths for $\MgII\,\lambda2796$, $\MgI\,\lambda
2852$, and $\FeII\,\lambda2600$.
 
As mentioned in \Sec{sec:sample:accretionsample}, the \emph{J1039}
galaxy-absorber pair at $z=0.9494$ is somewhat complicated by the presence of
another galaxy at 72~\kpc, i.e. 1.5 times the impact parameter of the primary
galaxy. Interestingly, the absorption system has two distinct components: a
weaker one ($\rewmgii\approx0.2\text{\AA}$) from $-40$ to $10\,\kms$ and a
stronger one ($\rewmgii\approx1.0\text{\AA}$) from $-80$ to
$-224\,\kms$~(Fig.~\ref{fig:rotcurve_vs_abs}). Given the anti-correlation
between impact parameter and \rewmgii\citep[e.g.][]{Chen:2010a, Nielsen:2013b},
it is more likely that the stronger component originates from the `primary`
galaxy's extended gas disk and the weaker component is due to an outflow from
the more distant galaxy, as further discussed in Schroetter et al. (in prep).

A further complication for this absorption system is that $\MgI\,\lambda2852$ is
contaminated by $\MgII\,\lambda2796$ of an absorber at $z=0.9875$. Using the
profile shape from the isolated $\MgII\,\lambda2803$ of the $z=0.9875$ system, we
could conclude that the \MgI{} peak absorption of the $z=0.9494$ absorber is
the reddest peak within the velocity range covered by the strong \MgII{}
component.

Finally, there is also a complication for the $\MgI{}\,\lambda2852$ absorption of
\emph{J1358}. At an observed wavelength of $6897\,\text{\AA}$ it is in a
wavelength region strongly affected by telluric absorption. We used the {\it
molecfit} software \citep{Smette:2015a, Kausch:2015a} to create a model
telluric transmittance spectrum in the region from $6860\,\text{\AA}$ to
$6940\,\text{\AA}$. Then we divided the science spectrum by this model telluric
transmission. The $\MgI$ spectrum shown in \Fig{fig:rotcurve_vs_abs} is the
telluric corrected version.

\section{Line-of-sight kinematics in comparison to galaxy kinematics}
\label{sec:result}

With our sample of \nsampletext{} galaxies geometrically selected to be
likely to probe extended disk-like structures (cf.
\Sec{sec:sample:accretionsample}), we now perform a direct comparison between
galaxy and absorber kinematics in order to investigate the existence and
properties of a large gaseous structure. In~\Sec{sec:comparison_gal_abs_kin},
we qualitatively compare the \MgII{} and \MgI{} absorptions with the galaxy
kinematics. In~\Sec{sec:results:corotation}, we make a quantitative comparison
of the absorption kinematics with simple models for the kinematics of an
extended gaseous structure with no radial motion. In~\Sec{sec:results:vrot_vr},
we discuss the absorption kinematics adding a radial component.
In~\Sec{sec:angmomentum}, we discuss the implication for the angular momentum.
Finally, in \Sec{sec:mass_accretion_rate} we discuss crude estimates of the
accretion rates onto our galaxies.

\subsection{Comparison of galaxy and absorber kinematics and qualitative test for co-rotation}
\label{sec:comparison_gal_abs_kin}

In order to compare the absorber and galaxy kinematics for each of the
\nsampletext{} galaxies in our sample,  we show, in \figrotcurvesanduves{},
the galaxy kinematics in a PVD. The PVD is obtained from a pseudo 2D spectrum
including the \OII{} doublet where the $x-$axis (upper) represents the
projected distance along the galaxy major-axis oriented with the quasar
line-of-sight on the left.  The $y$-axis represents the velocity scale set for
the \OII{}$\,\lambda3729$ line of the doublet, where the zero point of the
velocity scale is set by the galaxies' systemic redshift  (see
\Sec{sec:kin:redshifts}).  For each galaxy-absorption pair, the absorber
kinematics (as discussed in \Sec{sec:physprop:abskin}) is shown in the left
parts of the panels.

In the PVDs, the blue solid squares represent the observed rotation curve,
where the velocity is measured using a double Gaussian fit  on each spatial
pixel of the PVD with  the \mpdaf{} \citep{Piqueras:2017a} routine \emph{gauss\_dfit}.  The red solid
circles represent the seeing convolved rotation curve obtained from the \gpk{}
model cube exactly in the same way as for the data, i.e. by performing line
fitting on a pseudo PVD generated from the cube along the galaxy major-axis.
The red solid line represents the seeing corrected, intrinsic, rotation curve
obtained from the \gpk{} model (\Sec{sec:physprop:kinematics}).  
The red
dashed line represents the intrinsic rotation curve along the axis intercepting the quasar location.
As we selected our galaxy-absorber pairs to have $\alpha < 40\deg$, this axis
and the major axis are typically very close to each other, and solid and dashed
rotation curves differ in most cases little from each other. 
\figrotcurvesanduves{} show that, qualitatively, the majority of absorbers
tends to follow the rotation curve kinematics, i.e. tends to be blue or
red-shifted like the rotation curve is for the side towards the quasar
location.  

In  \Fig{fig:comp_abs_gal} (upper), we show all the rotation curves and the
corresponding absorptions profiles in a single plot. In this figure, we have
self-consistently flipped both the \gpk{} rotation curve (represented by the
solid lines) and the absorber velocity for those cases where the quasar is on
the side of the galaxy where the galaxy velocity field is negative (=
blue-shifted). The absorption profile is represented by the grey vertical bar
where darker regions indicate more absorption.  \Fig{fig:comp_abs_gal} (lower)
shows essentially the same as the upper panel, with the difference that here
the impact parameter and the velocity are scaled by their respective virial
values, $\rvir$ and $\vvir$.  In both panels, the stars indicate  the peak
absorption velocities as defined in \Sec{sec:physprop:abskin} and listed in
\Tab{tab:abs_ew_vel}.

From this figure, several important conclusions can be drawn. First and maybe
most importantly that the majority of absorbers share the
velocity-sign of the galaxy's velocity field. Indeed, seven out of the
\nsampletext{} galaxy-absorber pairs in our sample (i.e. all except
\emph{J2152} (\Colgali) and \emph{J1107} (\Colgale) share the velocity sign and
hence meet the minimal condition for co-rotation, if we use the peak velocity.
This shows that the gas traced by our \MgII{} absorbers is part of co-rotating
structures, supporting the basic  prediction from a pure co-rotating disk-like
structure \citep[e.g][]{Stewart:2013a,Stewart:2017a,Danovich:2015a}.

This co-rotation is in contrast to the expectation for gas clouds
on random orbits.  Assuming that the gas is randomly distributed, the
probability for 7 or more out of the \nsample{} sight-lines fulfilling the
co-rotation criterion is 9\%.  If we restrict the sample to the four galaxies
at $b<25\kpc$, the probability for four out of four sight-lines to be
consistent with co-rotation is 6\%.

In addition to this velocity sign test, a second order prediction can be
tested. The absorption velocities should not exceed $\vvir\sin(i)$.\footnote{
The product $\vmax\sin(i)\approx\vvir\sin(i)$ can be accurately measured even
if the inclination itself cannot be well constrained.}
\Fig{fig:comp_abs_gal}(b) shows that, reassuringly,  very little absorbing gas
exceeds $\vvir\sin i$, except \emph{J1039} which  has the most absorption with
$\vert\vlos\vert>\vvir\sin i$.

To better illustrate the global kinematic shift between the absorbing gas and the
systemic redshift, we proceed to stack the \nsampletext{}  absorption profiles
(normalized as in \Fig{fig:comp_abs_gal}b).  The stacked velocity profile are
shown in \Fig{fig:abs_stack}  for $\MgII\lambda2796$, $\MgI\lambda2852$,
$\FeII\lambda2600$.  In this Figure, gas that is co-rotating with the
galaxies' velocity fields is indicated in \colcorot{}, while gas that is
counter-rotating is coloured \colantirot{}. While  the co-rotating part is
significantly larger than in the counter-rotating part, this figure also shows
that there is almost no absorption below $\vlos = -0.5\vvir \sin i$ in the
counter-rotating direction. We will relate these features quantitatively to simple
disk models in \Sec{sec:results:corotation}.

\begin{figure*}
    \centering
    \includegraphics[width=0.9\textwidth]{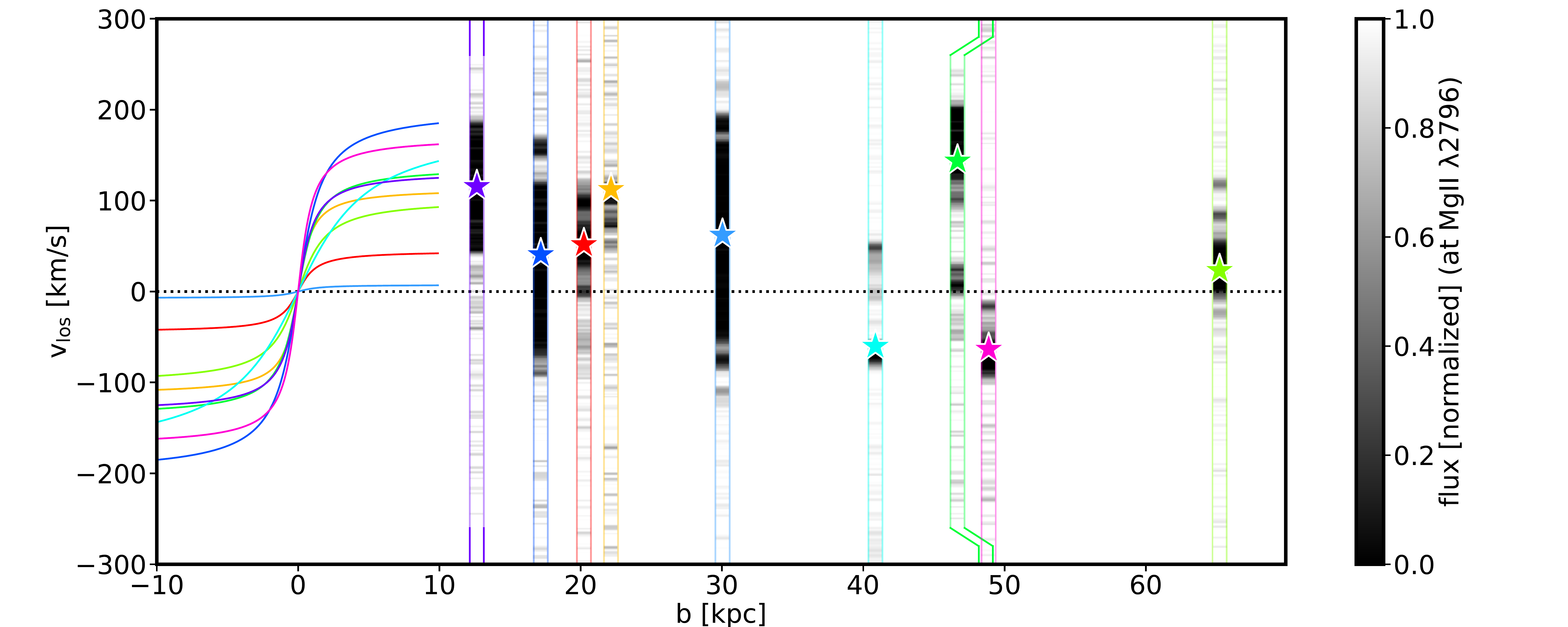}
    \includegraphics[width=0.9\textwidth]{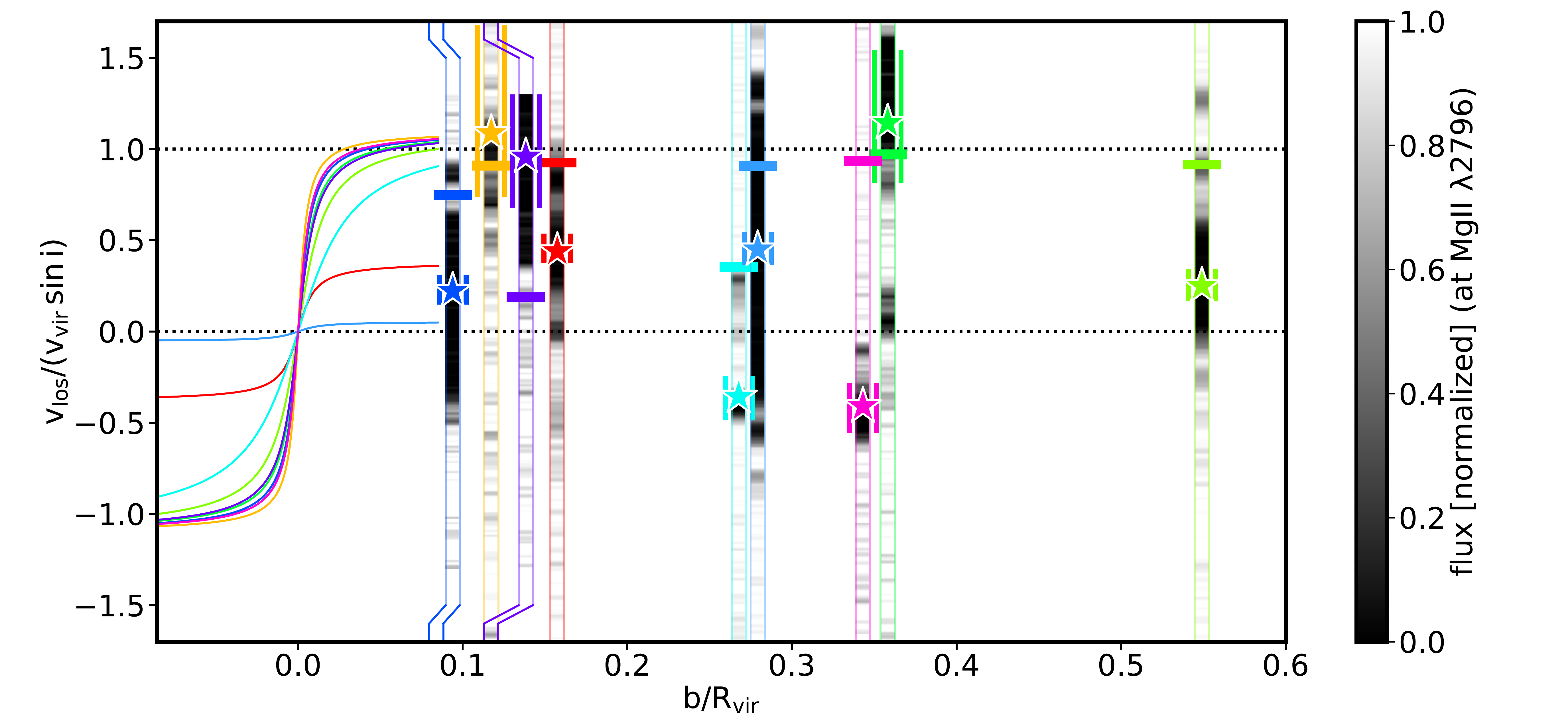}
    \caption{\label{fig:comp_abs_gal}  
    $\MgII\,\lambda2796$ absorption compared to galaxy rotation for each of the
    \nsampletext{} galaxies-absorber pairs of this study. Galaxy-absorber pairs
    can be identified by matching colours (as in    \Tab{tab:kinematics_and_morphology}).  
     The grey vertical bars represent the \MgII{}$\,\lambda2796$ profile where the darker regions indicate more absorption,
     and  shifted in the x-direction when they would overlap. 
 The stars indicate    the peak of the \MgI{} profile (see \Sec{sec:physprop:abskin}). 
The solid lines represent the   rotation curves  (at the observed inclination), as obtained from the
    $\textnormal{GalPaK}^{\textnormal{3D}}$ \OII{} fits, and measured along the galaxy
    major axis. 
The top panel shows the kinematics as a function of impact parameter $b$, not galacto-centric distance.
The bottom panel shows  the same as in the upper panel, but here $b$ and $\vlos$ have been normalized by
    $\rvir$ and $\vvir\,\sin i$, respectively. The coloured bold vertical bars
    indicate the impact of the uncertainties for $\vvir\,\sin i$ on the
    normalized peak absorption velocity.   
    The coloured horizontal bars indicate the velocity
    expected from extrapolating the rotating galaxy disk out to the galacto-centric radius of the quasar sight-line assuming a thin disk and rotation with $\vvir$.     }
\end{figure*}

\begin{figure*}
	\includegraphics[width=\textwidth]{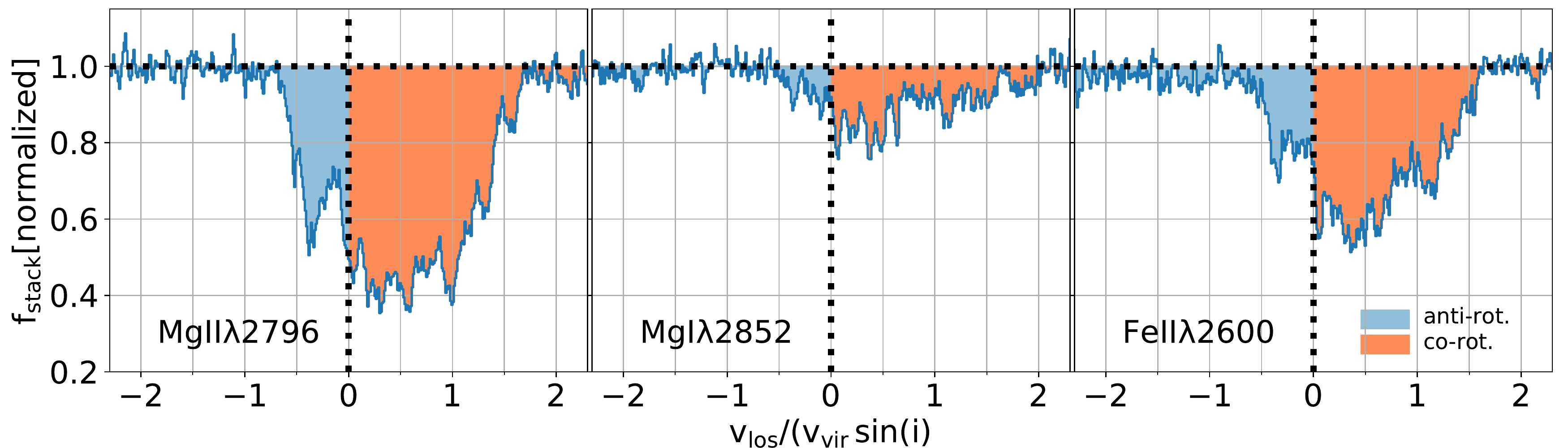}
\caption{\label{fig:abs_stack} Average absorption profile for our sample of
    \nsampletext{} galaxy-absorber pairs. The \MgII{} stack was obtained by
calculating the mean of the profiles shown in the lower panel of
\Fig{fig:comp_abs_gal}. The stacks for \MgI{} and \FeII{} were obtained in an
identical way.  As in \Fig{fig:comp_abs_gal}, for those pairs where the
galaxy-rotation was blue-shifted towards the quasar sight-line, we flipped both
galaxy and absorber velocities. Consequently, all co-rotation has
positive velocities in these stacks (\colcorot), while all counter-rotating gas
has negative velocities (\colantirot).}
\end{figure*}


\subsection{Thin disk with pure circular orbits}
\label{sec:results:corotation}

In the previous section, we have tested, in a relatively general way, whether
the gas probed by the low-$\alpha$ sight-lines approximately shares the
orientation of the angular momentum vector with the host galaxy.  Drawing more
quantitative conclusions from the line-of-sight (LOS) velocities is difficult,
for two main reasons. First, a line-of-sight velocity measurement does not
allow one to get the 3D velocity vector of the probed gas. Second, the
line-of-sight probes gas at different spatial positions with different
velocities.  Therefore, quantitative conclusions can only be drawn using a
parametrised model for the 3D density and kinematic of the gas.

The simplest possible model is to assume a thin disk which is spatially
perfectly co-aligned with the galaxy-disk, meaning that it can be understood as
an extension of the galaxy disk.  As the galaxy's disk orientation in space -
except for an ambiguity in the sign of the inclination - was determined by
means of \gpk{} modelling, we can predict where the sight-line would cross such
a (thin) disk. 

A quantitative comparison requires a description of the velocity field from
pure rotation at the galacto-centric radius, i.e. the distance from the galaxy
centre to the quasar location in the plane of the galaxy. A natural coordinate
system is the cylindrical coordinate system ($R,\,\phi$) where the cylinder is
perpendicular to the disk.  In this system, the radial coordinate is the
galacto-centric radius $\Rgal$ and $\phi$ is set arbitrarily to zero along the
projected major axis, i.e. $\phi=0^\circ$ where $\alpha=0^\circ$.  The
cylindrical coordinates $\phi$ and $\Rgal$ are related to the azimuthal angle
$\alpha$ (introduced in \Sec{sec:sample:accretionsample}) and impact
parameter $b$ through:

\begin{eqnarray}
    \tan \phi  &=& \frac{\tan \alpha}{\cos i} \hspace{5Ex} \label{eq:results:phi} \\
       R &=& b {(1 + \sin^2 \alpha     \tan^2 i)}^{1/2} \label{eq:results:R}
\end{eqnarray}
where $i$ is the galaxy inclination. The velocity vector of the gas in the disk
plane can then be described as:

\begin{eqnarray}
    \bmath{v_{gas}}(\phi,R) &=&  \vrot(R)\,\bmath{\hat{\phi}}(\phi) \label{eq:results:v_gas}
\end{eqnarray}
where $\vrot(R)$ is the tangential component of the rotation velocity, i.e.
there is no radial flow component and the gas is on (perfect) circular orbits.

Together with the unit vector along the quasar-sight-line, $\bmath{N}$, the
line-of-sight component of the velocity field can be determined by this simple
linear relation \citep[e.g.][]{Barcons:1995a}:

\begin{eqnarray}
    \vlos & = &\bmath{v_{gas}} \cdot \bmath{N} \label{eq:los_prod} \\ 
          & =& \frac{\vrot(R) \cos \alpha \sin i}{\sqrt {1 + \sin^2 \alpha \tan^2 i}}
           \label{eq:vlos_simplerot}
\end{eqnarray}
The linear relation between the line-of-sight velocity $\vlos$ and $\vrot$
means that we can either predict the $\vlos$ for gas with a certain $\vrot$, or
alternatively, determine the angular velocity, $\vrot$ from a measurement of
$\vlos$ of the gas, if the gas is indeed on perfectly circular orbits in the
hypothesized disk.

\Eq{eq:vlos_simplerot} shows that the line-of-sight velocity $\vlos$ will
always be smaller than the tangential velocity $\vrot$, i.e. $\vlos\leq\vrot$.
However, the range of values for the ratio $\vlos/\vrot$ can be estimated for
the range of allowed values in azimuthal angle $\alpha$ and inclination $i$
imposed by  our geometric selection (\Sec{sec:sample:accretionsample}). For
azimuthal angle  $\lesssim 30^{\circ}$, which corresponds to the range in our
sample, and inclinations in the range [40--70$^{\circ}$],
\Eq{eq:vlos_simplerot} implies that $\vlos$ ranges from $0.5\mbox{--}0.9\vrot$
($=0.5\mbox{--}1.0\,\vrot\sin i)$. Only for the highest inclinations
($i>70^{\circ}$), the range of values becomes very large, essentially from
$0.0\mbox{--}1.0\vrot$, with the sensitivity
on the inclination depending on $\alpha$. In other words, except for extreme
inclinations, $\vlos$ is expected to be $0.5\mbox{--}1.0\,\vrot\sin i$.

While this range of values is consistent with our observations, this simple
model would predict an absorption over a very narrow velocity range at $\vlos$,
whose width would be given by the disk turbulence ($\sim20-40$ \kms).  In
reality, we do not measure a single $\vlos$, but the absorption profiles (or
stacked profile) cover a relatively large range in velocities, from -0.5 to 1.5
$\vvir\sin(i)$ (Fig.~\ref{fig:abs_stack}).

One possibility to explain the spread in the absorption profiles is to assume
that the hypothesized disk has some thickness \citep[e.g.][]{Steidel:2002a,
Kacprzak:2010a, Ho:2017a}. In that case, the sight-line intercepts the gaseous
disk at different heights, corresponding to different radii $R$, leading to a
range of projected $\vlos$,  even if the gas would be on pure circular orbits
with a constant $\vrot$.  However, this broadening effect would be a few tens
of \kms, and is not sufficient to explain the range of velocities in
\Fig{fig:abs_stack}.

With the possible caveat of a very thick disk observed at high $i$, we assume that the peak velocity - the velocity where the optical
depth reaches its maximum -, as determined in \Sec{sec:physprop:abskin},  can
be used as an estimator for $\vlos$ at the mid-plane.  This line-of-sight
velocity $\vlos$ measured at $\phi$ and $R$ can then be used to estimate the
tangential velocity $\vrot$  using \Eq{eq:vlos_simplerot}.

Note that for sight-lines with large azimuthal angle $\alpha$ (i.e. close to
our selection limit of $40^\circ$) and for galaxies with very high inclination
($i>70^{\circ}$),  the galactocentric radius can be very large and very
uncertain. This affects especially \emph{J1107} and \emph{J1509}.

\begin{figure}
    \includegraphics[width=\columnwidth]{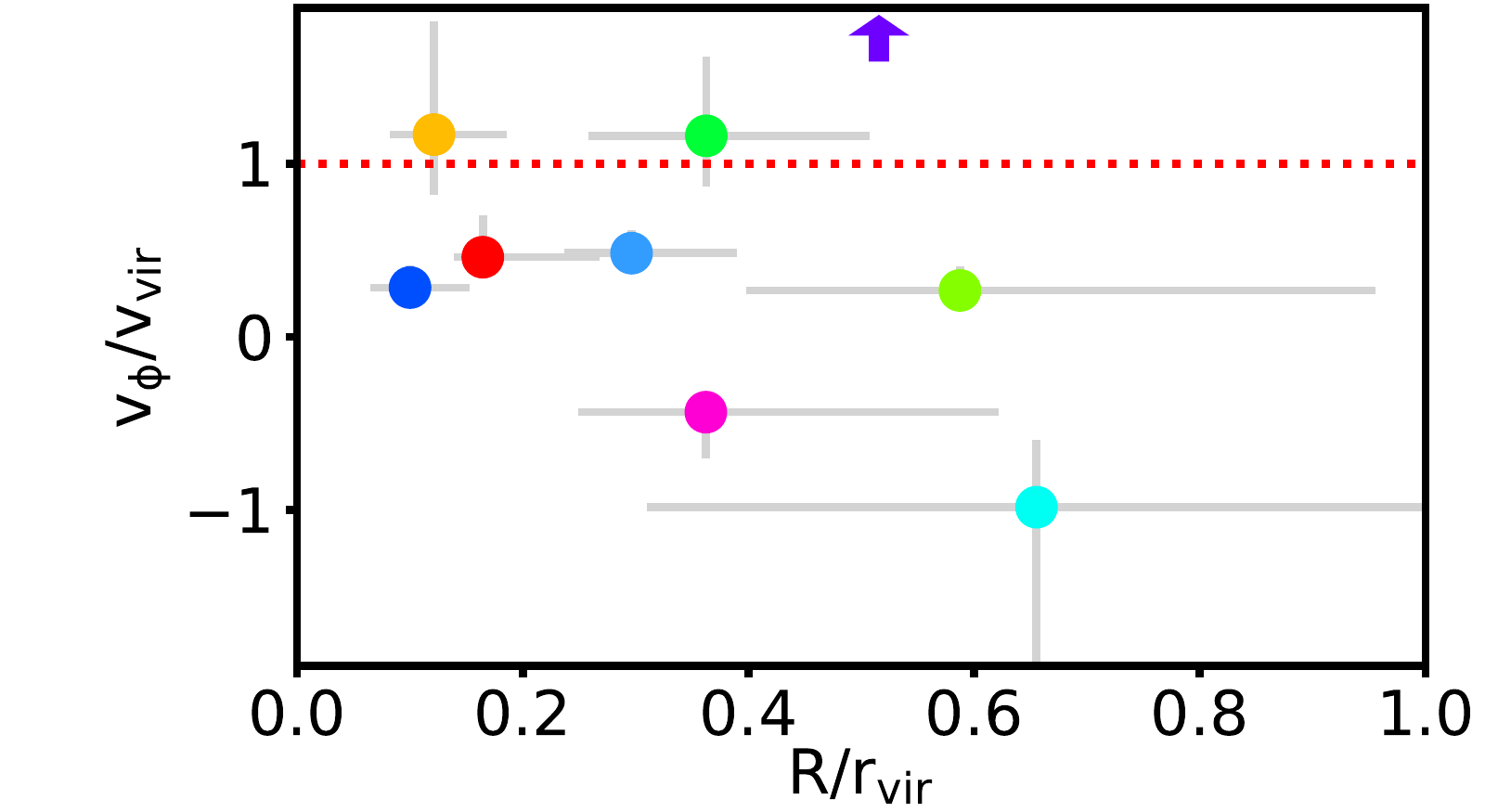}
    \caption{\label{fig:results:vrot_simple} Tangential velocity $\vrot$
        inferred from $\vlos$ under the assumption that the gas is on pure
        circular orbits in the disk plane. For gas on stable circular orbits
        the expectation is that $\vrot\approx\vvir$. The two cases at the
        extreme ends of the $\vrot/\vvir$ distribution (\emph{J1107} cyan; \emph{J1509}
        purple, indicated by an arrow as outside of plot range) have both the
    highest $\alpha$'s and inclinations among our \nsampletext{}
galaxy-absorber pairs. The values of the plotted points are listed in \Tab{sup:tab:geo_and_kin_absgas} of the Supplementary Appendix.
}
\end{figure}
    
This deprojected tangential velocity $\vrot$ can then be compared to the
circular velocity $\vcirc$, taken to be $\approx\vvir$, in order to assess
whether the gas motion is consistent with stable circular orbits.
\Fig{fig:results:vrot_simple} shows the ratio $\vrot/\vvir$ as a function of
the galactocentric radius $\Rgal$ (normalized by $\rvir$). The error bars are
large for \emph{J1107} and \emph{J1509}, as expected from the discussion above.
Excluding these two objects, there seems to be a trend with galacto-centric
radius, in a sense that galaxies at smaller $R/\rvir$ are closer to being on
stable circular orbits in the extended galaxy disk. 

This trend is also visible by simply looking at the absorption profiles in
\Fig{fig:comp_abs_gal}. We indicate in the lower panel of this Figure the
expected $\vlos$ for circular orbits with $\vvir$ at disk-mid plane crossing.
While in most cases the peak absorption is not coincident with the expected velocity,
there is in many cases at least some absorption at the position of
the exact co-rotation.

\subsection{Disk with angular and radial motion -  Cold accretion disks}
\label{sec:results:vrot_vr}

While the comparison between galaxy and halo gas absorption velocities for our
sample strongly (\S~\ref{sec:comparison_gal_abs_kin}) indicates that the halo
gas probed by the low ionization lines shares at least approximately the
direction of the galaxy's angular momentum vector, the previous section showed
that the data is in most cases not consistent with the hypothesis that the gas is on stable
circular orbits in a perfect extension of the galaxy disks.

Here,  we try to gain further insight by extending our thin-disk toy model with
a non-zero radial velocity component $\vr$. The velocity vector of the gas,
$\bmath{v_{gas}}$, at a certain position in the disk can then be described by:

\begin{equation}
    \bmath{v_{gas}}(\phi,R) =  \vrot(R)\,\bmath{\hat{\phi}}(\phi)  + \vr(R)\bmath{\hat{r}}(R),
\end{equation}
while the observed line-of-sight velocity  $v_\mathrm{los}$ follows (using  \Eq{eq:los_prod}):
\begin{equation}
    v_\mathrm{los}   =  \frac{\vrot \cos \alpha \sin i + \vr \sin \alpha \tan i}{\sqrt{1 + \sin^2 \alpha \tan^2 i}}.
                    \label{eq:vlos_full}
\end{equation}

Unfortunately, removing the constraint of perfect circular orbits and adding
the possibility of inflow means that it is no longer possible to infer the
velocity vector of the gas from the measured $\vlos$, as there are in this
scenario two unknowns for one measurement. 

However, as discussed in \Sec{sec:results:corotation} for the case of pure
circular rotation, we can estimate the range of values for the tangential and
radial terms in \Eq{eq:vlos_full} imposed by our geometrical selection
(\Sec{sec:sample:accretionsample}). For the tangential term, the conclusion is
similar as in \Sec{sec:results:corotation}, i.e. the numerical value ranges
from $0.5\mbox{--}0.9\,\vrot$ for azimuthal angles $\alpha\lesssim30^{\circ}$ and
inclinations from 40 to 70$^{\circ}$. For the radial term, the absolute value
of the numerical factor ranges for this $\alpha$ and $i$ range from $0.0 \mbox{--}
0.8$. Unless the radial velocity $|\vr|$ is
{\it larger} than the tangential component $|\vrot|$, the tangential component
will \emph{typically} dominate the line-of-sight velocity $\vlos$, regardless of
geometrical factors, except when both $i$ and $\alpha$ are large.

\begin{figure}
    \includegraphics[width=\columnwidth]{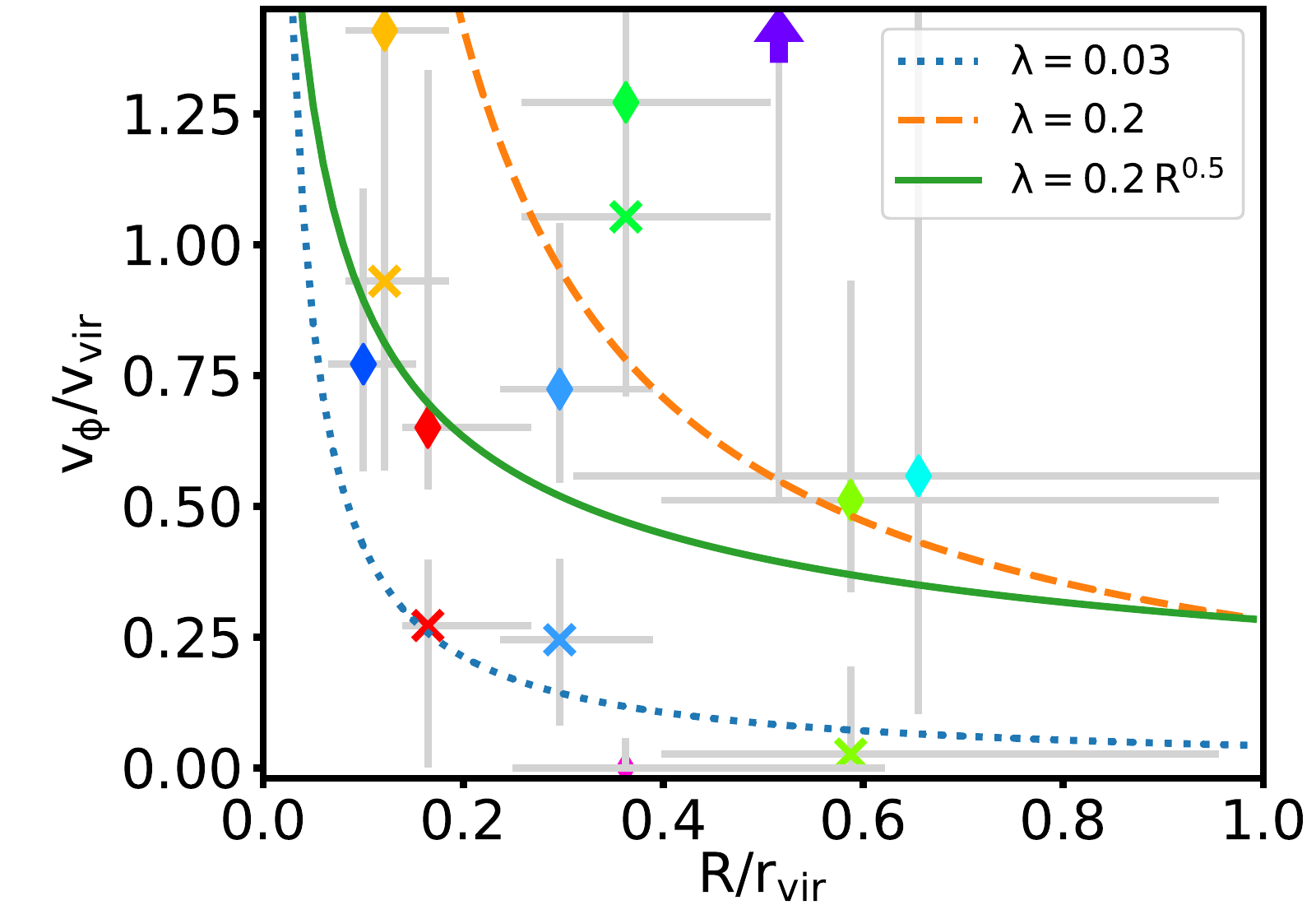}
    \caption{\label{fig:results:vrot_vr} Rotation velocities, $\vrot$, inferred
        from measured $\vlos$ under the assumption of a radial velocity
        component with $\vr=-0.6\vvir$ and co-rotation with the galaxy for each
        of the \nsampletext{} absorbers. There can be up to two solutions per
        absorber.  The projected LOS components of $\vr$ and $\vrot$ have the
        same sign for solutions indicated by stars, while they have opposite
        signs for solutions indicate by diamonds. The lines indicate different
    spin parameters ($\lambda$, see \Eq{eq:spin_lambda}), with their values
indicated in the legend. The values of the plotted points are listed in
\Tab{sup:tab:geo_and_kin_absgas} of the Supplementary Appendix. }
\end{figure}

In order to break the degeneracy between radial and circular velocity, it is
necessary to impose some additional information to reduce the dimensionality.
Motivated by results from simulations one justifiable assumption is a constant
$v_r$ \citep[e.g.][]{Rosdahl:2012a,Goerdt:2015a}.  By measuring $\vr$ directly
from different simulations with different codes as a function of radius,
redshift, and halo mass, \citet{Goerdt:2015a} find that typical inflow
velocities are approximately constant with radius at $\vr \approx -0.6\,\vvir$ for
redshift $z\approx1$.

We now test the impact of the assumption of
$\vr=-0.6\,\vvir$ and the possible signatures that this would have on our data.
Such a radial component means that the magnitude of the radial term in
\Eq{eq:vlos_full} is thus at most $|\vr|<0.6\,\vvir$, while the tangential
component is $\approx 0.5\mbox{--}0.9\,\vvir$, for inclinations less than 70$^{\circ}$
and $\alpha<30^{\circ}$.  This implies that disks with such a radial flow allow a wide
range of line-of-sight velocities $\vlos$, from $\approx0 \mbox{--} 1.1$
$\vvir$   in most general situations,
i.e. except for the most extreme inclinations $i>70^{\circ}$, and thus can
better reproduce the range of velocities observed in \Fig{fig:abs_stack}.

Now, we use the $\vlos$ observations of our individual cases to put constraints
on the tangential velocity with the assumption of a radial flow at
$\vr=-0.6\vvir$. Given that we do not know whether the line-of-sight is
crossing the disk on the near or far side, this means the radial contribution
to the line-of-sight velocity $\vlos$ in \Eq{eq:vlos_full} can have a sign
opposite to the tangential component $\vrot$. Consequently, two solutions can
be found depending on the sign of the inclination, i.e.

\begin{equation}
    v_\mathrm{los}   =  \frac{\hbox{sign($v_{\rm rot}$)} |\vrot| \cos \alpha \sin |i| \pm \vr \sin \alpha \tan |i|}{\sqrt{1 + \sin^2 \alpha \tan^2 i}}, \label{eq:vlos_radial}
\end{equation}
where the $+(-)$ sign corresponds to whether the quasar intercepts the gaseous
disk on the far or near side. Given that we impose corotation between the gas
and host galaxy, i.e. sign$(\vrot)$ is given by sign$(v_{\rm rot})$, no
solution can be obtained, if the $\vlos$ has the opposite sign as the assumed
sign for $\vrot$, and the absolute value of $\vlos$ is larger than the absolute
value of the line-of-sight component from the radial velocity. With
\Eq{eq:vlos_radial}, it is possible to have either one, two or no solutions,
depending on $\vlos$.

In Figure~\ref{fig:results:vrot_vr}, we use \Eq{eq:vlos_radial} to put
constraints on $\vrot$ for the galaxies in our sample with our assumption of
$\vr=-0.6\vvir$. The resulting solutions for $\vrot$ normalized by $\vvir$ are
shown  as a function of $R/\rvir$. Solutions where the $\vlos$ contribution
from $\vr$ shares the sign with the contribution from $\vrot$, are indicated as
crosses, while solutions where they have the opposite sign are shown as
diamonds.  A case where the measured value does not give a co-rotating solution,
but a co-rotating solution is possible within the uncertainties, is indicated
by the small pink diamond at  0.36 for \emph{J2152}.

A first important conclusion from Figure~\ref{fig:results:vrot_vr} is that,
under the assumption of this infall velocity, there is for each absorber at
least one solution consistent with $0<\vrot/\vvir<1$ in the co-rotating direction, even
for the two cases which would in the absence of infall be interpreted to have a
counter-rotating circular velocity. 

\subsection{Consequences for the angular momentum}
\label{sec:angmomentum}

Using the constraints on the tangential velocity $\vrot$ from the previous
section, we now analyse the implications for the gas and galaxy angular
momentum. Indeed, a crucial aspect for any theory of galaxy evolution is to
understand the amount of angular momentum which the accreted gas carries with
it. A useful quantity, which expresses the specific angular momentum of the
gas, $j=\vrot \Rgal$, in a form independent of the virial parameters, is
through the dimensionless spin-parameter, $\lambda$. We use the definition of
$\lambda$ from \citet{Bullock:2001b}:

\begin{equation}
    \lambda = \frac{1}{\sqrt{2}}\frac{j}{\rvir\vvir} = 
\frac{1}{\sqrt{2}}\frac{R}{\rvir}\frac{\vrot}{\vvir} \label{eq:spin_lambda}
\end{equation}

Pure N-body simulations have shown that $\lambda$ of the dark matter component
integrated over the halo is approximately distributed with a log-normal
distribution with $\vert \lambda \vert = 0.035$  and $\sigma=0.5$, almost
independent from redshift and halo mass \citep[e.g.][]{Bullock:2001b,
    Bett:2007a}.  Some hydrodynamical simulations predict that the cold gas
    should have a higher $\lambda$ on its way to the galaxy than the dark
    matter \citep[e.g.][]{Stewart:2011a, Teklu:2015a, Danovich:2015a}. E.g.
    \citet{Danovich:2015a}, using a similar simulation as \citet{Goerdt:2015a},
    predict that the spin parameter $\lambda$ for the cold gas, which we
    presumably probe with our data, scales with $(\Rgal/\rvir)^{0.5}$, and has
    approximately a $\lambda=0.06$ at $R/\rvir=0.1$ and $\lambda=0.2$ at
    $\Rgal/\rvir=1.0$.

We indicate in \Fig{fig:results:vrot_vr} the curves of constant $\lambda=0.03$
(blue dotted line) and $\lambda=0.2$ (orange dashed line). In addition, we show
the curve $\lambda=0.2 (R/\rvir)^{0.5}$ (green solid line). All galaxy-absorber
pairs except \emph{J2152} have within the uncertainties at least one solution
within $0.03 < \lambda < 0.2$. For most of the absorber pairs a solution with
$\lambda=0.2(R/\rvir)^{0.5}$ seems plausible, but also a solution with constant
$\lambda\approx0.03$ is consistent with several of the absorbers.

\subsection{Amount of accretion compared to star formation}
\label{sec:mass_accretion_rate}

\begin{table}
\centering
\begin{tabular}{crrr}
\hline
ID &   $\Rgal/\rvir$ &  {\small $\log_{10}(\cos(i) N_\mathrm{HI})$}& $\dot{M}_\mathrm{in}(R)$ \\
 (1) & (2) & (3) & (4)   \\
\hline
\cellcolor{GalColJ0103}J0103 & $0.16_{-0.03}^{+0.10}$ & $18.9{\scriptstyle \pm0.5}$ & $0.9_{-0.6}^{+2.0}$ \\
\cellcolor{GalColJ0145}J0145 & $0.12_{-0.04}^{+0.06}$ & $18.6{\scriptstyle \pm0.5}$ & $0.7_{-0.5}^{+1.5}$ \\
\cellcolor{GalColJ0800}J0800 & $0.59_{-0.19}^{+0.37}$ & $18.7{\scriptstyle \pm0.5}$ & $1.8_{-1.3}^{+4.0}$ \\
\cellcolor{GalColJ1039}J1039 & $0.36_{-0.10}^{+0.14}$ & $18.9{\scriptstyle \pm0.5}$ & $2.7_{-2.0}^{+6.0}$ \\
\cellcolor{GalColJ1107}J1107 & $0.66_{-0.35}^{+1.27}$ & $17.9{\scriptstyle \pm0.5}$ & $0.7_{-0.6}^{+1.8}$ \\
\cellcolor{GalColJ1236}J1236 & $0.10_{-0.03}^{+0.05}$ & $19.8{\scriptstyle \pm0.5}$ & $12.5_{-9.1}^{+27.5}$ \\
\cellcolor{GalColJ1358}J1358 & $0.30_{-0.06}^{+0.09}$ & $19.9{\scriptstyle \pm0.5}$ & $17.7_{-12.3}^{+38.4}$ \\
\cellcolor{GalColJ1509}J1509 & $0.52_{-0.33}^{+1.52}$ & $18.8{\scriptstyle \pm0.5}$ & $2.2_{-1.9}^{+6.5}$ \\
\cellcolor{GalColJ2152}J2152 & $0.36_{-0.11}^{+0.26}$ & $18.4{\scriptstyle \pm0.5}$ & $1.0_{-0.7}^{+2.1}$ \\
\hline
\end{tabular}
\caption{\label{tab:results:accretion} Accretion measurements (see \Sec{sec:mass_accretion_rate}) (2) Galacto-centric radius (see \Eq{eq:results:R}) normalised by virial radius, $\rvir$; (3) 
    \HI{} column density perpendicular to disk (based on \MgII{} equivalent width) [$cm^{-2}$] ;
 (4) Mass accretion rate [$\mpy$].}
\end{table}

Galaxies need to accrete gas from their halo. A plausible reservoir might be
the extended cold gas disks probed by our sample.  In section
\ref{sec:results:vrot_vr} we have shown that the $\vlos$ measured for this gas
are indeed consistent with infall rates and angular momentum predicted by
simulations. Motivated by this, we can estimate the mass accretion rates from
the \MgII{} equivalent widths and the assumed infall velocities ($\vr = -0.6
\vvir$).

For this estimate, first the $\MgII\,\lambda 2796$ equivalent widths need to be
converted to approximate \HI{} column densities using the relation from
\citet{Menard:2009a}: \begin{equation} N_\HI = (2.45\pm0.38)\times 10^{19}
    \mathrm{cm}^{-2}
\left(\frac{\rewmgii}{\textnormal{\AA}}\right)^{2.08\pm0.24} \end{equation} The
$\rewmgii-N_\mathrm{HI}$ relation is not a tight relation. We assume in the
following an approximate statistical uncertainty of $0.5\,\mathrm{dex}$ on the
$N_\mathrm{HI}$ estimates. However, significantly stronger outliers are not
unlikely. 

Then, the accretion rate, $\dot{M}_\mathrm{in}(R)$ through a cylinder at the
galacto-centric radius $\Rgal$ can be calculated as in \cite{Bouche:2013a}:

\begin{equation} 
    \dot{M}_\mathrm{in}(R) \approx
    2\pi\,\Rgal\,v_\mathrm{r}\,m_p\,\mu\,\cos(i)\,N_\mathrm{HI}
    \label{eq:m_inflow} 
\end{equation} 

Here the assumption is made that the disk is thin enough so that the column
density perpendicular to disk can be estimated from the column density along
the inclined view through $\cos(i)\,N_\mathrm{HI} $. Further, $m_p$ is the
proton mass, and $\mu$ is the mean molecular weight, assumed to be 1.6. A
further assumption is that all $\vrot$, $\vr$, and the perpendicular column
density do at fixed galacto-centric radius not depend on the disk azimuthal
angle.  The resulting $\dot{M}_\mathrm{in}$ are listed in
\Tab{tab:results:accretion}.

\begin{figure}
    \centering
	\includegraphics[width=0.75\columnwidth]{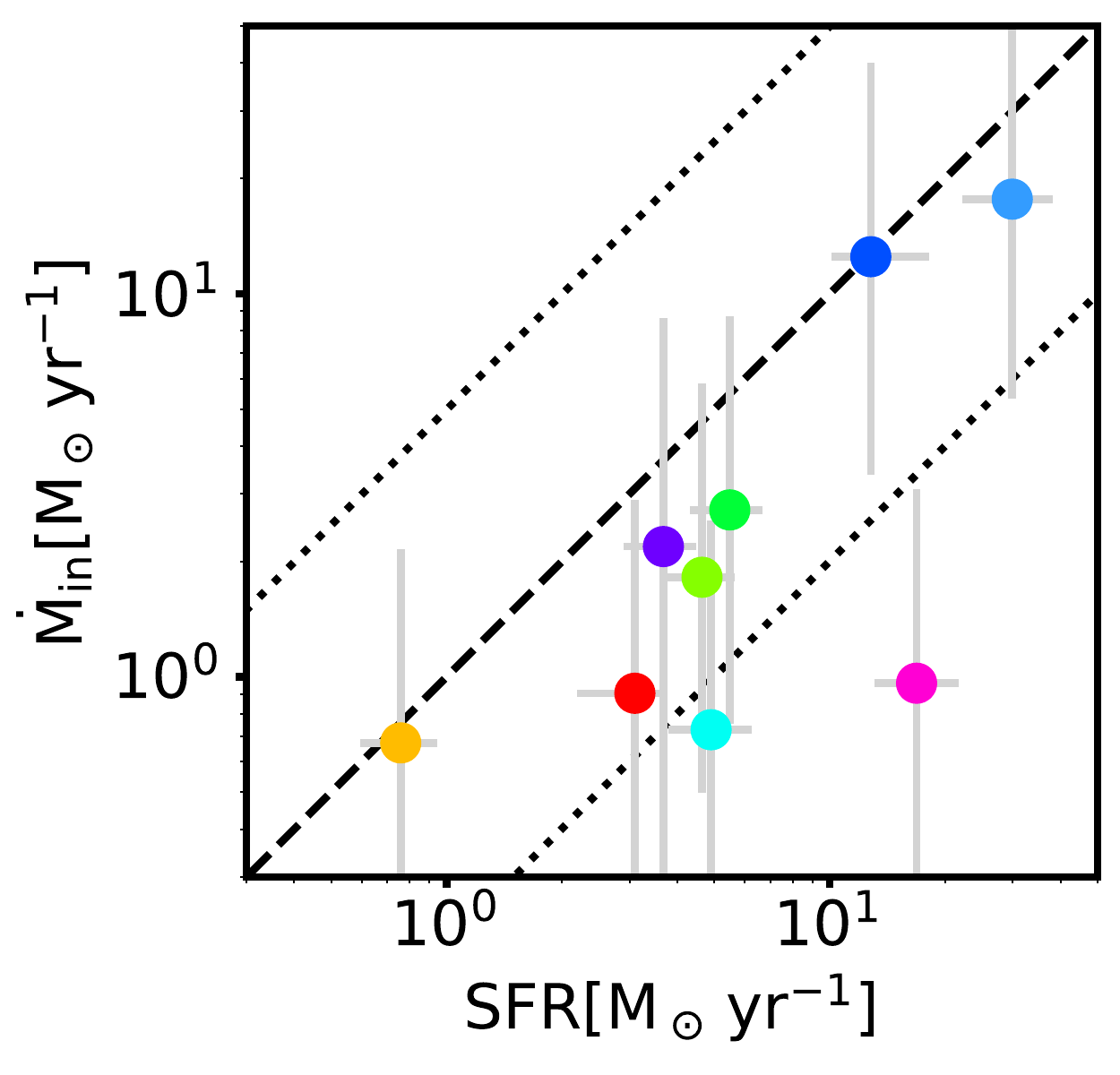}
	\caption{\label{fig:infall_vs_sfr}
    Mass inflow rate estimated from $\MgII$ equivalent width plotted against
    $\mathrm{SFR}$ estimate obtained from \OII{} flux corrected for reddening using the $E(B-V)$ estimate based on the \citet{Garn:2010a} $M_*$-extinction relation. 
The dashed line represents a 1:1 relation. The dotted lines represent deviations from the
1:1 correspondence by a factor five.  
}
	
\end{figure}

In \Fig{fig:infall_vs_sfr}, we compare the estimated $\dot{M}_\mathrm{in}$   to
the SFRs computed  using the \OII{} luminosities with the extinction estimated
from the $M_\star$-extinction relation of \citet{Garn:2010a} as described in
\Sec{sec:oiiflux}.  The dashed line shows the 1:1 relation, and the dotted
lines are deviations from the 1:1 relation  by a factor five.  This figure
shows that  the estimated amount of $\dot{M}_\mathrm{in}$ seems sufficient to
balance the SFRs.  Given that our estimate from \Eq{eq:m_inflow} only takes
account for the neutral component, the total $\dot{M}_\mathrm{in}$ could be
sufficiently higher if an ionised or molecular component would contribute  to
the inflow. 

This result seems consistent with the expectation of self-regulation model
\citep[as in][]{BoucheN:2010a,Lilly:2013a} where
\begin{equation}
\hbox{SFR}=\frac{\epsilon_{\rm in}\,f_B\,\dot{M}_{\rm h}}{1-R+\eta} \approx \frac{\dot{M}_{\rm in}}{1-R+\eta}
\end{equation}
where $\epsilon_{\rm in}$ is the accretion efficiency, $f_B$ the baryon
fraction, $\dot{M}_{\rm h}$ the halo accretion rate at $\rvir$, $R$ the
`recycling fraction` which massive stars return to the ISM, and $\eta$ the
outflow loading factor, $\eta\equiv\dot{M}_{\rm out}/$SFR. In the mass range
for the galaxies in our sample $\eta$ is measured to be small $\lesssim 1$
\citep{Martin:2012a,Schroetter:2015a,Schroetter:2016a} and $R$ is around
$0.3\mbox{--}0.5$ depending on the IMF \citep[e.g.][]{Madau:2014a}.

\section{Summary \& Conclusions}
\label{sec:conclusions}

Using our ongoing \mfl{} survey \citep[][Bouch\'e et al. (in prep.), Schroetter
et al. (in prep.)]{Schroetter:2016a}, which targets galaxies around 79 strong
\MgII{} absorbers (with $\rewmgii\gtrsim0.3\text{\AA}$) at $z\approx1$ with the
VLT/MUSE spectrograph, we have investigated the distribution and kinematic
properties of this low-ionization gas.  Remarkably, the distribution of the
azimuthal angle $\alpha$ for the galaxy-absorbers pairs within $100\,\kpc$
shows a clear bi-modality (\Fig{fig:sel:alphavsincl}, top panel), which is
highly suggestive of a CGM geometry with biconical outflows and extended gas
disks. Our result confirms previous ones by
\citet{BoucheN:2012a,Kacprzak:2012a} and \citet{Bordoloi:2011a}.

In light of this bimodal distribution, we have selected for this study the
\nsampletext{} galaxy-absorber pairs which have the right orientation for the
purpose of studying extended gaseous disks at galactocentric distances of
$20\mbox{--}100\kpc$. This is the first statistical sample with both galaxy and
absorber kinematics at $z\approx1$ specifically selected to study the cold gas
disks.

Through a comparison of absorber and galaxy kinematics, we derived the following main conclusions:
\begin{itemize}

    \item There is little gas with $\vlos > \vvir \sin(i)$, which suggests
        that the gas in the disk is gravitationally bound to the halo.
        (\Sec{sec:comparison_gal_abs_kin}; \Figs{fig:comp_abs_gal} and
        \ref{fig:abs_stack})

    \item For seven out of the \nsampletext{} pairs the absorption velocity
        shares the velocity sign with the extrapolation of the galaxy rotation
        curve to the position of the quasar. This is the case for all four
        absorbers at $b/\rvir < 0.2$ (\Sec{sec:comparison_gal_abs_kin};
        \Fig{fig:comp_abs_gal}); 

    \item The inferred rotation velocity using the peak absorption velocity is
        in many cases smaller than $\vvir$. This indicates that the gas is not on purely
        circular orbits. There seems to be a tendency for the discrepancy to be 
        larger at larger impact parameters, with two cases at $b/\rvir > 0.25$
        even having counter-rotating velocities (\Sec{sec:results:corotation};
        \Fig{fig:results:vrot_simple}); 

    \item We investigated a scenario where the disk gas has a radial inflow
        (accretion component) in addition to the circular component, as
        predicted by simulations and required by indirect evidence. We showed
        that the data are consistent with inflow rates
        ($\vr\approx-0.6\vvir$) and angular momentum distributions
        (\Sec{sec:results:vrot_vr} \& \Sec{sec:angmomentum};
        \Fig{fig:results:vrot_vr}) from simulations;

    \item The inferred accretion rates are consistent with the expectation from the gas-regulator 
        \citep{BoucheN:2010a,Lilly:2013a,Dave:2012a}   (\Sec{sec:mass_accretion_rate}; \Fig{fig:infall_vs_sfr}) and from hydrodynamical calculations \citep[e.g.][]{FaucherGiguere:2011a,Nelson:2015a,Correa:2018a}.

\end{itemize}

For some of the conclusions listed above we needed to make relatively strong
assumptions. E.g. we assumed that the gas is in a perfect extension of the
galaxy disk and the angular momentum vector of the gas is perfectly aligned
with the galaxy. This is not exactly what simulations predict for the cold
streams falling into the halo, and only to some extent the case for the
cold-accretion disks into which the streams supposedly settle in the
inner-halo. In addition, extended gas disks might be warped. While the gas
needs to finally align with the galaxy disk, how this alignment happens and out
to where it persists is far from a solved question. In simulations it seems to
depend on the code and the feedback implementations
\citep[e.g.][]{Stewart:2017a}. However, reassuringly our $\alpha$ histogram
shows that the gas that we probe with \MgII{} needs to be at least in a disky
structure.

The finding of gravitationally bound co-rotating gas is consistent with a range
of quasar-absorption studies for individual objects with the right geometry
ranging up to $z\approx2$ \citep[e.g.][]{Barcons:1995a, Steidel:2002a, Bouche:2013a,
Bouche:2016a, Bowen:2016a, Rahmani:2018a}. The only other study that
systematically selected a statistical sample with the right geometry for the
purpose of studying extended cold gas disks was recently performed by
\citet{Ho:2017a}. Importantly, their work is targeting galaxies at
significantly lower redshift, $z\approx0.2$, and hence can be considered
complementary to our study. Qualitatively they find at this lower redshift a
similar result as we do: The majority of the \MgII{} absorption profiles
matches the sign of the galaxy rotation (8 out of their 12 robust
galaxy-absorber identifications). Similarly to us, they also find that the
majority of the gas has $\vrot<\vcirc$. Combining this with the fact that for
some systems part of the absorption counter-rotates, they conclude also - for
$z\approx0.2$ - that the gas likely has a radial infall component.

\section*{Acknowledgements}


 We thank the anonymous referee for a careful and constructive report, which helped to improve the quality of the manuscript.
This work has been carried out thanks to the support of the ANR FOGHAR (ANR-13-BS05-0010-
02), the ANR 3DGasFlows (ANR-17-CE31-0017), the OCEVU Labex (ANR-11-LABX-0060), and the A*MIDEX project
(ANR-11-IDEX-0001-02) funded by the ``Investissements d'avenir'' French
government program.
LW acknowledges support by the Competitive Fund of the Leibniz Association through grant SAW-2015-AIP-2.

This  work  made  use  of  the  following  open  source
software:
 GalPak3D  \citep{Bouche:2013a},
 ZAP \citep{SotoK:2016b},
 MPDAF \citep{Piqueras:2017a},
 matplotlib \citep{Hunter:2007a},
 NumPy \citep{vandderWalt:2011a, Oliphant:2007},  Astropy \citep{Robitaille:2013a}.



\bibliographystyle{mnras}
\bibliography{main_jz} 


\newpage
\clearpage
\appendix


\section{Uncertainties of inclination and azimuthal angle estimates}
\label{sup:sec:alpha-incl-uncerts}

In this appendix, we review the uncertainties (statistical and systematics) on
the three key parameters necessary for this paper, namely the galaxies'
inclinations, $i$, position angles, $\alpha$ (or $PA$), and maximum rotation
velocities, $\vmax$.  

The statistical error estimates obtained from  $\gpk$  are the Bayesian
uncertainties under the assumption of the model. From a comparison with the
dispersion of the differences between input and measured values in a matched
sample of simulated galaxies taken from \citet{Bouche:2015a} ($\gpk$ reference
paper), we inferred that the statistical uncertainties stated by \gpk{} are
mainly accurate, but potentially underestimated by about 20\%. Therefore, we
conservatively increased the Bayesian uncertainties by 20\%. 
 
The systematic errors can be caused by   mismatches between the parameterised
model and the data (especially at high SNR) and/or a slightly imperfect
characterisation of the PSF.  We  have estimated the level of systematic error
for the morphological parameters ($PA$ and $i$) in two different ways. First,
we compare two independent methods ( \gpk{}   and \galfit) on our seeing
limited data, and second, we compare the seeing-limited \gpk{} morphological
parameters on a separate sample with ancillary {\it HST} morphology.
 
The first method compares the  \gpk{}  (see \Sec{sec:kin:galpak}) and the
\galfit{} (2D) (see \Sec{sec:photometry})  morphological measurements ($PA$, $i$)
for  our sample galaxies.  We also used  two different measurements of the PSF
parameters, as described in \Sec{sec:data:quality}, namely the \mpdaf{} based
estimates of the Moffat profile for the \gpk{} measurements,  and  the
\pampelmuse{} based estimates of the PSF for the \galfit{} measurements.  Thus,
by comparing measurements from the two estimates we capture uncertainties all
due to the fitting code, the way we extract the PSF, and differences due to
using continuum light and line emission. 

The result  of this comparison is shown both for $\incl$ and $PA$ in
\Fig{fig:comp_parameters_galpak_vs_galfit}. In general, the agreement between
the two estimates is good, importantly without any strong outliers. The rms
difference between the \galfit{} and \gpk{} is $\sim9\deg$  for both $i$ and
$PA$.  Assuming that  the \galfit{} and \gpk{} measurements contribute about
equally, we attribute a $7\deg$ systematic uncertainty to the \gpk{}
measurements (for $\alpha$ and $i$). Note this  is always   larger than the
\gpk{} statistical uncertainties, and thus can be safely used to cover both
statistical and systematic uncertainties. 

For the second method, we compared the morphological parameters ($i$ in
particular) obtained with \gpk{} for a larger sample of about 60 \OII{}
emitters (Bouch\'e et al. in prep.) in seeing-limited MUSE observations in the
UDF \citep{BaconR:2017a} againt the $i$ measurements from the {\it HST} H-band
from \citet{VanderWel:2012a}. This second method gives a scatter of about
$7\deg$ between the two estimates, confirming our error budget estimate.

We end this section with a quantitative assessment of the systematic
uncertainties for the $\vmax$  parameter.  Using the  same UDF \OII{} emitter
sample, we compared $\vmax$ measurements from \gpk{} measurements to $\vmax$
determined with the \camel{} code \citep{Epinat:2012a} discussed in Contini et
al. (in prep.) and found a relative systematic error of about 10\%.  Thus, to
the $\vmax$ error budget stated in \Tab{tab:kinematics_and_morphology} of the
main manuscript, we added this 10\% relative error to the statistical errors in
quadrature.


\begin{figure}
    \begin{center}
    \includegraphics[width=0.47\columnwidth]{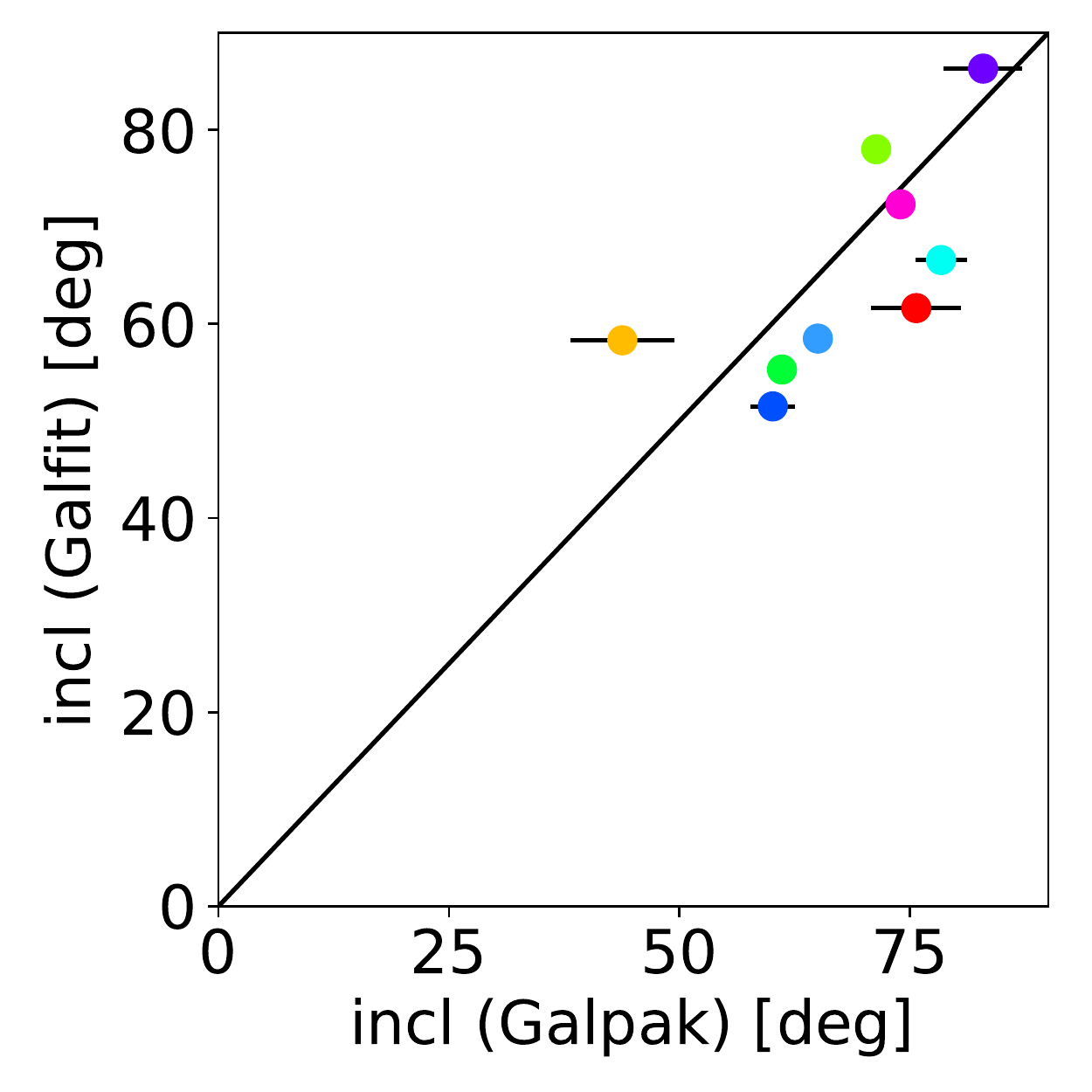}
    \includegraphics[width=0.47\columnwidth]{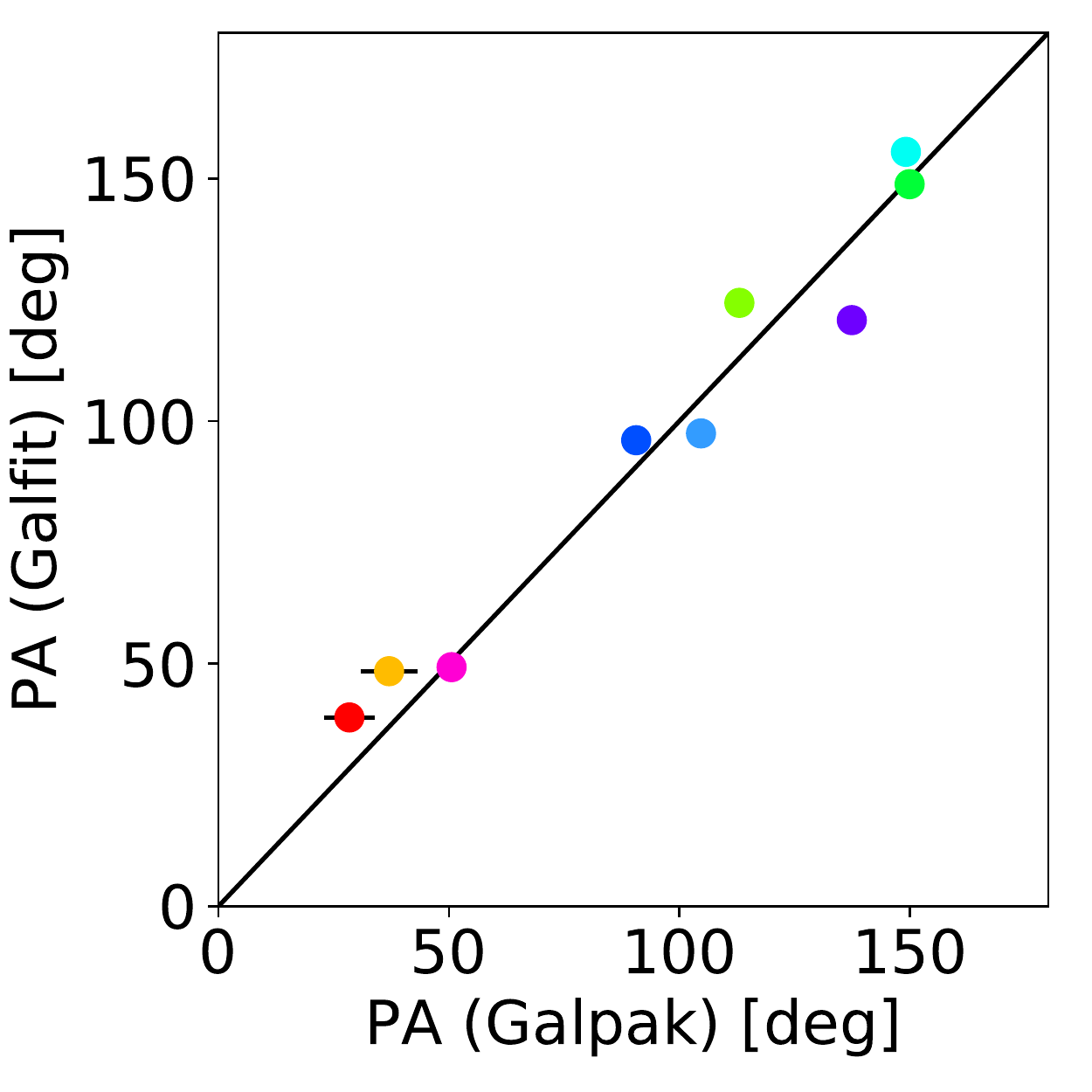}
\end{center}
\caption{\label{fig:comp_parameters_galpak_vs_galfit} Comparison between \gpk{}
    and \galfit{} based estimates for the inclination, $\incl$, (left) and position angle,
    $PA$, (right) for each of the \nsample{} galaxies in our sample. The statistical
 errors obtained from \gpk{} are shown as error bars, and are in some cases
 not visible, as they are smaller than the marker size. }
\end{figure}

\section{Field, spectral, and kinematic plots for each of the absorbers}

The same information as available in \Fig{fig:img_and_kin:J0145p1056_0554} and
\Fig{fig:spec:J0145p1056_0554} for \emph{J0145}, meaning an \OII{} NB FoV image
around the absorber redshift, kinematics of the primary galaxies, and the
spectral SED, is shown in \Fig{fig:img_and_kin:J0103p1332_0788} to
\ref{fig:img_and_kin:J2152p0625_1053} for the eight other galaxy-absorber pairs
of our sample.

\begin{figure*}
\includegraphics[width=\textwidth]{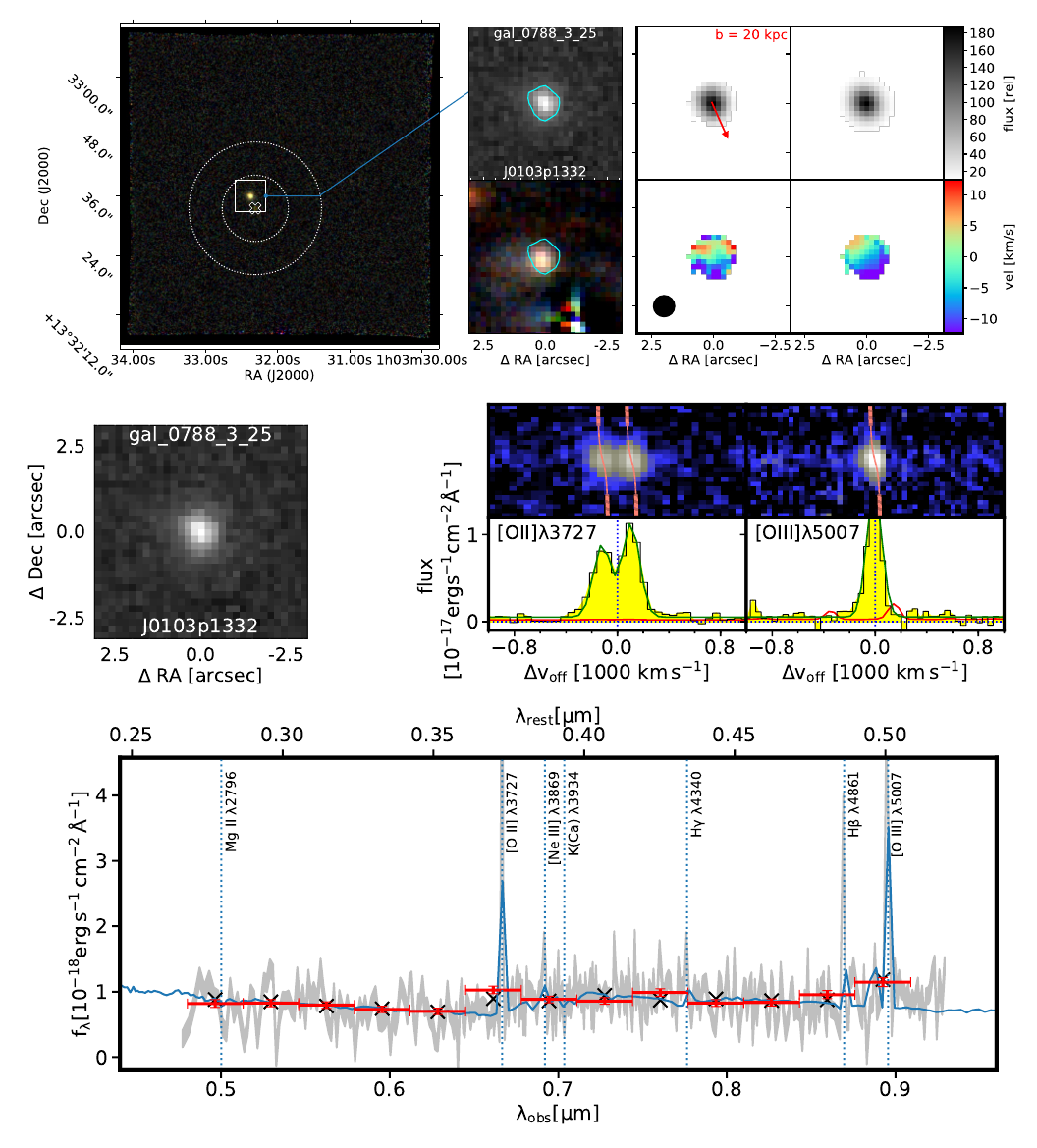}
    \caption{\label{fig:img_and_kin:J0103p1332_0788} Same as in Figs. \ref{fig:img_and_kin:J0145p1056_0554} and \ref{fig:spec:J0145p1056_0554}, but here for the main galaxy associated to the absorber abs\_J0103p1332\_0788.}
\end{figure*}

\begin{figure*}
\includegraphics[width=\textwidth]{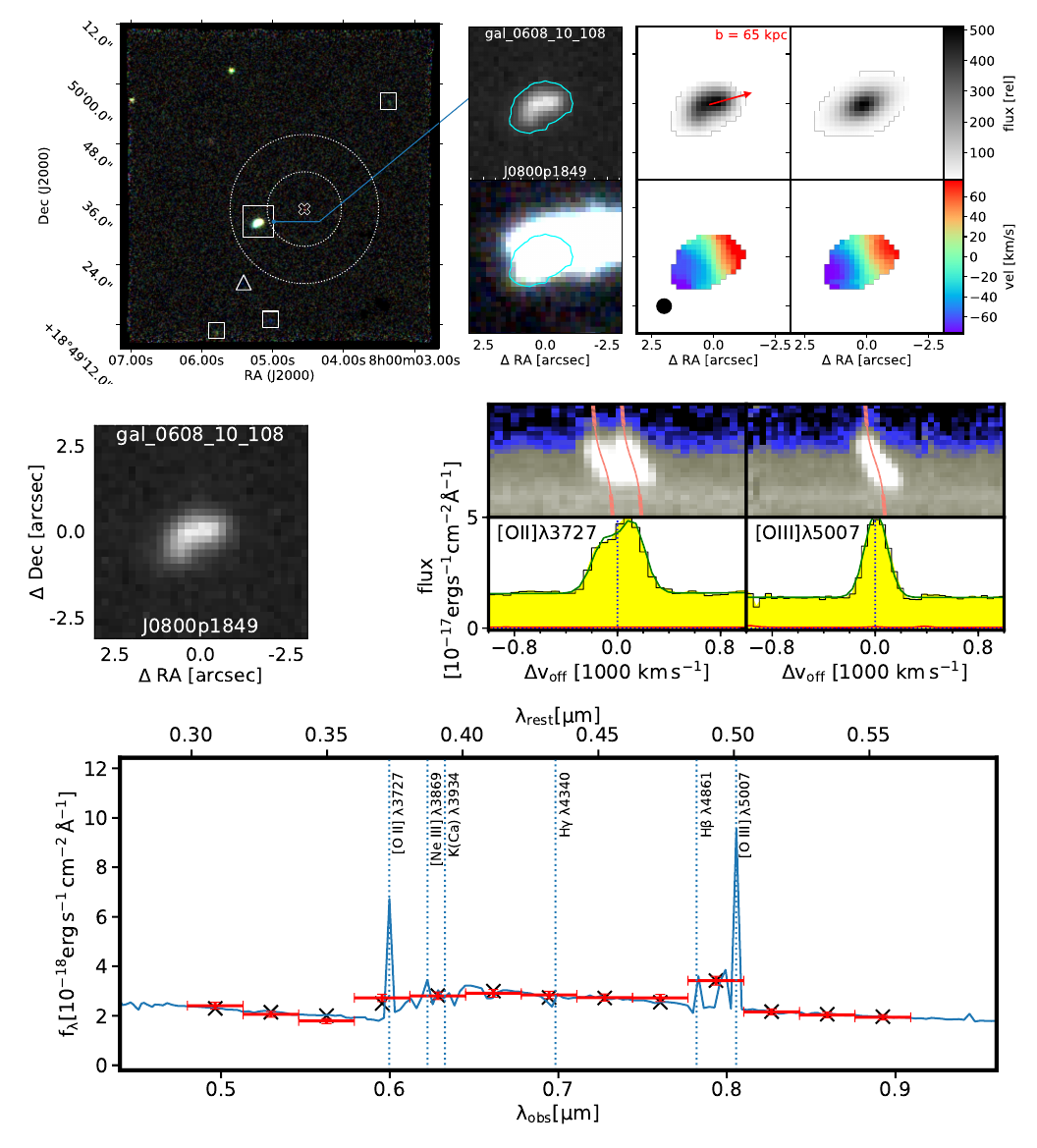}
\caption{\label{fig:img_and_kin:J0800p1849_0608} Same as in \Figs{fig:img_and_kin:J0145p1056_0554} and \ref{fig:spec:J0145p1056_0554}, but here for the main galaxy associated with the absorber abs\_J0800p1849\_0608. A pure absorption line galaxy is indicated as a white triangle. Here the grey line (aperture spectrum) is not visible, as the contamination from an unrelated foreground source is large.}
\end{figure*}

\begin{figure*}
\includegraphics[width=\textwidth]{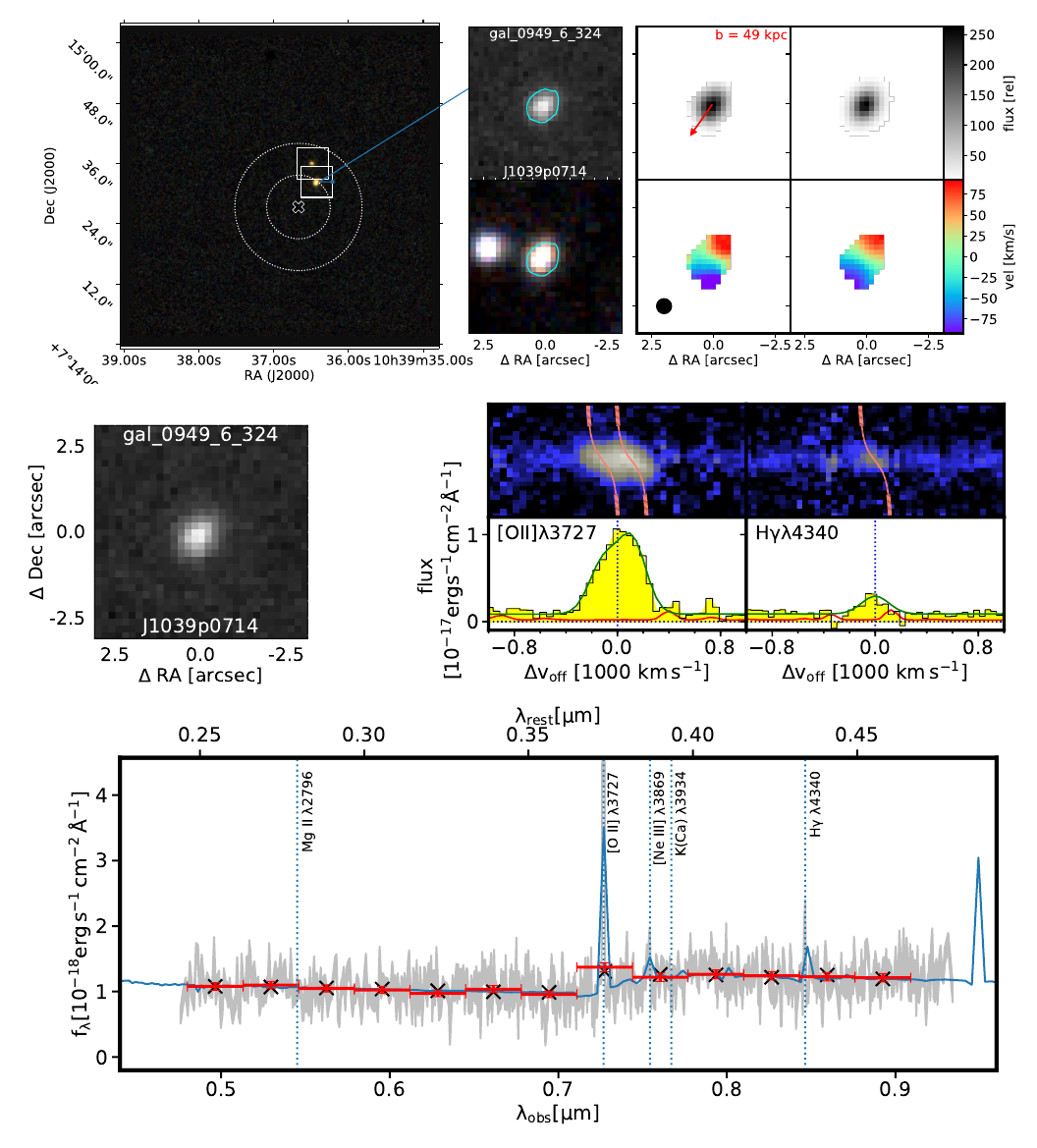}
    \caption{\label{fig:img_and_kin:J1039p0714_0949} Same as in \Figs{fig:img_and_kin:J0145p1056_0554} and \ref{fig:spec:J0145p1056_0554}, but here for the main galaxy associated with the absorber abs\_J1039p0714\_0949.}
\end{figure*}

\begin{figure*}
\includegraphics[width=\textwidth]{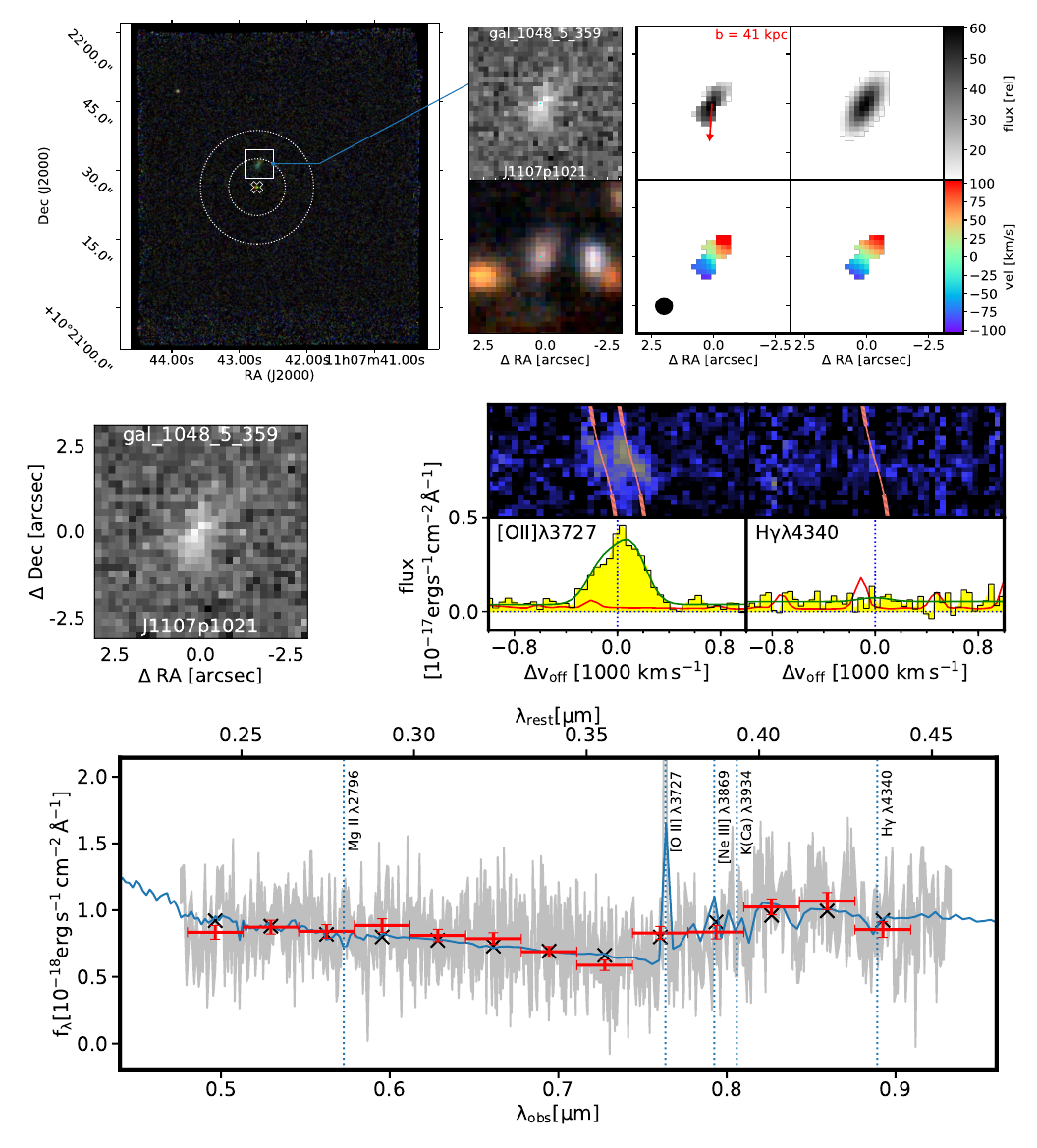}
    \caption{\label{fig:img_and_kin:J1107p1021_1048} Same as in \Figs{fig:img_and_kin:J0145p1056_0554} and \ref{fig:spec:J0145p1056_0554}, but here for the main galaxy associated with the absorber abs\_J1107p1021\_1048.}
\end{figure*}

\begin{figure*}
\includegraphics[width=\textwidth]{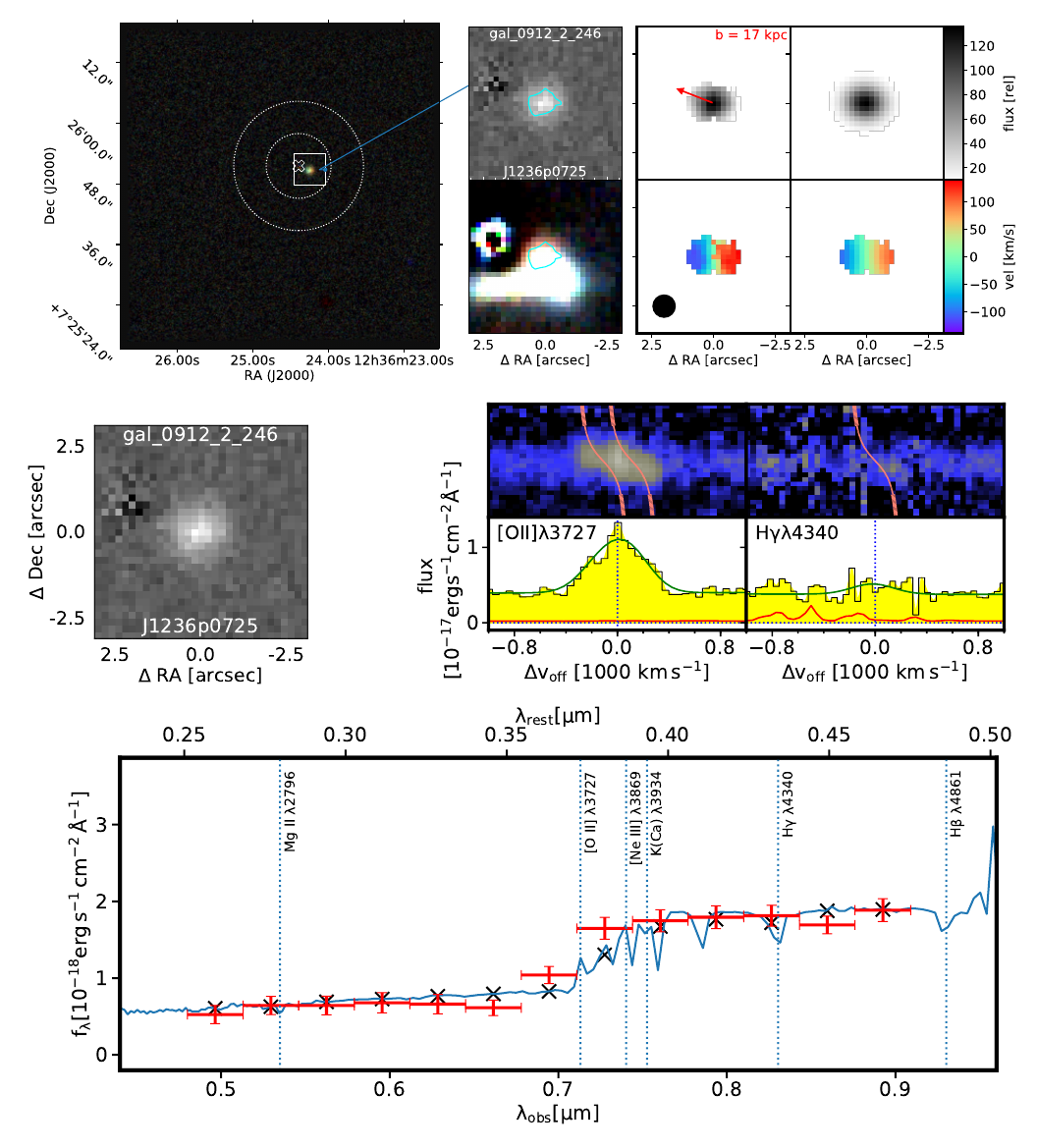}
    \caption{\label{fig:img_and_kin:J1236p0725_0912} Same as in \Figs{fig:img_and_kin:J0145p1056_0554} and \ref{fig:spec:J0145p1056_0554}, but here for the main galaxy associated with the absorber abs\_J1236p0725\_0912. Here the grey line (aperture spectrum) is not visible, as the contamination from an unrelated foreground source is large.}
\end{figure*}

\begin{figure*}
\includegraphics[width=\textwidth]{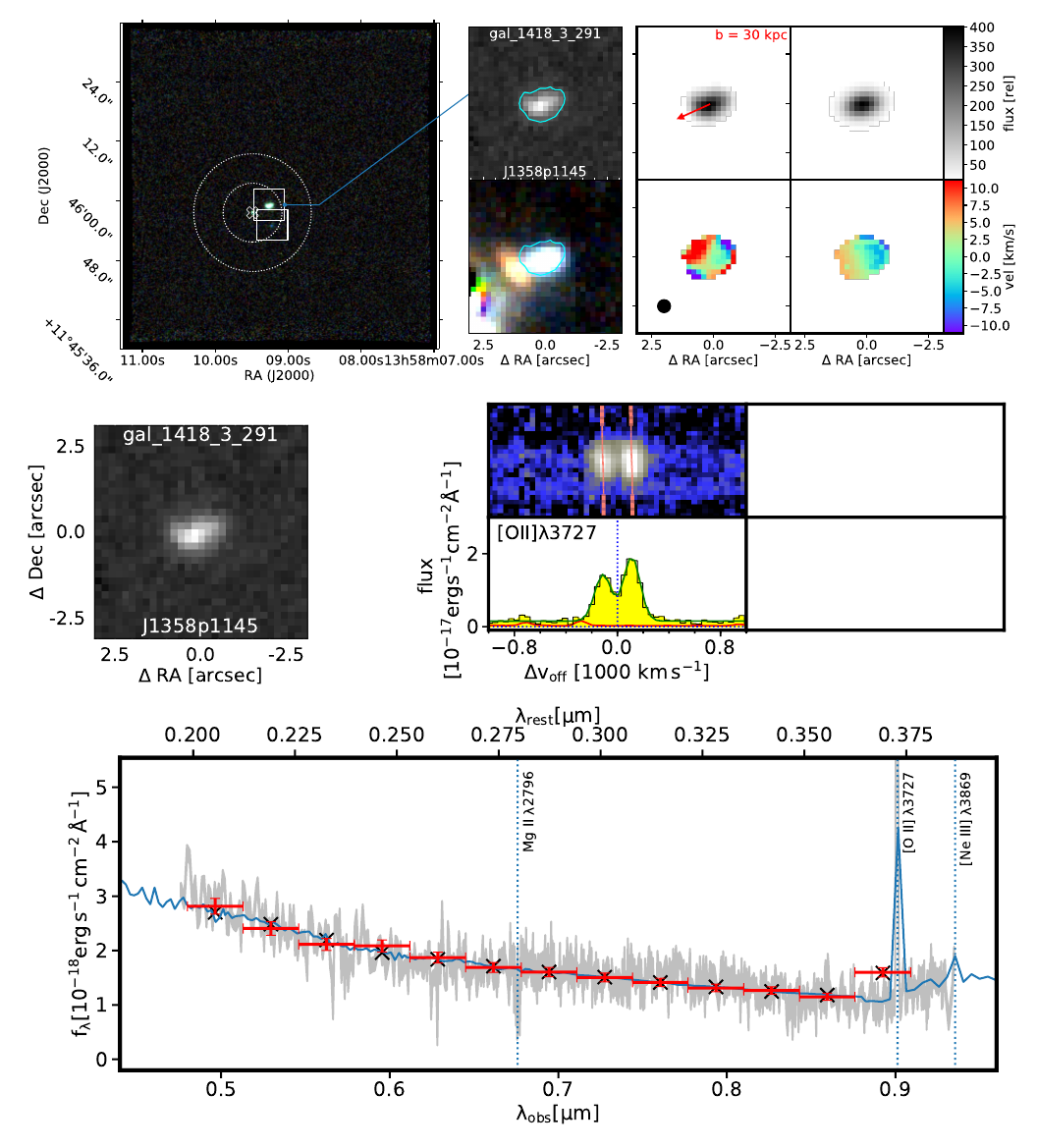}
    \caption{\label{fig:img_and_kin:J1358p1145_1418} Same as in \Figs{fig:img_and_kin:J0145p1056_0554} and \ref{fig:spec:J0145p1056_0554}, but here for the main galaxy associated with the absorber abs\_J1358p1145\_1418.}
\end{figure*}

\begin{figure*}
\includegraphics[width=\textwidth]{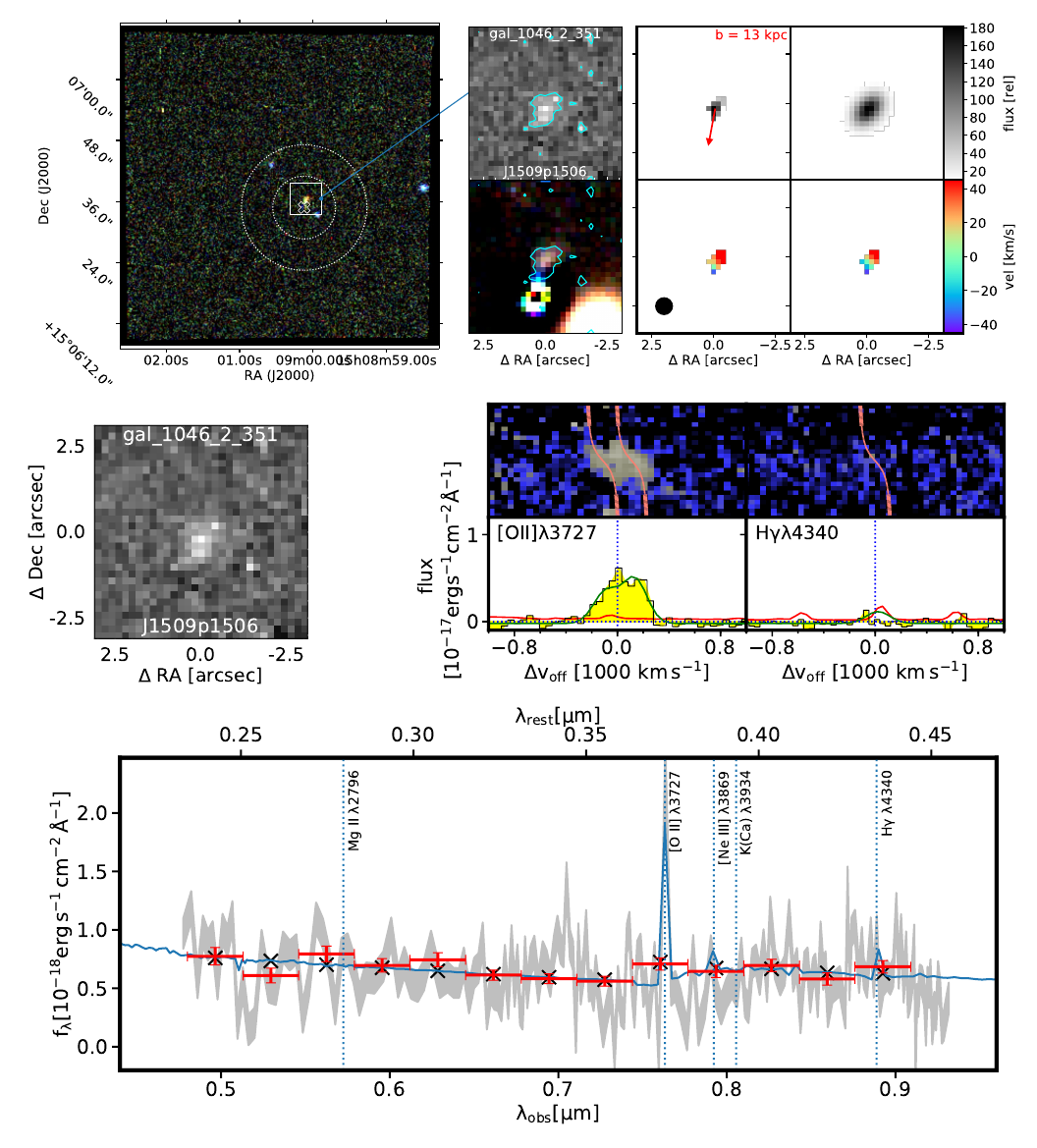}
    \caption{\label{fig:img_and_kin:J1509p1506_1046} Same as in \Figs{fig:img_and_kin:J0145p1056_0554} and \ref{fig:spec:J0145p1056_0554}, but here for the main galaxy associated with the absorber abs\_J1509p1506\_1046.}
\end{figure*}

\begin{figure*}
\includegraphics[width=\textwidth]{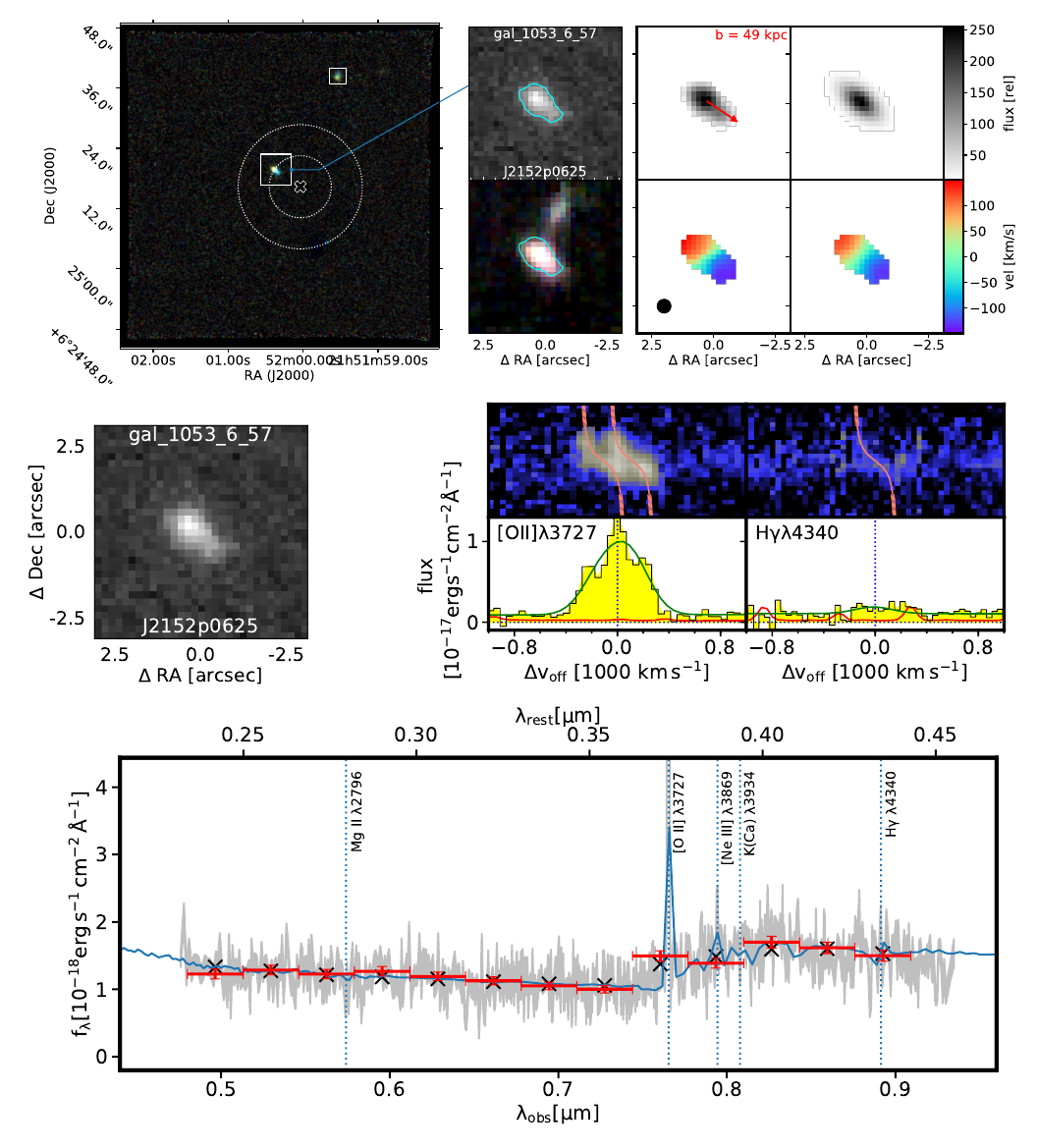}
    \caption{\label{fig:img_and_kin:J2152p0625_1053} Same as in \Figs{fig:img_and_kin:J0145p1056_0554} and \ref{fig:spec:J0145p1056_0554}, but here for the main galaxy associated with the absorber abs\_J2152p0625\_1053.}
\end{figure*}

\section{Supplementary Tables}

\begin{table*}
    \begin{tabular}{ccccccccc}
    \multicolumn{9}{c}{MUSE Observations} \\
    \hline
        Quasar & R.A. & Dec. &  T$_{\rm exp}$ & Seeing (G.) & Seeing (M.) & Date-Obs & Prog. IDs  & Refs\\
        (1) & (2) & (3) & (4) &  (5) & (6) & (7) & (8) & (9) \\
        \hline

    J0014m0028 & 00:14:53.4 & -00:28:28 & 6.3 & 0.97 & 0.78 & \makecell[t]{2015-08-23,\\2015-09-10,\\2015-10-12} & \makecell[t]{095.A-0365(A),\\096.A-0164(A)} & This work \\
    J0014p0912 & 00:14:53.2 & +09:12:18 & 10.8 & 1.06 & 0.85 & \makecell[t]{2014-10-19,\\2014-10-20,\\2014-10-24} & \makecell[t]{094.A-0211(B)} & This work \\
    J0015m0751 & 00:15:35.2 & -07:51:03 & 9.0 & 0.98 & 0.80 & \makecell[t]{2015-10-09,\\2015-10-10,\\2016-08-31} & \makecell[t]{096.A-0164(A),\\097.A-0138(A)} & This work \\
    J0058p0111 & 00:58:55.8 & +01:11:29 & 7.2 & 0.97 & 0.77 & \makecell[t]{2015-11-08,\\2016-08-29} & \makecell[t]{096.A-0164(A),\\097.A-0138(A)} & This work \\
    J0103p1332 & 01:03:32.3 & +13:32:34 & 7.2 & 1.05 & 0.84 & \makecell[t]{2015-11-11,\\2015-11-12} & \makecell[t]{096.A-0164(A)} & This work \\
    J0131p1303 & 01:31:36.4 & +13:03:31 & 7.2 & 1.04 & 0.81 & \makecell[t]{2014-10-27} & \makecell[t]{094.A-0211(B)} & This work \\
    J0134p0051 & 01:34:05.8 & +00:51:09 & 10.2 & 0.94 & 0.73 & \makecell[t]{2015-10-14,\\2015-10-15,\\2016-08-31} & \makecell[t]{096.A-0164(A),\\097.A-0138(A)} & This work \\
    J0145p1056 & 01:45:13.1 & +10:56:27 & 6.0 & 1.03 & 0.85 & \makecell[t]{2015-11-12,\\2016-08-29} & \makecell[t]{096.A-0164(A),\\097.A-0138(A)} & This work \\
    J0800p1849 & 08:00:04.6 & +18:49:35 & 7.2 & 0.73 & 0.56 & \makecell[t]{2014-12-24} & \makecell[t]{094.A-0211(B)} & This work \\
    J0838p0257 & 08:38:52.1 & +02:57:04 & 12.0 & 0.73 & 0.54 & \makecell[t]{2016-02-01,\\2016-02-02,\\2017-02-02} & \makecell[t]{096.A-0164(A),\\098.A-0216(A)} & This work \\
    J0937p0656 & 09:37:49.6 & +06:56:56 & 7.2 & 0.83 & 0.67 & \makecell[t]{2015-04-15,\\2015-04-17} & \makecell[t]{095.A-0365(A)} & This work \\
    J1039p0714 & 10:39:36.7 & +07:14:27 & 12.0 & 0.78 & 0.61 & \makecell[t]{2016-04-07,\\2016-04-08,\\2017-01-27} & \makecell[t]{097.A-0138(A),\\098.A-0216(A)} & This work \\
    J1107p1021 & 11:07:42.7 & +10:21:26 & 12.0 & 0.92 & 0.70 & \makecell[t]{2016-03-11,\\2017-01-28,\\2017-02-01} & \makecell[t]{096.A-0164(A),\\098.A-0216(A)} & This work \\
    J1107p1757 & 11:07:35.3 & +17:57:31 & 7.2 & 1.05 & 0.88 & \makecell[t]{2015-04-22,\\2015-04-23} & \makecell[t]{095.A-0365(A)} & This work \\
    J1236p0725 & 12:36:24.4 & +07:25:52 & 6.0 & 1.13 & 0.91 & \makecell[t]{2016-03-12} & \makecell[t]{096.A-0164(A)} & This work \\
    J1314p0657 & 13:14:05.6 & +06:57:22 & 6.0 & 0.68 & 0.53 & \makecell[t]{2016-04-06,\\2016-04-07} & \makecell[t]{097.A-0138(A)} & This work \\
    J1352p0614 & 13:52:17.7 & +06:14:33 & 6.0 & 1.22 & 0.98 & \makecell[t]{2017-04-22,\\2017-04-23} & \makecell[t]{099.A-0059(A)} & This work \\
    J1358p1145 & 13:58:09.5 & +11:45:58 & 6.0 & 0.72 & 0.54 & \makecell[t]{2016-04-09} & \makecell[t]{097.A-0138(A)} & This work \\
    J1425p1209 & 14:25:38.1 & +12:09:19 & 3.6 & 1.19 & 0.96 & \makecell[t]{2016-05-11} & \makecell[t]{097.A-0138(A)} & This work \\
    J1509p1506 & 15:09:00.1 & +15:06:35 & 3.0 & 0.85 & 0.70 & \makecell[t]{2017-04-22} & \makecell[t]{099.A-0059(A)} & This work \\
    J2137p0012 & 21:37:48.4 & +00:12:20 & 3.6 & 0.94 & 0.74 & \makecell[t]{2014-09-23} & \makecell[t]{094.A-0211(B)} & paper I \\
    J2152p0625 & 21:52:00.0 & +06:25:16 & 7.2 & 0.74 & 0.58 & \makecell[t]{2014-09-24} & \makecell[t]{094.A-0211(B)} & paper I \\
     \hline
    \end{tabular}
    \caption{Details of MUSE observations for the 22 MEGAFLOW fields as used in this study. 
    (1) Quasar/Field identifier;
    (2) Right ascension of the QSO [hh:mm:ss; J2000];
    (3) Declination of the QSO [dd:mm:ss; J2000];
    (4) Total MUSE exposure time [s];
    (5) Seeing FWHM  measured at $7050\text{\AA}$ by fitting a Gaussian [arcsec];
    (6) Seeing FWHM  measured at $7050\text{\AA}$ by fitting a Moffat profile with $\beta=2.5$ [arcsec];
    (7) Date of Observations;
    (8) ESO Program IDs;
    (9) Reference.
    \label{sup:tab:obs_full}}
\end{table*}

\begin{table*}
    \begin{tabular}{cccccccccc}
    \multicolumn{9}{c}{UVES Observations} \\
    \hline
        Quasar & R.A. & Dec. & $z_{\rm em}$ & T$_{\rm exp}$  & Seeing & Date-Obs & Setting & Prog. IDs  & Refs\\
        (1) & (2) & (3) & (4) &  (5) & (6) & (7) & (8) & (9) & (10) \\
        \hline
J0014m0028 & 00:14:53.4 & -00:28:28 & 1.92 & \makecell[t]{9015} & \makecell[t]{0.8} & \makecell[t]{2015-10-03} & \makecell[t]{HER\_5 \& SHP700} & \makecell[t]{096.A-0609(A)} & This work \\
J0014p0912 & 00:14:53.2 & +09:12:18 & 2.34 & \makecell[t]{6010,\\1483} & \makecell[t]{0.8,\\0.5} & \makecell[t]{2015-11-09,\\2016-10-28} & \makecell[t]{HER\_5 \& BK7\_5,\\HER\_5 \& SHP700} & \makecell[t]{096.A-0609(A),\\098.A-0310(A)} & This work \\
J0015m0751 & 00:15:35.2 & -07:51:03 & 0.87 & \makecell[t]{12020} & \makecell[t]{0.7} & \makecell[t]{2016-10-29,\\2016-12-27,\\2016-12-28} & \makecell[t]{HER\_5 \& SHP700} & \makecell[t]{098.A-0310(A)} & This work \\
J0058p0111 & 00:58:55.8 & +01:11:29 & 1.22 & \makecell[t]{1483,\\1483} & \makecell[t]{0.9,\\--$^{a}$} & \makecell[t]{2016-12-29,\\2016-12-30} & \makecell[t]{HER\_5 \& BK7\_5,\\HER\_5 \& SHP700} & \makecell[t]{098.A-0310(A)} & This work \\
J0103p1332 & 01:03:32.3 & +13:32:34 & 1.66 & \makecell[t]{9015} & \makecell[t]{0.6} & \makecell[t]{2016-10-28,\\2016-10-29,\\2016-11-01} & \makecell[t]{HER\_5 \& SHP700} & \makecell[t]{098.A-0310(A)} & This work \\
J0131p1303 & 01:31:36.4 & +13:03:31 & 1.59 & \makecell[t]{6010} & \makecell[t]{1.0} & \makecell[t]{2015-10-14} & \makecell[t]{HER\_5 \& SHP700} & \makecell[t]{096.A-0609(A)} & This work \\
J0134p0051 & 01:34:05.8 & +00:51:09 & 1.52 & \makecell[t]{5710,\\1483} & \makecell[t]{0.6,\\0.4} & \makecell[t]{2016-10-29,\\2016-12-03} & \makecell[t]{HER\_5 \& BK7\_5,\\HER\_5 \& SHP700} & \makecell[t]{098.A-0310(A)} & This work \\
J0145p1056 & 01:45:13.1 & +10:56:27 & 0.94 & \makecell[t]{12020} & \makecell[t]{0.6} & \makecell[t]{2015-11-11,\\2016-09-03,\\2016-10-28} & \makecell[t]{HER\_5 \& SHP700} & \makecell[t]{096.A-0609(A),\\097.A-0144(A),\\098.A-0310(A)} & This work \\
J0800p1849 & 08:00:04.6 & +18:49:35 & 1.29 & \makecell[t]{6010} & \makecell[t]{0.9} & \makecell[t]{2015-12-10} & \makecell[t]{SHP700} & \makecell[t]{096.A-0609(A)} & This work \\
J0838p0257 & 08:38:52.1 & +02:57:04 & 1.77 & \makecell[t]{1483,\\1483} & \makecell[t]{0.6,\\1.0} & \makecell[t]{2015-11-20,\\2016-12-22} & \makecell[t]{HER\_5 \& SHP700,\\SHP700} & \makecell[t]{096.A-0609(A),\\098.A-0310(A)} & This work \\
J0937p0656 & 09:37:49.6 & +06:56:56 & 1.82 & \makecell[t]{9015} & \makecell[t]{0.7} & \makecell[t]{2015-12-20,\\2016-01-11,\\2016-03-07} & \makecell[t]{HER\_5 \& SHP700} & \makecell[t]{096.A-0609(A)} & This work \\
J1039p0714 & 10:39:36.7 & +07:14:27 & 1.53 & \makecell[t]{9015} & \makecell[t]{0.8} & \makecell[t]{2016-04-03} & \makecell[t]{HER\_5 \& SHP700} & \makecell[t]{097.A-0144(A)} & This work \\
J1107p1021 & 11:07:42.7 & +10:21:26 & 1.92 & \makecell[t]{6010} & \makecell[t]{1.0} & \makecell[t]{2016-02-09,\\2016-03-07} & \makecell[t]{HER\_5 \& SHP700} & \makecell[t]{096.A-0609(A)} & This work \\
J1107p1757 & 11:07:35.3 & +17:57:31 & 2.15 & \makecell[t]{9015} & \makecell[t]{1.0} & \makecell[t]{2016-01-11,\\2016-03-06,\\2016-03-07} & \makecell[t]{HER\_5 \& BK7\_5} & \makecell[t]{096.A-0609(A)} & This work \\
J1236p0725 & 12:36:24.4 & +07:25:52 & 1.61 & \makecell[t]{6010,\\1483} & \makecell[t]{0.4,\\0.9} & \makecell[t]{2016-03-06,\\2016-04-06} & \makecell[t]{HER\_5 \& BK7\_5,\\SHP700} & \makecell[t]{096.A-0609(A),\\097.A-0144(A)} & This work \\
J1314p0657 & 13:14:05.6 & +06:57:22 & 1.88 & \makecell[t]{1483} & \makecell[t]{0.4} & \makecell[t]{2016-04-06} & \makecell[t]{HER\_5 \& SHP700} & \makecell[t]{097.A-0144(A)} & This work \\
J1352p0614 & 13:52:17.7 & +06:14:33 & 1.80 & \makecell[t]{1483} & \makecell[t]{0.7} & \makecell[t]{2016-05-31} & \makecell[t]{HER\_5 \& BK7\_5} & \makecell[t]{097.A-0144(A)} & This work \\
J1358p1145 & 13:58:09.5 & +11:45:58 & 1.48 & \makecell[t]{1483,\\1483} & \makecell[t]{0.5,\\0.5} & \makecell[t]{2016-04-06} & \makecell[t]{HER\_5 \& OG590,\\HER\_5 \& SHP700} & \makecell[t]{097.A-0144(A)} & This work \\
J1425p1209 & 14:25:38.1 & +12:09:19 & 1.62 & \makecell[t]{1483,\\1483} & \makecell[t]{0.6,\\0.5} & \makecell[t]{2016-04-06,\\2016-05-31} & \makecell[t]{HER\_5 \& SHP700,\\SHP700} & \makecell[t]{097.A-0144(A)} & This work \\
J1509p1506 & 15:09:00.1 & +15:06:35 & 2.24 & \makecell[t]{6010} & \makecell[t]{0.6} & \makecell[t]{2016-04-03,\\2016-04-06} & \makecell[t]{SHP700} & \makecell[t]{097.A-0144(A)} & This work \\
J2137p0012 & 21:37:48.4 & +00:12:20 & 1.67 & \makecell[t]{4487} & \makecell[t]{1.0} & \makecell[t]{2014-10-18} & \makecell[t]{HER\_5 \& SHP700} & \makecell[t]{293.A-5038(A)} & Paper~I \\
J2152p0625 & 21:52:00.0 & +06:25:16 & 2.38 & \makecell[t]{9015} & \makecell[t]{1.2} & \makecell[t]{2014-10-20,\\2014-10-23,\\2014-11-17} & \makecell[t]{HER\_5 \& SHP700} & \makecell[t]{293.A-5038(A)} & Paper~1 \\
     \hline
    \end{tabular}
    \caption{Details of UVES observations for the 22 \mfl{} Quasars.
        (1) Quasar identifier;
        (2) Right ascension of QSO [hh:mm:ss; J2000];
        (3) Declination of QSO [dd:mm:ss; J2000];
        (4) Emission redshift of the QSO;
        (5) Total UVES exposure time split into settings [s];
        (6) Seeing FWHM measured by DIMM split into settings [arcsec] ;  
        (7) Date of Observations;
        (8) UVES settings
        (9) ESO Program IDs;
        Comments: a) For part of the observations DIMM measurements unavailable.
    \label{sup:tab:uvesobs_full}}
\end{table*}

\begin{table}
\centering
\begin{tabular}{cccccc}
\hline
    & Gal ID & $z$ & $\Delta v\;\textbf{I}$  &  $\Delta v\; \textbf{II}$ & $\Delta v \; \textbf{III}$ \\
      &  (1)    & (2) & (3) & (4) & (5) \\
\hline
\hline
\cellcolor{GalColJ0103} & J0103 & 0.7882 & 5 & -5 & -7 \\
\cellcolor{GalColJ0145} & J0145 & 0.5500 & 40 & 53 & 7 \\
\cellcolor{GalColJ0800} & J0800 & 0.6082 & 11 & 26 & 7 \\
\cellcolor{GalColJ1039} & J1039 & 0.9494 & 14 &  & 0 \\
\cellcolor{GalColJ1107} & J1107 & 1.0481 & -7 &  & -5 \\
\cellcolor{GalColJ1236} & J1236 & 0.9128 & 14 &  & -21 \\
\cellcolor{GalColJ1358} & J1358 & 1.4171 & 2 &  & -1 \\
\cellcolor{GalColJ1509} & J1509 & 1.0469 & -7 &  & 22 \\
\cellcolor{GalColJ2152} & J2152 & 1.0530 & -19 &  & -12 \\
\hline
\end{tabular}
\caption{\label{sup:tab:redshift_comparison} Comparison of different redshift measurements:
    (2): $z$ is the redshift obtained with \gpk on the \OII{} doublet.
    (3)  $\Delta v \; \textbf{I}$ is the velocity difference between the redshift obtained by visual inspection of the \OII{} PVD and $z$.
    (4) $\Delta v\;\textbf{II}$ is the same as for  $\Delta v\;\textbf{I}$, but here based on \OIII{} $\lambda5007$. MUSE covers \OIII{} only for the three lowest redshift galaxies.
    (5) $\Delta v\;\textbf{III}$ is the velocity difference between the redshift obtained by the 1D line fit and the \gpk{} redshift ($z$). (3\mbox{--}5) are in $\kms$.
}
\end{table}

\begin{table*}
\begin{tabular}{crrrrrrrr}
\hline
ID &  $\phi$ & $\Rgal$ & $\Rgal/\rvir$ & $\vrot/\vvir$  & $\vrot/\vvir$  & $\vrot/\vvir$  & {\small $\log_{10}(\cos(i) N_\mathrm{HI})$}& $\dot{M}_\mathrm{in}(R)$ \\
   &  &  &  & (no infall) & (infall same)  & (infall opposite) &  &   \\
 (1) & (2) & (3) & (4) & (5) & (6) & (7) & (8) & (9)   \\
 \hline
\cellcolor{GalColJ0103}J0103 & $17_{-11}^{+30}$ & $21_{-1}^{+8}$ & $0.16_{-0.03}^{+0.10}$ & $0.46_{-0.04}^{+0.24}$ & $0.27_{-0.27}^{+0.13}$ & $0.65_{-0.12}^{+0.68}$ & $18.9{\scriptstyle \pm0.5}$ & $0.9_{-0.6}^{+2.0}$ \\
\cellcolor{GalColJ0145}J0145 & $22{\scriptstyle \pm9}$ & $23{\scriptstyle \pm1}$ & $0.12_{-0.04}^{+0.06}$ & $1.17_{-0.35}^{+0.65}$ & $0.93_{-0.36}^{+0.65}$ & $1.41_{-0.36}^{+0.69}$ & $18.6{\scriptstyle \pm0.5}$ & $0.7_{-0.5}^{+1.5}$ \\
\cellcolor{GalColJ0800}J0800 & $22_{-14}^{+21}$ & $70_{-4}^{+17}$ & $0.59_{-0.19}^{+0.37}$ & $0.27_{-0.06}^{+0.14}$ & $0.03_{-0.03}^{+0.17}$ & $0.51_{-0.18}^{+0.42}$ & $18.7{\scriptstyle \pm0.5}$ & $1.8_{-1.3}^{+4.0}$ \\
\cellcolor{GalColJ1039}J1039 & $10_{-7}^{+15}$ & $49_{-1}^{+3}$ & $0.36_{-0.10}^{+0.14}$ & $1.16_{-0.29}^{+0.46}$ & $1.05_{-0.34}^{+0.40}$ & $1.27_{-0.27}^{+0.55}$ & $18.9{\scriptstyle \pm0.5}$ & $2.7_{-2.0}^{+6.0}$ \\
\cellcolor{GalColJ1107}J1107 & $69_{-13}^{+12}$ & $100_{-34}^{+123}$ & $0.66_{-0.35}^{+1.27}$ & $-0.98_{-1.40}^{+0.39}$ & \mbox{--} & $0.56_{-0.46}^{+0.98}$ & $17.9{\scriptstyle \pm0.5}$ & $0.7_{-0.6}^{+1.8}$ \\
\cellcolor{GalColJ1236}J1236 & $39_{-11}^{+12}$ & $20_{-2}^{+3}$ & $0.10_{-0.03}^{+0.05}$ & $0.29_{-0.08}^{+0.13}$ & \mbox{--} & $0.77_{-0.20}^{+0.34}$ & $19.8{\scriptstyle \pm0.5}$ & $12.5_{-9.1}^{+27.5}$ \\
\cellcolor{GalColJ1358}J1358 & $22_{-14}^{+16}$ & $32_{-2}^{+4}$ & $0.30_{-0.06}^{+0.09}$ & $0.48_{-0.07}^{+0.13}$ & $0.24_{-0.16}^{+0.15}$ & $0.72_{-0.18}^{+0.32}$ & $19.9{\scriptstyle \pm0.5}$ & $17.7_{-12.3}^{+38.4}$ \\
\cellcolor{GalColJ1509}J1509 & $79_{-11}^{+7}$ & $56_{-27}^{+109}$ & $0.52_{-0.33}^{+1.52}$ & $4.97_{-2.58}^{+10.97}$ & $1.92_{-1.40}^{+4.48}$ & $8.02_{-4.08}^{+17.39}$ & $18.8{\scriptstyle \pm0.5}$ & $2.2_{-1.9}^{+6.5}$ \\
\cellcolor{GalColJ2152}J2152 & $19_{-12}^{+26}$ & $52_{-2}^{+16}$ & $0.36_{-0.11}^{+0.26}$ & $-0.43_{-0.26}^{+0.10}$ & \mbox{--} & $0.00_{-0.00}^{+0.06}$ & $18.4{\scriptstyle \pm0.5}$ & $1.0_{-0.7}^{+2.1}$ \\
\hline
\end{tabular}
\caption{\label{sup:tab:geo_and_kin_absgas} Geometrical and kinematical
    constraints for the absorbing gas inferred under the assumption of an
    extended gas disk. (2) Position angle of quasar sight-line in disk plane as
    defined in \Eq{eq:results:phi} [deg]; (3) Galacto-centric radius of quasar
    sight-line as defined in \Eq{eq:results:R} [kpc]; (4) $\Rgal$ normalised by
    $\rvir$; (5) Tangential velocity (normalised by $\vvir$) as shown in
    \Fig{fig:results:vrot_simple} estimated under the assumption of $\vr=0$
    (see \Eq{eq:vlos_simplerot}); (6 \& 7) Tangential velocity (normalised by
    $\vvir$) as shown in \Fig{fig:results:vrot_vr} estimated under the
    assumption of $\vr=-0.6\vvir$ (see \Eq{eq:vlos_radial}).  Depending on the
    disk's unknown sign of inclination the signs of the projected components of
    $\vrot$ and $\vr$ can be the same or the opposite, leading to the two
    solutions 6) and 7) for $\vrot{}$, respectively. Due to the requirement of
    co-rotation with the galaxy disk, it is possible that no solution can be
    found (indicated by "$\mbox{--}$"). Error intervals that are consistent
    with zero indicate that only part of the uncertainty interval is consistent
    with co-rotation; (8) \HI{} column density perpendicular to disk (based on
\MgII{} equivalent width) [$cm^{-2}$]; (9) Mass accretion rate [$\mpy$].  } 
\end{table*}

\section{Supplementary Figures}

\begin{figure}
	\includegraphics[width=\columnwidth]{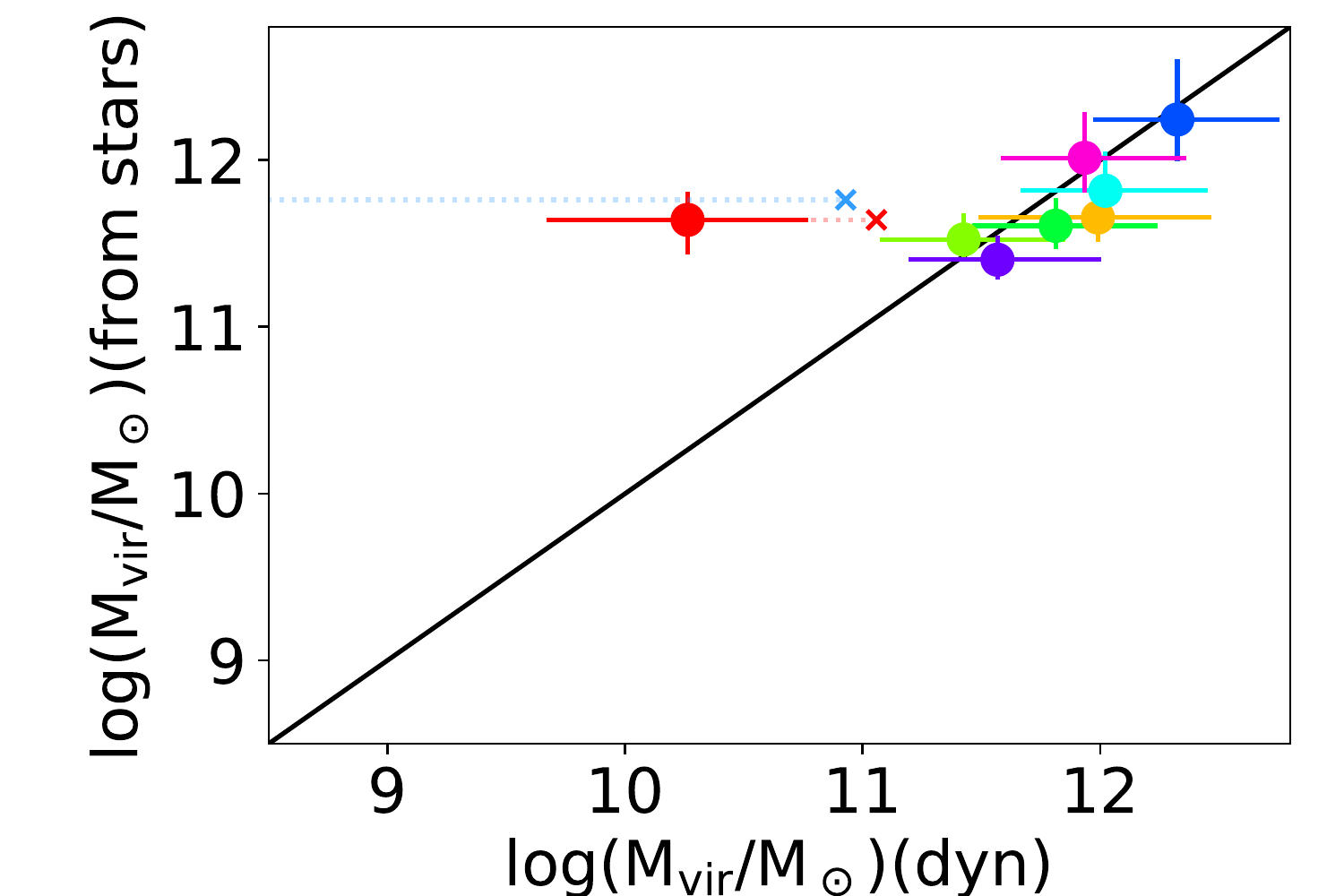}
    \caption{\label{suppl:fig:galprob:mvir_diff_est} Comparison between halo mass
        estimates ($M_\mathrm{vir}$) from two different methods plotted against
        each other. $M_\mathrm{vir}(\text{from stars})$ was obtained from the
        stellar mass - halo mass relationship, while $M_\mathrm{vir}(dyn)$ was
        determined based on the dynamics of the galaxies. For two of the
        galaxies, \emph{J0103} (red) and \emph{J1358} (light blue), the latter is plotted
        based on two different estimates of the virial velocity, once taking
        account for the pressure support in the galaxy dynamics (little cross)
        and once not taking this into account (full circle; for \emph{J1358} the circle is outside of the plotted range.).
}
\end{figure}

\begin{figure*}
\begin{tabular}{cc}
abs\_J0103p1332\_0788 & abs\_J0145p1056\_0554\\
\includegraphics[width=0.5\textwidth]{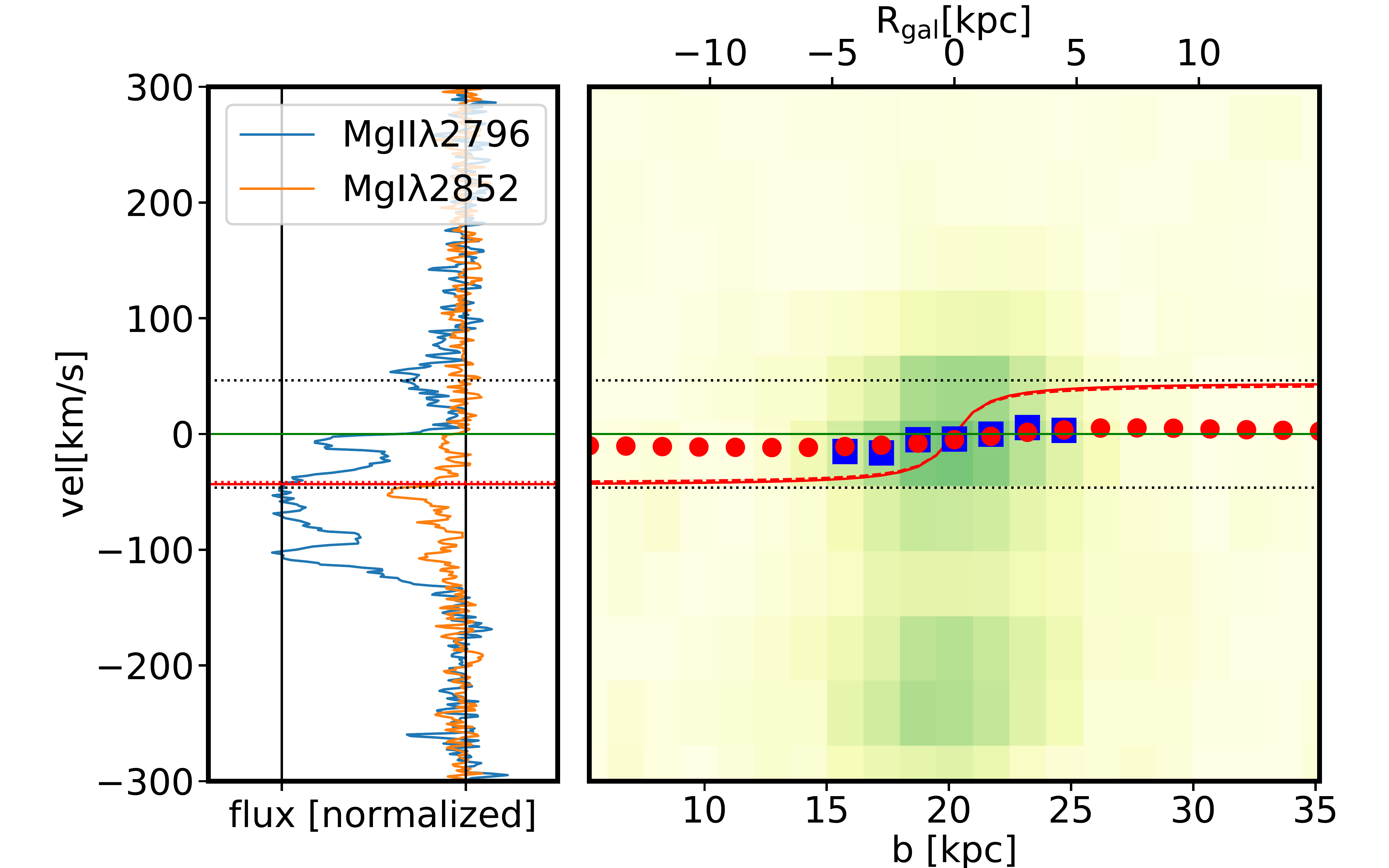} & \includegraphics[width=0.5\textwidth]{abs_J0145p1056_0554_uves_kinematics_type2.pdf} \\
abs\_J0800p1849\_0608 & abs\_J1039p0714\_0949\\
\includegraphics[width=0.5\textwidth]{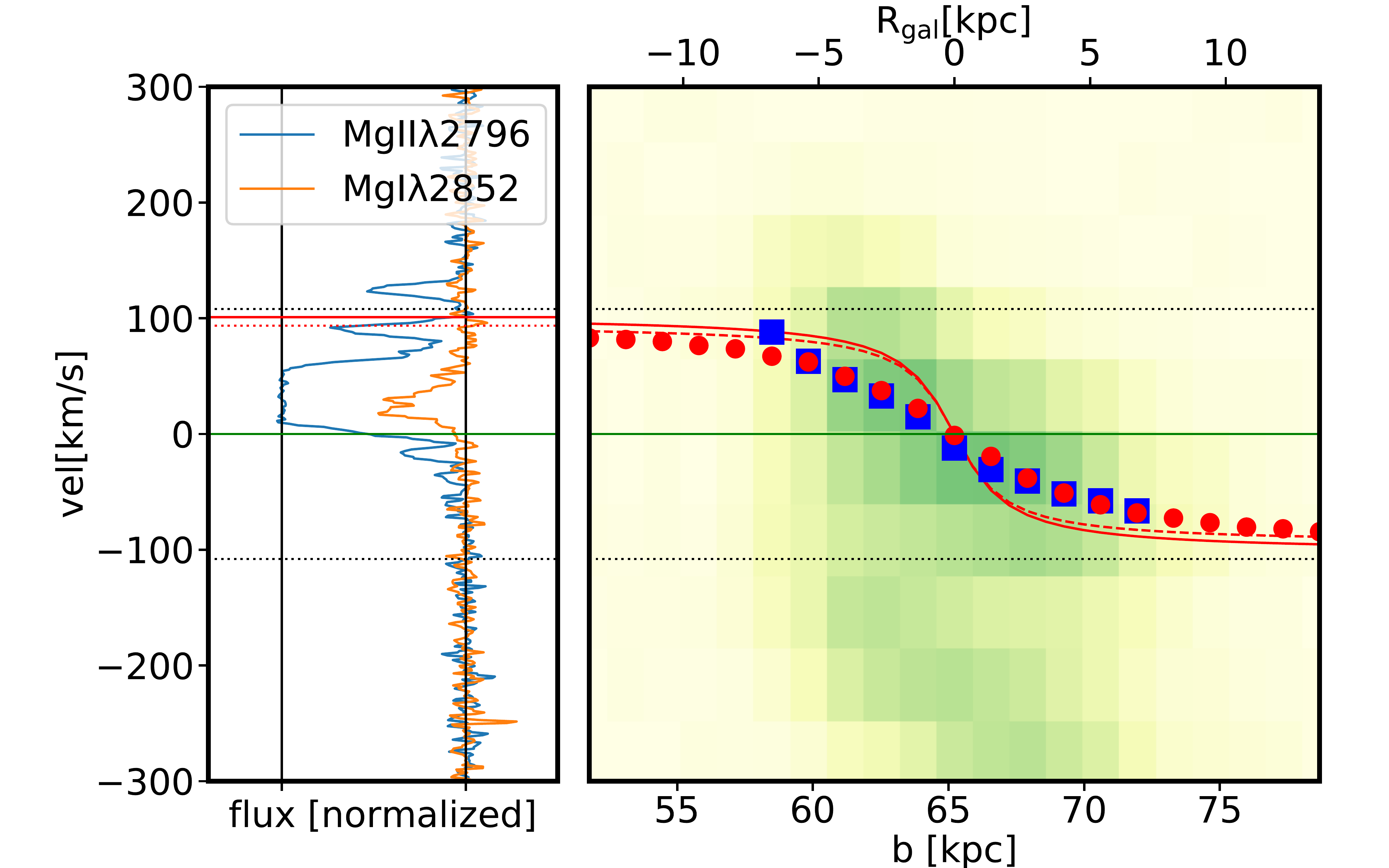} & \includegraphics[width=0.5\textwidth]{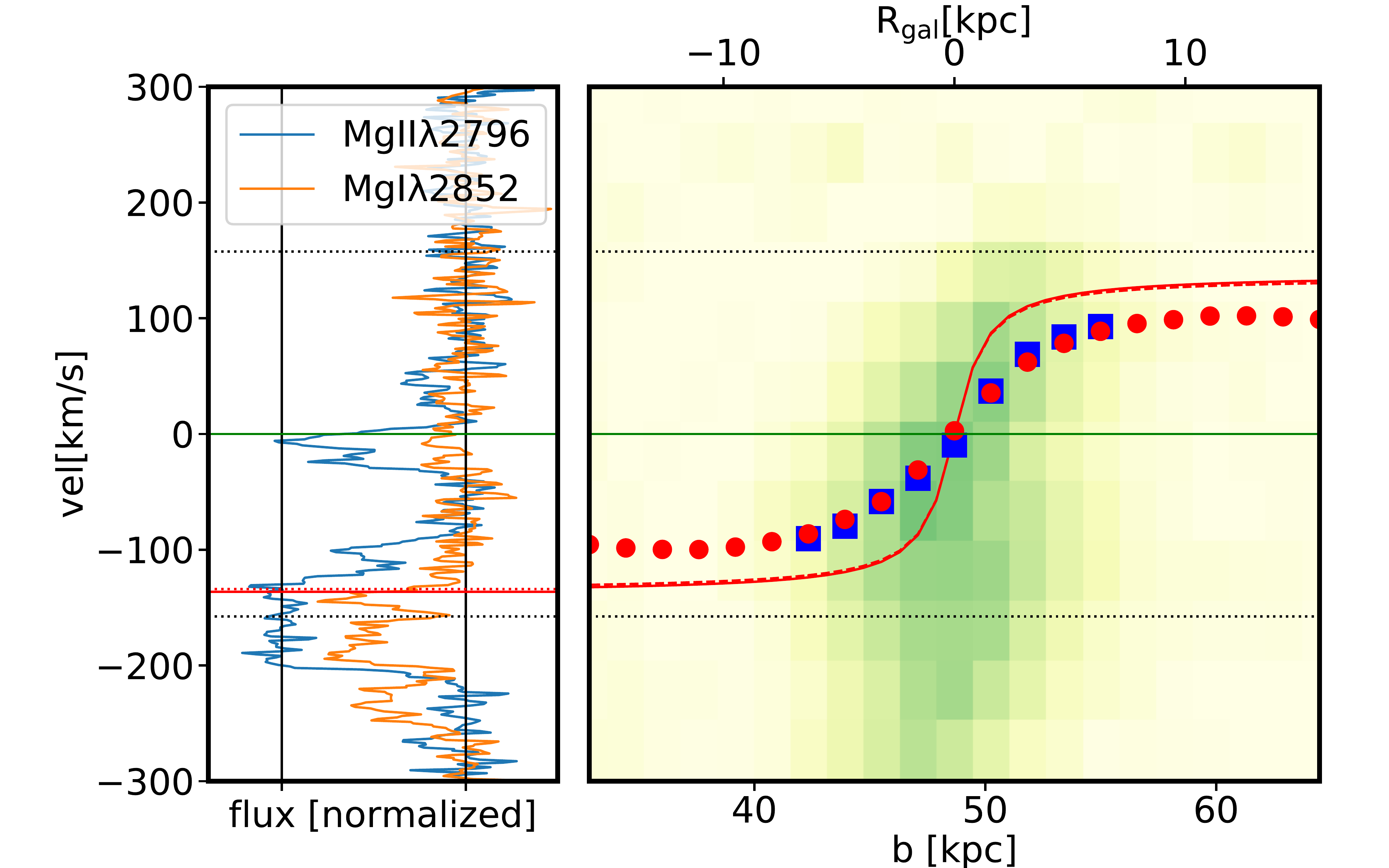} \\
abs\_J1107p1021\_1048 & abs\_J1236p0725\_0912\\
\includegraphics[width=0.5\textwidth]{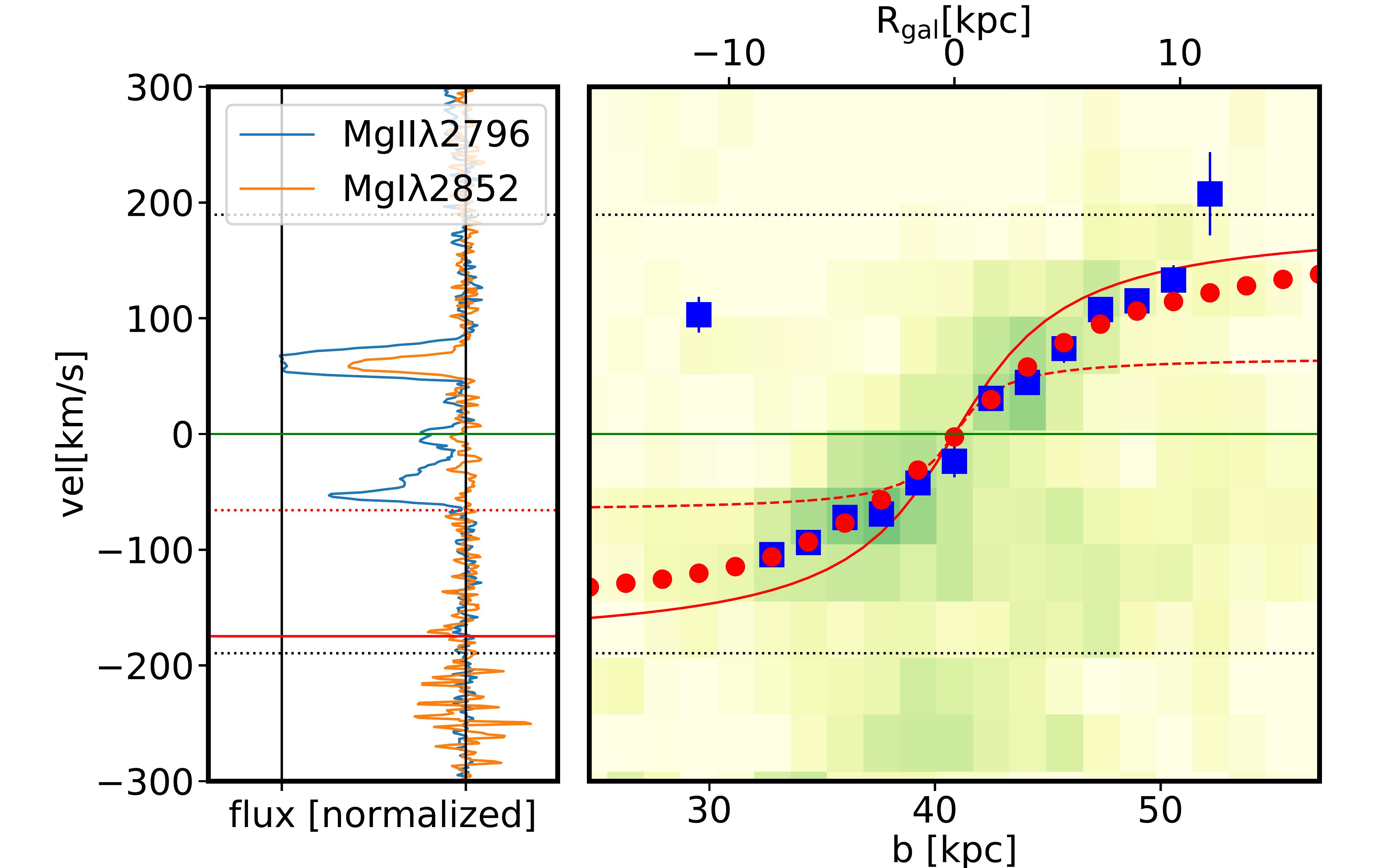} & \includegraphics[width=0.5\textwidth]{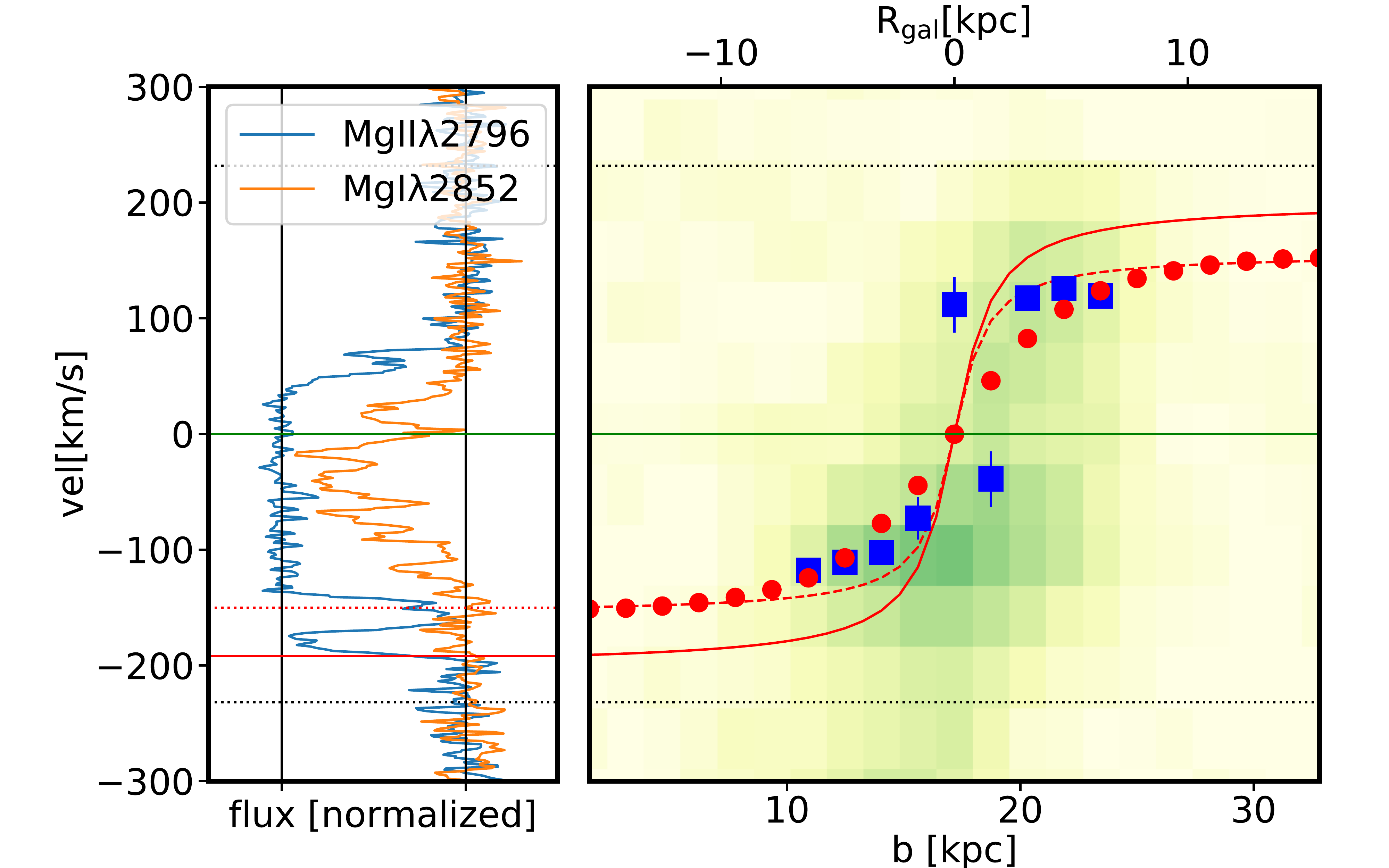} \\

\end{tabular}
\caption{\label{fig:all_rotcurve_vs_abs}
    Comparison of galaxy and absorber kinematics for each of the \nsampletext{}
    galaxy-absorber pairs. The \textbf{right panel} shows the 1D galaxy
    rotation curve (blue points) obtained from the 2D PVD diagram (shown as
    background image) on the \OII{} doublet (see
    \Sec{sec:comparison_gal_abs_kin}). Among the blue points are a few obvious
    outliers where the automatic doublet fitting algorithm failed. The red
    points are obtained by reproducing this measurement procedure on the seeing
    convolved best-fit \gpk{} model.  The solid red line represents the
    intrinsic  \gpk{} rotation curve along the galaxy major-axis.  The dashed
    red line represents the modeled rotation curve along the line connecting
    the galaxy and quasar positions on the sky.  The lower x-axis represents
    the distance $b$ from the quasar along this connecting line.  The upper
    x-axis  shows the galacto-centric distance along the galaxy's major axis.
    In the \textbf{left panel} the \MgII{}$\,\lambda2796$ and
    \MgI{}$\,\lambda2852$ absorption profiles are shown on the same velocity
    scale as the galaxy rotation curve.  The solid red line in this panel
    indicates $V_\mathrm{max}$ at the observed inclination, which is a
    continuation of the red curve in the right panel.  Similarly, the red
    dashed line is the continuation of the rotation curve along the
galaxy-quasar axis. Further, the black dotted line shows $\vmax$ at
$incl=90^\circ$ and the green line is the systemic redshift as obtained from
\gpk{} ($v=0\,\kms$).  } 
\end{figure*}

\begin{figure*}
\begin{tabular}{cc}
abs\_J1358p1145\_1418 &  abs\_J1509p1506\_1046 \\
 \includegraphics[width=0.5\textwidth]{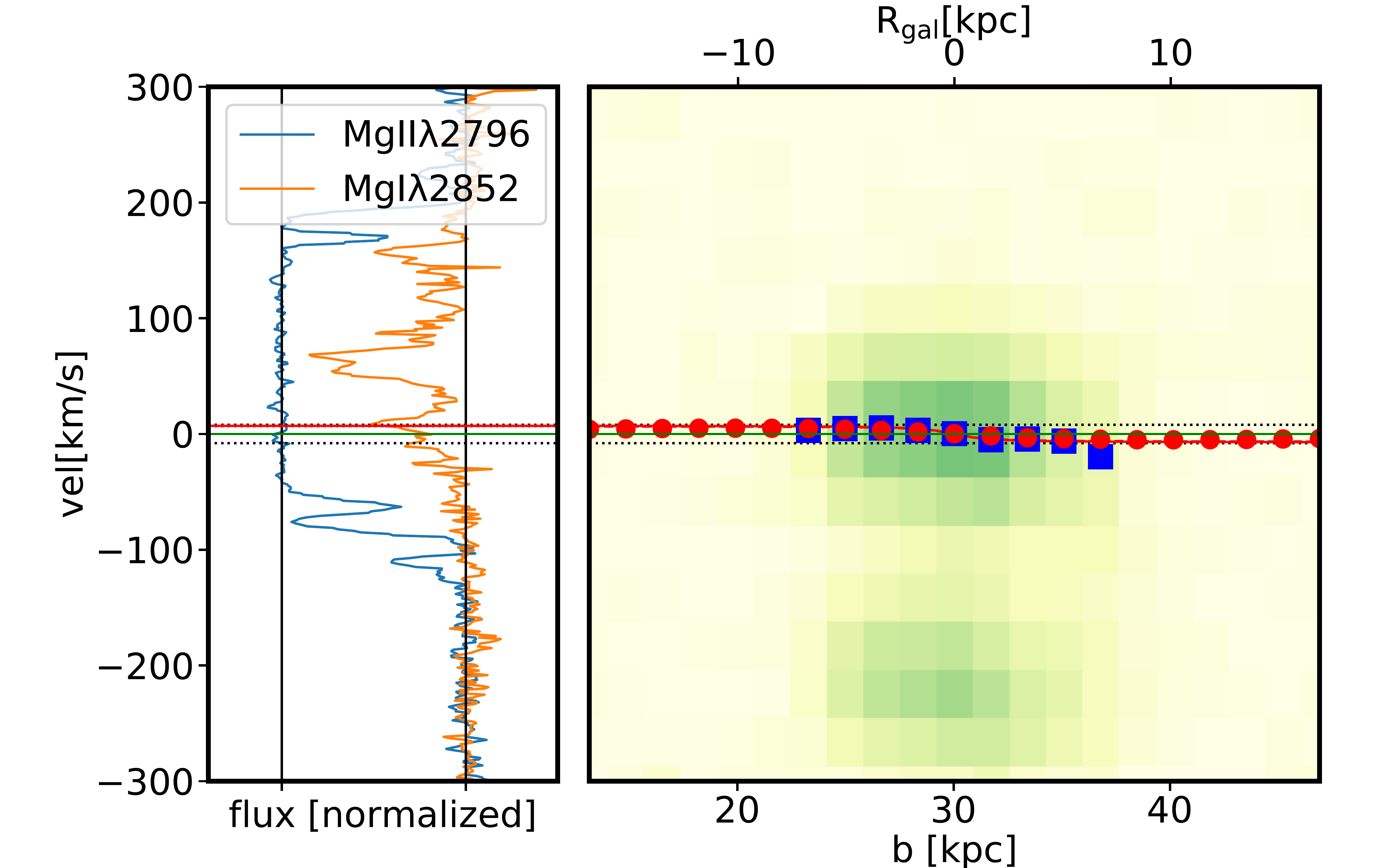} & \includegraphics[width=0.5\textwidth]{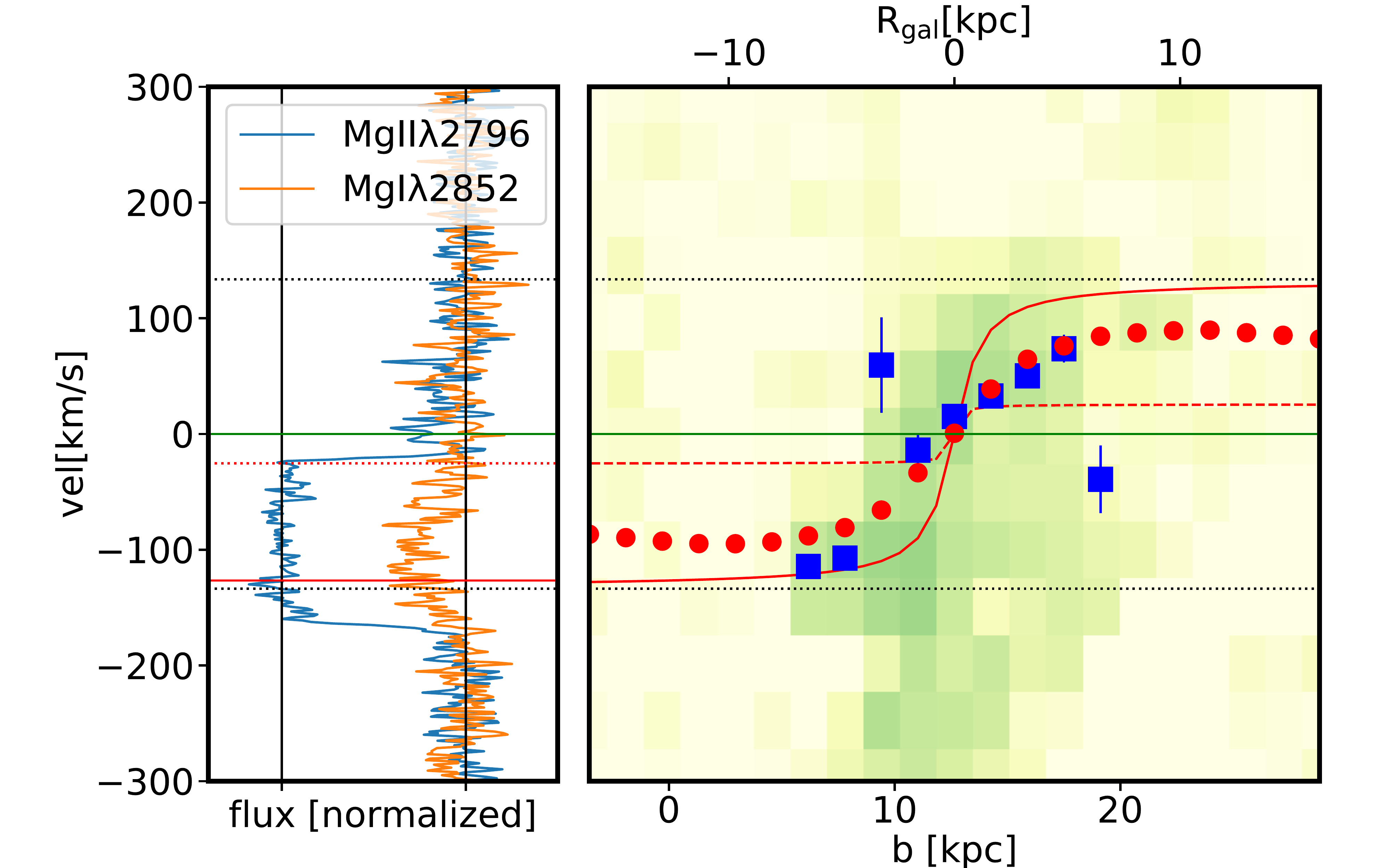}\\
abs\_J2152p0625\_1053\\
\includegraphics[width=0.5\textwidth]{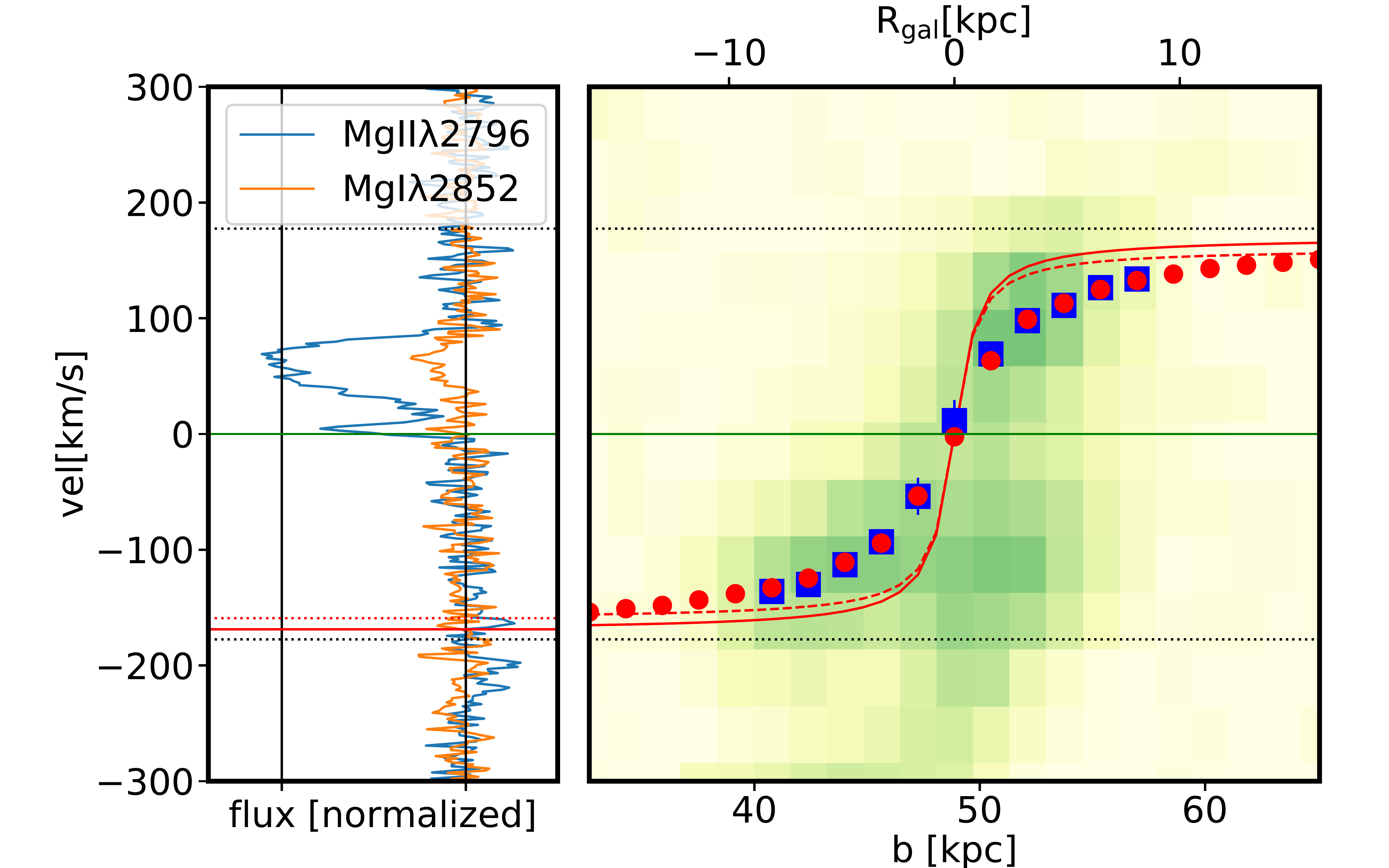} \\
\end{tabular}
\contcaption{}
\end{figure*}

\begin{figure}
    \includegraphics[width=\columnwidth]{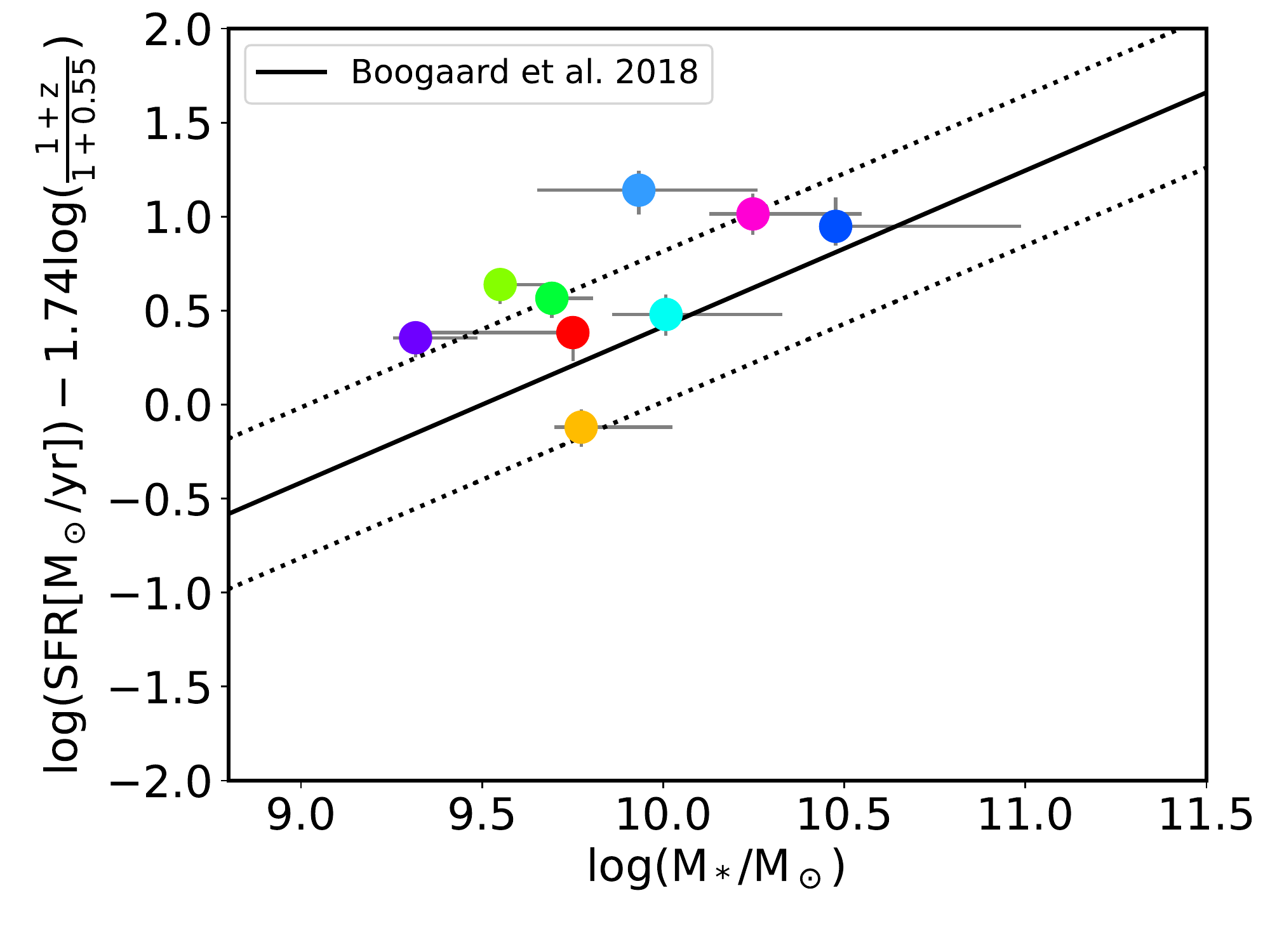}
    \caption{\label{fig:MS} Position of our \nsampletext{} galaxies compared to the main-sequence of star formation (MS). The shown MS (solid line) and its $1\sigma$ scatter (dotted lines) is from \citet{Boogaard:2018a}. Redshift evolution of the MS ($\mathrm{SFR}\propto(1+z)^{1.74}$) is removed from the plot through the choice of y-axis, which normalises all $\mathrm{SFR}$'s to $z=0.55$. }
\end{figure}


\bsp	
\label{lastpage}
\end{document}